\documentclass[11pt,a4paper]{article}

\usepackage{amsmath}
\numberwithin{equation}{section} 
\newcommand{\sect}[1]{\section{#1}}

\pdfoutput=1 
\usepackage[colorlinks=true, linkcolor=black!50!blue, urlcolor=blue, citecolor=blue, anchorcolor=blue]{hyperref}
\usepackage[font=small,labelfont=bf,margin=0mm,labelsep=period,tableposition=top]{caption}
\usepackage[a4paper,top=3cm,bottom=2.5cm,left=2.5cm,right=2.5cm,bindingoffset=0mm]{geometry}
\usepackage[T1]{fontenc} 

\usepackage[utf8]{inputenc}
\usepackage[english]{babel}

\usepackage{pifont}
\pdfoutput=1
\usepackage{graphicx,mathtools}
\usepackage{epsfig,cite}
\usepackage{amssymb}
\usepackage{color}
\usepackage{amstext,alltt,setspace}
\usepackage{amsbsy}
\usepackage{array}
\usepackage{booktabs, makecell}
\usepackage{hyperref}
\usepackage{slashed}
\usepackage{url}
\usepackage{geometry}
\usepackage{amssymb,amsthm,cancel,hyperref,graphicx,xcolor}
\usepackage{picinpar,graphicx,xy}
\usepackage[subnum]{cases}
\usepackage{booktabs}
\usepackage{mathrsfs}
\usepackage{amsfonts}
\usepackage{latexsym}
\usepackage{booktabs,graphicx,hyperref,epsfig,multirow}
\usepackage{bm}
\usepackage{comment}
\def\bea{\begin{eqnarray}}
\def\eea{\end{eqnarray}}
\def\beq{\begin{equation}}
\def\eeq{\end{equation}}
\def\ba{\begin{eqnarray}}
\def\ea{\end{eqnarray}}
\def\be{\begin{equation}}
\def\ee{\end{equation}}
\definecolor{darkgreen}{HTML}{008000}
\newcommand{\sss}{\scriptscriptstyle\rm}

\renewcommand{\Re}{\mathrm{Re}}

\newcommand{\abs}[1]{\left|\,#1\,\right|}

\newcommand{\as}{\alpha_s}

\def\({\left(}
\def\){\right)}
\def\[{\left[}
\def\]{\right]}
\def    \hepph  #1 {{\tt hep-ph/#1}}
\def    \hepex  #1 {{\tt hep-ex/#1}}
\long\def\symbolfootnote[#1]#2{\begingroup%
\def\thefootnote{\fnsymbol{footnote}}\footnote[#1]{#2}\endgroup}

\def\lapprox{\lower .7ex\hbox{$\;\stackrel{\textstyle <}{\sim}\;$}}
\def\gapprox{\lower .7ex\hbox{$\;\stackrel{\textstyle >}{\sim}\;$}}

\renewcommand{\(}{\left(}
\renewcommand{\)}{\right)}
\renewcommand{\Re}{\mathrm{Re}\:}

\newcommand{\Ca}{C_{\rm\sss A}}
\newcommand{\Cf}{C_{\rm\sss F}}

\graphicspath{{Images/}}
\usepackage{physics}
\allowdisplaybreaks

\usepackage{xspace} 
\usepackage{enumitem}

\usepackage{dowith}

\newcommand{\UpperCase}[1]{
  \expandafter\newcommand\csname bb#1\endcsname{{\mathbb{#1}}}
  \expandafter\newcommand\csname cal#1\endcsname{{\mathcal{#1}}}   
  \expandafter\newcommand\csname rm#1\endcsname{{\mathrm{#1}}}
  \expandafter\newcommand\csname bf#1\endcsname{{\mathbf{#1}}}
  \expandafter\newcommand\csname bold#1\endcsname{{\boldsymbol{#1}}}
  \expandafter\newcommand\csname hat#1\endcsname{\hat{#1}}
  \expandafter\newcommand\csname tilde#1\endcsname{\widetilde{#1}}
  \expandafter\newcommand\csname bar#1\endcsname{\overline{#1}}
  \expandafter\newcommand\csname frak#1\endcsname{\mathfrak{#1}}
  }
\DoWith\UpperCase ABCDEFGHILJKMNOPQRSTUVWXYZ\StopDoing

\newcommand{\LowerCase}[1]{
  \expandafter\newcommand\csname rm#1\endcsname{{\mathrm{#1}}} 
  \expandafter\newcommand\csname bf#1\endcsname{{\mathbf{#1}}} 
  \expandafter\newcommand\csname bold#1\endcsname{{\boldsymbol{#1}}}
  \expandafter\newcommand\csname hat#1\endcsname{\hat{#1}}
  \expandafter\newcommand\csname tilde#1\endcsname{\tilde{#1}}
  \expandafter\newcommand\csname bar#1\endcsname{\bar{#1}}
  \expandafter\newcommand\csname frak#1\endcsname{\mathfrak{#1}}
  }
\DoWith\LowerCase abcdefghijklmnopqrstuvwxyz\StopDoing

\renewcommand{\rmd}{d}
\newcommand{\dk}{\rmd k}
\newcommand{\dpzero}{\rmd p_0}
\newcommand{\dpz}{\rmd p_z}
\newcommand{\dpT}{\rmd p_\rmT}

\newcommand{\dt}{\rmd t}

\newcommand{\dw}{\rmd w}
\newcommand{\dx}{\rmd x}

\newcommand{\dz}{\rmd z}

\newcommand{\dalpha}{\rmd\alpha}

\newcommand{\dmu}{\rmd \mu}
\newcommand{\dphi}{\rmd \phi}
\newcommand{\dOmega}{\rmd \Omega}

\newcommand{\muF}{\mu}
\newcommand{\muH}{\mu_\rmH}

\newcommand{\muS}{\mu_\rmS}

\newcommand{\Q}{M}

\newcommand{\amu}[1]{\frac{\as(#1)}{4\pi}}

\newcommand{\F}{\widetilde{\calF}}
\newcommand{\I}{\widetilde{\calI}}
\newcommand{\Ileft}{\tildeI^{\,(1)}}
\newcommand{\T}{\widetilde{\calT}}
\newcommand{\PAP}{P^{(0)}}

\newcommand{\sproc}{\tildes}
\newcommand{\sDY}{\tildes_{\rm DY}}

\newcommand{\tildebeta}{2\beta}

\newcommand{\gZeroscet}{g_0^{\rm scet}}

\newcommand{\Fscet}{F^{\rm scet}}

\newcommand{\xa}{z_1}
\newcommand{\xb}{z_2}

\newcommand{\gammaE}{{\gamma_\rmE}}

\newcommand{\nf}{n_\rmf}
\newcommand{\TR}{T_\rmR}

\newcommand{\vecpT}{\vec{p}_\rmT}

\newcommand{\barpT}{\barp_\rmT}

\newcommand{\kT}{k_\rmT}
\newcommand{\pT}{p_\rmT}
\newcommand{\qb}{\barq}

\newcommand{\cusp}{{\rm cusp}}

\newcommand{\eq}{Eq.~}

\newcommand{\pdf}{{\rm pdf}}

\begin{document}
\begin{flushleft}
\begin{figure}[h]
\includegraphics[width=.2\textwidth]{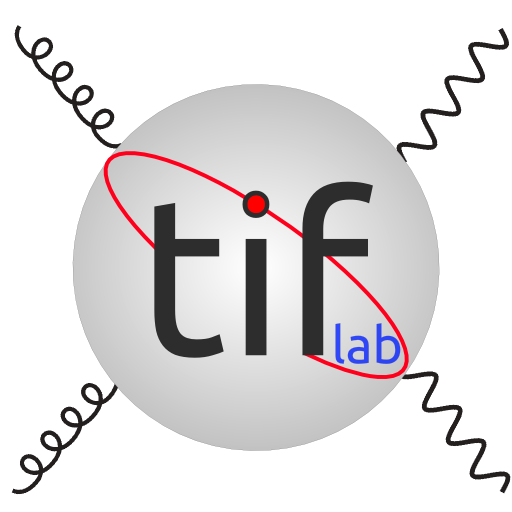}
\end{figure}
\end{flushleft}
\vspace{-5.0cm}
\begin{flushright}
TIF-UNIMI-2025-24
\end{flushright}

\vspace{2.0cm}

\begin{center}
{\Large \bf Threshold resummation of rapidity distributions\\
  at fixed partonic rapidity}
\end{center}

\vspace{1.3cm}

\begin{center}
Lorenzo De Ros$^{1,2}$, Stefano Forte$^3$, Giovanni Ridolfi$^4$, Davide Maria Tagliabue$^{3,5}$ \\
\vspace{.3cm}
{\it
{}$^1$Physik Department T31, Technische Universit\"at M\"unchen, \\
James-Franck-Stra\ss e 1, 
D-85748 Garching, Germany \\
{}$^2$\text{PRISMA}$^+$ Cluster of Excellence \& Mainz Institute for Theoretical Physics, \\
Johannes Gutenberg-Universität Mainz, 55099 Mainz, Germany \\
{}$^3$Tif Lab, Dipartimento di Fisica, Universit\`a di Milano and \\
INFN, Sezione di Milano, Via Celoria 16, I-20133 Milano, Italy\\
{}$^4$Dipartimento di Fisica, Universit\`a di Genova and \\
INFN, Sezione di Genova, Via Dodecaneso 33, I-16146 Genova, Italy \\
{}$^5$Institute for Theoretical Particle Physics, KIT, \\
Wolfgang-Gaede-Straße 1, 76131, Karlsruhe, Germany. \\
}
\vspace{0.9cm}


{\bf \large Abstract}
\end{center}
We derive a general expression for the resummation of
rapidity distributions for processes with a colorless final 
state, such as Drell-Yan or Higgs production, in the limit in which
the center-of-mass energy goes on  threshold, but with fixed
rapidity of the Higgs or gauge boson in the partonic center-of-mass frame.
The result is obtained
by suitably generalizing the renormalization-group based
approach to threshold resummation previously pursued by  us. The
ensuing expression is valid to all logarithmic orders but the
resummation coefficients must be determined by comparing to fixed
order results. We perform this comparison for the Drell-Yan process
using the fixed-order next-to-next-to-leading (NNLO) result, thereby
determining resummation coefficients up to next-to-next-to-leading
logarithmic (NNLL) accuracy, for the 
quark-antiquark coefficient function in the quark nonsinglet channel.
We provide a translation to
direct QCD of a result for this resummation previously obtained using
SCET methods, and we show that it agrees with our own.

\clearpage
\tableofcontents


\sect{Introduction}
\label{sec:introduction}
The resummation of rapidity distributions for the Drell-Yan (DY) process,
and more generally for the production of colorless final states, is
as old as soft (Sudakov) resummation itself: indeed, at the next-to-leading
logarithmic (NLL) level it was presented in Ref.~\cite{Catani:1989ne},
one of the original papers that first performed Sudakov resummation
for hard processes~\cite{Sterman:1986aj,Catani:1989ne}. In this paper, results are presented using
the then-customary Feynman $x_F$ variable to parametrize the rapidity
of the final-state boson, suitable for the phenomenology of
fixed-target DY. The results of Ref.~\cite{Catani:1989ne} were
subsequently rewritten using variables more suited to collider
kinematics in Refs.~\cite{Ravindran:2006bu,Ravindran:2007sv} (see also Refs.~\cite{Westmark:2017uig,Westmarkphd}), and more
recently extended
  up to next-to-next-to-leading log (NNLL) matched to fixed next-to-next-to
  leading order (NNLO)~\cite{Banerjee:2018vvb}  and even to
  next-to-leading power~\cite{Ravindran:2022aqr}.

  However, all these results are obtained in the limit in which the
  center-of-mass energy of the partonic process $\sqrt{s }$ goes on threshold,
  i.e.\ $s \to M^2$, where $M$ is the mass of the final-state colorless
  object, which henceforth we will call gauge boson for short (as for
  DY). In this limit the rapidity goes to 
  zero: the gauge boson is produced at rest. The meaning of this limit
  can be understood by introducing a pair of scaling variables $x_1$,
  $x_2$ such that $x_1x_2=\frac{M^2}{s}\equiv \tau$ and
  $\frac{x_1}{x_2}=\exp(2 y)$, where $y$ is the gauge boson rapidity,
  defined by $\tanh   y=\frac{p_z}{E}$ with $E$ and $p_z$ respectively
  the energy of the gauge boson and its momentum  along the partonic collision axis in the
  partonic center of mass frame. In the limit $s \to M^2$ of course $p_z\to 0$,
  thus $y\to 0$, and both scaling variables $x_i\to 1$. We will
  call this a double-soft limit: both scaling variables tend to their
  common kinematic threshold.

  In this limit, the partonic differential rapidity distribution
  acquires at all perturbative orders logarithmically enhanced
  contributions, that can be put in to one-to-one correspondence to
  the contributions to the total partonic
  cross section~\cite{Catani:1989ne}. Namely, they can be obtained
  from the resummation of the inclusive cross section, which resums
  contributions of the form $\ln(1-\tau)^2$, by the simple
  substitution $\ln(1-\tau)^2\to\ln[(1-x_1)(1-x_2)]$. While of course
  the dependence on $y$ is logarithmically subleading~\cite{Bolzoni:2006ky}, because
  $\ln(1-x_i)=\ln(1-\tau)+O\left((1-\tau)^0\right)$, the
    resummation of Ref.~\cite{Catani:1989ne} predicts it fully, in the
    sense that if the limit $\tau\to1$ is reached for fixed
    $\frac{x_1}{x_2}$, the dependence on this ratio  (and thus on
    $y$) is predicted, to given  logarithmic accuracy in $\tau$, up to
    terms that are either constant or suppressed by powers of $1-\tau$.\footnote{The fact that the rapidity
    dependence is subleading was proven with increasing degrees of
    detail in Refs.~\cite{Laenen:1992ey,Bolzoni:2006ky}, and used in
    Refs.~\cite{Mukherjee:2006uu,Becher:2007ty,Bonvini:2010tp} to
    perform soft resummation of the rapidity distribution by reducing
    it to that of the total cross section. The validity of the
    arguments of Ref.~\cite{Laenen:1992ey,Bolzoni:2006ky} was
    questioned in Ref.~\cite{Lustermans:2019cau} and established more
    rigorously in Ref.~\cite{Bonvini:2023mfj}, though the issue of a
    formal proof is somewhat academic, given that the explicit form of
    the rapidity dependence in the limit is known from
    Refs.~\cite{Catani:1989ne,Westmark:2017uig}, so whether it
    is subleading can be checked from the known result without need of
    a formal proof, see Eq.~(\ref{eq:N_1n2nsq}) below.}

 In this paper we consider the more general situation in which the
 partonic center of mass energy reaches its threshold value, but now
 for fixed nonzero $p_z$, or, equivalently, at fixed rapidity. In
 terms of the scaling variables $x_1$, $x_2$ this corresponds to the
 case in which $x_1\to1$ but $x_2$ is fixed. We will consequently call
 this a single-soft limit. Of course, results are
 symmetric upon the interchange of $x_1$ and $x_2$: we will henceforth
 take $x_1$ as the scaling variable that tends to the soft limit.  In this limit resummation should include
 all contributions that are logarithmically enhanced in $\ln (1-x_1)$,
 even if they are not logarithmically enhanced in the other scaling variable.

 Sudakov resummation in this limit is  akin to the resummation of
 transverse momentum
 distributions~\cite{deFlorian:2005fzc,Forte:2021wxe}, but with the
 role of the longitudinal and transverse momenta of the final-state
 gauge boson interchanged. We will consequently attack it here with
 methods similar to those used in Ref.~\cite{Forte:2021wxe} to perform
 resummation of transverse-momentum distributions, in turn based on a
 multi-scale generalization~\cite{Bolzoni:2005xn} of the
 renormalization-group based approach to
 Sudakov resummation~\cite{Forte:2002ni}.

 The Sudakov resummation for the inclusive
 Drell-Yan process of Refs.~\cite{Sterman:1986aj,Catani:1989ne}
 was more recently re-derived using soft-collinear effective field
 theory (SCET)
 methods~\cite{Idilbi:2005ky,Becher:2006nr,Becher:2007ty}. Specifically,
 in Ref.~\cite{Becher:2007ty} the rapidity distribution was also
 considered, but, as mentioned, its resummation was reduced to that of
 the inclusive cross section based on the observation that the
 rapidity dependence is subleading. The result of
 Ref.~\cite{Catani:1989ne} for resummation of the rapidity
 distribution in the double-soft limit was reobtained using SCET
 methods in Ref.~\cite{Lustermans:2019cau}, where for the first time
 resummed results in the single-soft limit where also derived, up to
 NNLL. This SCET resummation of rapidity distributions in the single
 soft limit was recently re-derived in
 Ref.~\cite{Mistlberger:2025lee}, where it was also extended to N$^3$LL.

 The comparison of direct QCD (dQCD)
 and SCET results is generally nontrivial: specifically, in the
 approach of Ref.~\cite{Lustermans:2019cau} resummed results are given
 in the space of physical momentum variables, and are expressed as the
 solution to renormalization group equations that can only be worked
 out in closed form order by order (in Ref.~\cite{Lustermans:2019cau}
 results up to order $\alpha_s^3$ are explicitly given). A more direct
 comparison can be performed by using results of
 Ref.~\cite{Becher:2007ty}, in which SCET resummation can be expressed
 in Mellin-transform space, thereby allowing for a direct comparison
 to the dQCD result. Even so, the comparison is not immediate: for DY,
 the dQCD and SCET results were shown to be equivalent in
 Refs.~\cite{Bonvini:2012az,Bonvini:2014qga}.

 The purpose of this paper is to perform resummation of rapidity
 distributions in the single-soft limit. In particular, we will derive a
 general expression, valid at any logarithmic order, using the methods of
 Ref.~\cite{Forte:2002ni}. We will then obtain explicit expressions
 for DY up to NNLL accuracy, by comparing to the fixed NNLO result of
 Ref.~\cite{Anastasiou:2003ds} for the quark-antiquark coefficient
 function in the quark-quark nonsinglet channel.
 We will finally provide a translation
 to dQCD of the results of Ref.~\cite{Lustermans:2019cau} and show
 their equivalence to our own. Note that our results correspond to
 what is usually referred to as N$^k$LL' accuracy in the SCET
 literature, which is half a perturbative more accurate than what is
 usually called NNLL in that context.

  \sect{The kinematic structure of rapidity distributions}
\label{sec:kin}
 
 We will study the rapidity distributions for processes of the form
\beq
\label{HHZ}
H_1+H_2\rightarrow Z+X,
\eeq
where $Z$ is a massive colorless particle, which we denote by $Z$
because in the sequel we will specifically work out explicit
expressions up to NNLL for Drell-Yan
production, but which could equally be a Higgs boson (and in fact
was denoted by $H$ in Ref.~\cite{Forte:2021wxe} where NNLL
expressions for the Higgs $\pT$ distribution were given). For
definiteness, and without intended loss of generality, we will
henceforth refer to this final-state particle as the $Z$.

 As explained in the introduction, we want to perform
resummation of a rapidity distribution in the case in which the
kinematics of the underlying partonic sub-process is characterized by
the fact that the minimum energy of the partonic sub-process is higher
than the $Z$ mass, because the $Z$ is assumed to be boosted, i.e.\ to have a non-vanishing
longitudinal momentum in the center-of-mass frame of the partonic
collision. In this situation, unlike
in the case of  transverse momentum distributions,
discussed in Ref.~\cite{Forte:2021wxe}, the kinematic constraint on
the partonic subprocess (namely the requirement that the $Z$ is boosted)
is not decoupled from the partonic kinematics itself, as they are both
longitudinal. Because of this, it is crucial to discuss the kinematics
of the process at the partonic level. Indeed, using standard QCD
factorization for a hadronic process, the partonic center-of-mass frame
is boosted with respect to the hadronic center of mass frame whenever
the momentum fractions of the incoming partons  are unequal. The
nontrivial boost at the parton level, that changes the form of the
soft limit in the partonic cross section, gets then mixed with this
trivial boost at the level of the kinematics of the hadronic process.

Consequently, we will  henceforth focus on purely partonic kinematics,
and specifically study the process 
\beq
\label{eq:ffZ}
f_1(p_1)+f_2(p_2)\rightarrow Z(p)+X(k) \,,
\eeq
where $f_i$ are incoming partons with momenta $p_i$, and $p$ and $k$
are respectively the
momenta of the $Z$ and of the system $X$ that recoils against it, all given in the partonic
center-of-mass frame. For explicit expressions of the factorized
distribution at the hadronic level, as well as a discussion of the
relation between hadronic and partonic kinematics (neither of which are
relevant here) we refer e.g. to Sect.~4 of Ref.~\cite{Bonvini:2010tp}.

\subsection{Kinematics}
\label{subsec_Kinematics}
In the partonic center of mass frame 
\begin{align}\label{eq:cm}
&p_1 = \frac{\sqrt{s}}{2} (1,0,0,1),\qquad p_2 = \frac{\sqrt{s}}{2} (1,0,0,-1),
\\
\label{eq:p_H}
& p = \Big(\sqrt{M^2+\pT^2+p_z^2}, \vecpT,p_z\Big)
    =\Big(\sqrt{M^2+\pT^2} \cosh y, \vecpT, \sqrt{M^2+\pT^2} \sinh{ y}\Big),
\end{align}
where $\vecpT$ and $ y$ are respectively the transverse momentum and rapidity of the $Z$. The final-state kinematics
is fully parametrized by the scale $M^2$, the rapidity $y$, and the scaling variable
\begin{equation}\label{eq:taudef}
  \tau =\frac{M^2}{s}.
\end{equation}

The rapidity range for fixed $\tau$ is restricted by the requirement
that $|p_z|$ Eq.~(\ref{eq:p_H}) does not exceed the value allowed by the maximum
available energy. Because
\begin{align}\label{eq:sexp}
\sqrt{s}=\sqrt{M^2+\pT^2+p_z^2}+\sqrt{m_X^2+\pT^2+p_z^2}\ge\sqrt{M^2+p_z^2}+|p_z|
\end{align}
the minimum value of $s$ corresponds to the minimum
value of both the energy of the $Z$ and of the system $X$ with
invariant mass $m_X$ that recoils
against it, reached when  $\pT=0$ and $m_X=0$:
\begin{equation}\label{eq:smin}
s\ge s^{\rm min}(p_z) =\left(\sqrt{M^2+p_z^2}+\sqrt{p_z^2}\right)^2 \,.
\end{equation}
This gives
\begin{equation}
    |p_z|\le |p_z|^{\rm max}
    =\frac{s-M^2}{2\sqrt{s}}
    = M\frac{1-\tau}{2\sqrt{\tau}} \,.
\end{equation}
The maximum is attained when $\pT=0$, so using Eq.~(\ref{eq:p_H}) the
condition is $|p_z|^{\rm
  max}=M|\sinh y^{\rm max}|$ leading to 
\begin{equation}\label{eq:yrange}
    \ln\sqrt{\tau}\leq y\leq\ln\frac{1}{\sqrt{\tau}} \,,
\end{equation}
which can be equivalently obtained by
rewriting Eq.~(\ref{eq:sexp}) in terms of rapidity
\begin{align}\label{eq:sexprap}
\sqrt{s}=\sqrt{M^2+\pT^2}\cosh y+\sqrt{m_X^2+\pT^2+(M^2+\pT^2)\sinh^2 y}\ge M\cosh y+M|\sinh y| \,.
\end{align}

Resummation is performed in conjugate space, in which convolutions
reduce to ordinary products. For the rapidity distribution, the appropriate integral transform is a double Mellin transform.
We define a coefficient function in terms of the partonic differential cross section:
\begin{equation}\label{eq:cfdeflo}
C(x_1,x_2, M^2)=\frac{1}{\sigma_0}\frac{1}{\tau}\frac{d^2\sigma}{dM^2 dy},
\end{equation}
where 
\begin{equation}
x_1=\sqrt{\tau}e^y \,, \qquad x_2=\sqrt{\tau}e^{-y} \,,
\label{eq:variables}
\end{equation}
so that
\begin{equation}
\label{eq:variables_inv}
\tau=x_1 x_2 \,, \qquad y=\frac{1}{2}\ln\frac{x_1}{x_2},
\end{equation}
and $\sigma_0$ is defined so that
\begin{equation}
C(x_1,x_2, M^2)=\delta(1-x_1)\delta(1-x_2)+\mathcal{O}(\as).
\end{equation}
It is easy to check that the jacobian of the transformation from
$(x_1, x_2)$ to $(y,\tau)$ equals one, and that 
\begin{equation}
    \label{domains}
    0\leq x_i \leq 1,
    \qquad 
    i=1,2.
\end{equation}

The coefficient function is a dimensionless function with a power
expansion in $\as$; as such, it also depends on the strong coupling
$\as$ evaluated at a renormalization scale $\mu_R^2$, and on the ratio
of $M^2$ to renormalization and factorization scales. Furthermore,
both the coefficient function and $\sigma_0$ depend on the choice of
initial state partons and thus carry parton indices.
 In this section, for simplicity, we omit all these functional
dependences and indices, and we only display the dependence on kinematic
variables. The full set of arguments will be restored in the next
section. 
The coefficient function in
Mellin-Mellin space
is given by
\begin{align}
C(N_1, N_2, M^2)=
    \int_0^1 dx_1\, x_1^{N_1-1}  \int_0^1 dx_2\, x_2^{N_2-1}  C(x_1,x_2, M^2),
    \label{eq:sigma_mellin-mellin}
\end{align}
where by slight abuse of notation we denote both the original
function and its transform with the same symbol $C$.

Using Eq.~\eqref{eq:variables}, the integral transform Eq.~\eqref{eq:sigma_mellin-mellin} can be equivalently viewed as a Fourier-Mellin transform~\cite{Sterman:2000pt}
of the coefficient function with respect to $y$ and $\tau$. Indeed, defining
\begin{equation}
    \label{eq:N_1}
    N_1=N+i\frac{b}{2};\qquad
    N_2=N-i\frac{b}{2},
\end{equation}
so that
\begin{equation}
    \label{eq:invN_1}
    N(N_1,N_2)=\frac{1}{2}\left(N_1+N_2\right)\,, \qquad
    b(N_1,N_2)=\frac{N_1-N_2}{i},
\end{equation}
and
\begin{align}\label{eq:signtox}
     \hat C(N,b,M^2)&= C(N_1(N,b), N_2(N,b), M^2) \,, \\
     \label{eq:sigxton}
\hat C(\tau,y, M^2)&=C(x_1(\tau,y), x_2(\tau,y),M^2) \,.
\end{align}
Eq.~\eqref{eq:sigma_mellin-mellin} can be written
\begin{align}
   \hat C(N,b,M^2)&= 
    \int_0^1 d \tau \, \tau^{N-1} \int_{\ln\sqrt{\tau}}^{-\ln\sqrt{\tau}} d y \, e^{iby} \hat C(\tau,y, M^2).
    \label{sigma_mellin-fourier}
\end{align}


The soft limits that we consider are defined as follows. The
double-soft limit is the limit in which
\begin{equation}\label{eq:doublesoft}
s\to  s^{\rm min}(0)=M^2,
\end{equation}
with $s^{\rm min}(p_z)$ given in Eq.~(\ref{eq:smin}).
This means that
\begin{equation}\label{eq:doublesoft1}
  \pT\to0;\quad  p_z\to0.
\end{equation}
  In this limit
\begin{equation}\label{eq:doublesoft2}
\tau\to1;\quad  y\to 0,
\end{equation}
i.e.
\begin{equation}\label{eq:doublesoft3}
 x_1\to 1, \quad x_2\to1.
\end{equation}
In Mellin-Mellin space this means
\begin{equation}\label{eq:doublesoft4}
  \Re N_1\to\infty;\quad \Re N_2\to\infty.
\end{equation}
In Fourier-Mellin space this corresponds to
\begin{equation}\label{eq:doublesoft5}
 \Re  N\to\infty.
\end{equation}

The single-soft limit is instead the limit in which
\begin{equation}\label{eq:singlesoft}
  s\to  s^{\rm min}(p_z)
\end{equation}
for  fixed $p_z$, or, equivalently for some fixed
rapidity $y$.
In terms of the variables $x_i$ Eq.~(\ref{eq:variables}) the
condition of fixed $y$ implies fixed ratio $\frac{x_1}{x_2}=e^{ 2y}$ and the
condition of minimum $s$ corresponds to maximum product $x_1 x_2$,
which (assuming without loss of generality $x_1\ge x_2$) for fixed
$e^{ 2y}$ means maximum $x_1 e^{-2y}$ i.e. $x_1=1$. So the single
soft limit is the limit in which
\begin{equation}\label{eq:singlesoft1}
 x_1\to1;\quad \hbox{fixed}\> x_2.
\end{equation}
In Mellin-Mellin space this means
\begin{equation}\label{eq:singlesoft2}
 N_1\to\infty;\quad \hbox{fixed}\> N_2.
\end{equation}

\subsection{Phase space}
      We now discuss the phase space measure for the process
      Eq.~(\ref{eq:ffZ}). We
      consider  the $Z$ rapidity distribution when the system
      $X$ includes   $n$ final-state massless partons, with momenta
$k_1,\ldots,k_n$:
\beq
\label{eq:consg}
p_1+p_2=p+k_1+\ldots +k_n.
\eeq
We are interested in the double-soft and single-soft limits.

\subsubsection{Double-soft limit}\label{sec:dsoft}
In the double-soft limit the center-of mass energy is going on
threshold Eq.~(\ref{eq:doublesoft}), hence the limit coincides with
the threshold limit of the total cross section, discussed by us in
Ref.~\cite{Forte:2002ni}. Namely,
squaring both sides of Eq.~(\ref{eq:consg}) 
\beq\label{eq:softlimit}
s(1-\tau)=\sum_{i,j=1}^n k_i\cdot  k_j+2\sum_{i=1}^n p\cdot k_i,
\eeq
but $p\cdot k_i$ is positive semi-definite, and vanishes only when
$k_i^0=0$, hence $k^0_i\to 0$ for all $i$ in the limit $\tau\to 1$, i.e. 
all emitted partons must be soft.

We write the phase space for the process as 
\begin{equation}
\label{eq:phsp}
\dphi_{n+1}(p_1+p_2;p,k_1,\dots,k_{n})
=
\int\frac{\dk^2}{2\pi}\,\dphi_2(p_1+p_2;p,k)\,
\dphi_{n}(k;k_1,\dots,k_{n}),
\end{equation}
with $n\ge0$.
Here  $\dphi_2$  is the phase space for production from the incoming
total momentum $p_1+p_2$ of a massive final state with mass $M$ and
momentum $p$, and a system with momentum
$k$ recoiling against it; $\dphi_{n}$ is the phase space for the
production, from incoming momentum $k$, of a final-state system
containing $n$ massless partons with momenta $k_i$. 

We now work out the two-body phase space  $\dphi_2$ in $d=4-2\epsilon$ dimensions in terms of
  the rapidity of the $Z$.
  In the partonic center of mass frame we have
\begin{align}
    \dphi_2(p_1+p_2;p,k)&=
    \frac{\rmd^{d-1}p}{(2\pi)^{d-1}2p^0} \frac{\rmd^{d-1}k}{(2\pi)^{d-1}2k^0}
    (2\pi)^d \delta^{(d)}(p_1+p_2-p-k)
\nonumber\\
\label{eq:phi2}
    &=\frac{\rmd^{d-1}p}{4(2\pi)^{d-2} p^0 k^0} \delta(\sqrt{s}-p^0-k^0).
\end{align}
The integration measure can be rewritten as
\begin{equation}\label{eq:phi2a}
    \rmd^{d-1}p=\frac{1}{2} \dpT^2 \,
    \abs{\vecpT}^{d-4} \dpz \, \dOmega_{d-2}=
    \frac{\pi^{1-\epsilon}}{\Gamma(1-\epsilon)} \dpT^2 \, \abs{\vecpT}^{-2\epsilon} \, \dpz,
\end{equation}
where we have performed the azimuthal integration using the identity $\Omega_d=\frac{2\pi^{d/2}}{\Gamma(d/2)}$.

Substituting in Eq.~(\ref{eq:phi2}) gives
\begin{equation}
    \label{eq:dphi2_predelta}
    \dphi_2(p_1+p_2;p,k )=\frac{(4\pi)^\epsilon \abs{\vecpT}^{-2\epsilon}}
    {16 \pi \Gamma(1-\epsilon)}
    \frac{\dpT^2 \, \dpz}{p_0 k_0}
    \delta(\sqrt{s}-p_0-k_0).
\end{equation}
We first trade $p_z$ for the rapidity $y$, using
\begin{equation}
p_z=\sqrt{\pT^2+M^2}\sinh y;\qquad \dpz=p_0 dy.
\end{equation}
We obtain
\begin{equation}
    \label{eq:dphi2_predeltay}
    \dphi_2(p_1+p_2;p,k )=\frac{(4\pi)^\epsilon \abs{\vecpT}^{-2\epsilon}}
    {16 \pi \Gamma(1-\epsilon)}
    \frac{\dpT^2 \, dy}{k_0}
    \delta(\sqrt{s}-p_0-k_0).
\end{equation}
Next, we perform the $\pT^2$ integration using the delta function. We have
\begin{equation}
 \delta(\sqrt{s}-p_0-k_0)=\frac{\delta(\pT^2-\barpT^2)}{J}
\end{equation}
where
\begin{equation}
J=\left|\frac{\rmd}{\dpT^2}(\sqrt{s}-p_0-k_0)\right|=\left(1+\frac{p_0}{k_0}\right)\frac{\dpzero}{\dpT^2}
=\frac{\sqrt{s}}{2p_0 k_0}\frac{\dpzero^2}{\dpT^2}
=\frac{\sqrt{s}}{2p_0 k_0}\cosh^2y,
\end{equation}
and $\barpT^2$ is the solution of
\begin{equation}\label{eq:encon}
\sqrt{s}=p_0+\sqrt{k^2+p_0^2-M^2}=\sqrt{M^2+\barpT^2}\cosh y+\sqrt{k^2+(M^2+\barpT^2)\cosh^2 y-M^2}.
\end{equation}
We get
\begin{equation}
    \label{eq:dphi2}
        \dphi_2(p_1+p_2;p,k )=\frac{(4\pi)^\epsilon (\barpT^2)^{-\epsilon}}
    {8\pi \Gamma(1-\epsilon)}
   \frac{p_0 }{\sqrt{s}\cosh^2y}\,dy.
\end{equation}
Expressing $p_0$ as a function of $s,M^2,k^2$ through the first equality in Eq.~(\ref{eq:encon}), we finally obtain
\begin{equation}\label{eq:phi2fin}
    \dphi_2(q;p,k )=\frac{(4\pi)^\epsilon (\barpT^2)^{-\epsilon}} 
    {16\pi \Gamma(1-\epsilon)}
\frac{s- k^2+M^2}{s} \frac{dy}{\cosh^2 y}.
\end{equation}
The phase space Eq.~(\ref{eq:phsp}) becomes thus
\begin{align}
    & \; \dphi_{n+1}(p_1+p_2;p,k_1,\dots,k_{n}) \notag 
    \\
    & = \frac{dy}{ \cosh^2 y}
    \frac{(4\pi)^\epsilon }{32\pi^2 \Gamma(1-\epsilon)}
    \int _{k^2_{\rm min}}^{k^2_{\rm max}} \dk^2\, 
    (\barpT^2)^{-\epsilon}\frac{s- k^2+M^2}{s} \dphi_{n} (k ;k_1,\dots,k_{n}).
    \label{eq:phspfin}
\end{align}

We can now work out the kinematic limits for the $ k^2$ integration. Of
course, $k^2_{\rm min}=0$: this happens whenever the final-state
parton momenta  $k_i$ are all collinear to each other (or there is a
single parton, of course). 
The upper bound is obtained by observing that
\begin{equation}
    k^2=s+M^2-2\sqrt{s}\sqrt{M^2+\pT^2+p_z^2}=s+M^2-2\sqrt{s}\sqrt{M^2+\pT^2}\cosh y,
\end{equation}
so for given rapidity the maximum value of $k^2$ is
obtained for $\pT^2 \to 0$:
\begin{align}\label{eq:kmaxy}
  k_{\rm max}^2&=\frac{M^2}{\tau} \left(1+\tau-2\sqrt{\tau}\cosh y\right)\\\label{eq:kmaxx1x2}
  &=M^2\frac{(1-x_1)(1-x_2)}{x_1x_2}.
    \end{align}
Introducing a dimensionless variable $v$ in order to interpolate
between $0$ and $k_{\rm max}^2$ we can set
\begin{equation}
    k^2=v k_{\rm max}^2;\qquad 0\leq v\leq 1,
\end{equation}
and rewrite the measure of integration over $  {k }^2$ in the phase space
Eq.~(\ref{eq:phspfin}) as
\begin{equation}\label{eq:vint} \dk^2 =\frac{M^2(1-x_1)(1-x_2)}{x_1x_2} d v.
\end{equation}
Furthermore, as in Ref.~\cite{Forte:2021wxe} the phase space
$\dphi_{n}(k;k_1,\ldots,k_{n})$ can be rewritten following
Ref.~\cite{Bolzoni:2005xn} (in turn based on Ref.~\cite{Forte:2002ni}) as
\begin{equation}
\dphi_{n}(k;k_1,\ldots,k_{n})=
2\pi\left[\frac{N(\epsilon)}{2\pi}\right]^{n-1}( k^2)^{n-2-(n-1)\epsilon}
\dOmega^{(n-1)}(\epsilon),
\label{eq:discase}
\end{equation}
where $N(\epsilon)=\frac{1}{2(4\pi)^{2-2\epsilon}}$ and
\begin{equation}\label{eq:angular}
\dOmega^{(n-1)}(\epsilon)=\dOmega_1\dots
\dOmega_{n-1}\int_0^1\dz_mz_m^{(n-3)-(n-2)\epsilon}
(1-z_{n-1})^{1-2\epsilon}\dots
\int_0^1\dz_2z_2^{-\epsilon}(1-z_2)^{1-2\epsilon}.
\end{equation}
The definition of the variables $z_i$ is irrelevant
here and can be found in Ref.~\cite{Forte:2002ni} (where they are
called $z'_i$).

We can finally consider the double-soft limit. Equation~(\ref{eq:kmaxx1x2}) implies that 
$ k^2\to0$ as both $x_1$ and $x_2$ approach 1. Furthermore, in Eq.~(\ref{eq:discase})  
 the phase space
$\dphi_{n}(k;k_1,\ldots,k_{n})$ is rewritten in terms of a
 dimensionless integration measure, with all the dimensional
 dependence contained in powers of a soft scale
 \begin{equation}\label{eq:dsscale}
   \Lambda^2_{\rm ds}=k_{\rm max}^2=M^2(1-x_1)(1-x_2)\left[1+ \mathcal{O}(1-x_1)+ \mathcal{O}(1-x_2)\right].
 \end{equation}
This is in fact the same as the phase space for 
deep-inelastic scattering with incoming momentum $ k $ whose
soft limit was discussed in Ref.~\cite{Forte:2002ni}.

\subsubsection{Single-soft limit}\label{sec:kinssl}
In this limit, the rapidity of the $Z$ is fixed, while $s\to s^{\rm
  min}(p_z)$, given by Eq.~(\ref{eq:smin}). At leading order, the process
Eq.~(\ref{eq:consg}) must thus include at least one parton that recoils
against the $Z$ with momentum $k_1$
and the soft limit corresponds to $\pT\to0$ with
fixed $y$ (fixed $p_z$) in Eq.~(\ref{eq:p_H}) so
\begin{equation}\label{eq:singsoft}
    k_1=k_1^{(0)}=(p_z,0,0,-p_z).
\end{equation}
Note that while a fixed  $y$ value of course corresponds to a fixed
$p_z$ value, approaching the  $s\to s^{\rm
  min}(p_z)$ limit at fixed  $y$ or fixed $p_z$ correspond to
different paths.

Using again Eq.~(\ref{eq:softlimit})
it is clear that with more than one parton in the final state the soft
limit is reached when $k_i\cdot k_j=0$ for all $i,j$, in which case
\begin{equation}\label{eq:singsoft1}
   \sum_{i=1}^n k_i=k_1^{(0)},
\end{equation}
i.e. all emitted partons must be collinear. They are not necessarily
soft, though of course some of them will: the only constraint being
that their longitudinal momentum fractions add in such a way that
Eq.~(\ref{eq:singsoft1}) is satisfied. 

The phase space can again be written in the form of
Eq.~(\ref{eq:phsp}), except that now $n\ge1$. In particular,
Eq.~(\ref{eq:phspfin}) still holds, and the upper bound of
$ k^2$  is still given by   $k_{\rm max}^2$
Eq.~(\ref{eq:kmaxy}). However, the limit is now given by
Eq.~(\ref{eq:singlesoft1}), so $1-x_2$ is now generic. Consequently,
  Eq.~(\ref{eq:discase}) now implies that the dimensional
 dependence is contained in powers of a soft scale
 \begin{equation}\label{eq:ssscale}
   \Lambda^2_{\rm ss}=k_{\rm max}^2=M^2(1-x_1)\left[1+ \mathcal{O}(1-x_1)\right].
 \end{equation}

It is interesting to contrast this with the case in which the $Z$ has
fixed $\pT$, discussed in Ref.~\cite{Forte:2021wxe}. In that case,
$\pT$ is fixed and 
$p_z\to0$ in the soft limit. Hence in that case Eq.~(\ref{eq:singsoft}) is
replaced by $k_1=k_1^{(0)}=(\pT,-\pT,0,0)$
(choosing without loss of generality the $x$ axis along $\vecpT$)
and the condition $k_i\cdot k_j=0$ when $i=1$  can be satisfied in two different
ways: either with $\kT^j\to 0$ (soft partons), or with   $\kT^j\to \kT^1>0$
(collinear partons). The phase space in this case must be written by
separating these two families of partons, that turn out to be
characterized by two different soft scales, due to the fact that soft
partons correspond to radiation from incoming partons (i.e. collinear
to the incoming partons), while collinear
partons correspond to radiation from the final-state parton that
recoils against the $Z$~\cite{deFlorian:2005fzc} (i.e. collinear to
the $Z$). In the case
considered here, it is impossible to distinguish initial- and
final-state radiation because $\pT\to0$ so partons collinear to the
$Z$ are also collinear to the initial state partons.

\sect{Resummation by renormalization group improvement}
\label{sec:rg}

The main result of Sect.~\ref{sec:kin} is that, due to the collinear
nature of the soft limits at fixed rapidity, the
rapidity distribution  in the soft limits depends on a single soft
scale. This is true both in the double-soft limit Eqs.~(\ref{eq:doublesoft3}-\ref{eq:doublesoft4}), with
soft scale  $\Lambda^2_{\rm ds}$ Eq.~(\ref{eq:dsscale}), and in the single-soft limit Eqs.~(\ref{eq:singlesoft1}-\ref{eq:singlesoft2}), with
soft scale  $\Lambda^2_{\rm ss}$ Eq.~(\ref{eq:ssscale}). Consequently,
resummation is simpler than that of transverse momentum distributions
that we derived in Ref.~\cite{Forte:2021wxe}, and can in fact be
performed by a mild generalization of the method we developed in
Ref.~\cite{Forte:2002ni} in the inclusive case. On the other hand,  
some technical complications are related to the need of taking a
double Mellin transform, and, in the single-soft  limit, of treating
the two incoming partons asymmetrically.

Resummation is performed based on a suitably adapted version of the
argument of Refs.~\cite{Forte:2002ni,Forte:2021wxe}.
Perturbative factorization implies that the Mellin-space hadronic cross
section can be written as 
\begin{equation}
\frac{d^2\sigma_H}{dM^2 dY}(N_1,N_2,M^2) =\tau_H\sum_{ij}\sigma_{0}^{ij}
C_{ij}\(N_1,N_2,\frac{M^2}{\mu^2},\frac{M^2}{\mu_R^2},\as(\mu_R^2)\)
f_i(N_1,\mu^2)f_j(N_2,\mu^2)
\label{eq:singlerg}
\end{equation}
where $\tau_H=M^2/S$, $S$ is the hadronic center-of-mass energy squared, and $f_i(N,\mu^2)$ are (Mellin-transformed) parton distributions (PDFs).
In view of the construction of a resummed result in the single-soft limit, in which the incoming partons are treated asymmetrically, it
is useful to study the dependence of the coefficient function on the factorization scale of each of the two partons independently.
Specifically, we write
\begin{equation}
\frac{d^2\sigma_H}{dM^2 dY}(N_1,N_2,M^2) =\tau_H\sum_{ij}\sigma_0^{ij}
C_{ij}\(N_1,N_2,\frac{M^2}{\mu_1^2},\frac{M^2}{\mu_2^2},\frac{M^2}{\mu_R^2},\as(\mu_R^2)\)
f_i(N_1,\mu_1^2)f_j(N_2,\mu_2^2),
\label{eq:doublerg}
\end{equation}
which is allowed because it is always possible to
express PDFs at one scale in terms of those at a different scale, and re-define the coefficient functions $C_{ij}$ accordingly. We are assuming here that $f_i(N,\mu^2)$ are evolution eigenstates, 
with eigenvalues $\gamma_i^{\rm AP}(N,\as( \mu^2))$, so that they do not mix upon evolution.

We perform resummation of the Mellin-transformed coefficient function
for the $ij$ partonic channel
\begin{equation}\label{eq:cfdefmel}
C_{ij}\(N_1,N_2,\frac{M^2}{\mu_1^2},\frac{M^2}{\mu_2^2},\frac{M^2}{\mu_R^2},\as(\mu_R^2)\)
=1+\mathcal{O}(\as).
\end{equation}
Observing that the hadronic cross section cannot depend on either $\mu_1^2$ or  $\mu_2^2$
we get a Callan-Symanzik-Altarelli-Parisi
equation of the form
\begin{align}\label{eq:ap}
& \mu_1^2\frac{d}{d \mu_1^2}C_{ij}\(N_1,N_2,\frac{M^2}{\mu_1^2},\frac{M^2}{\mu_2^2},\frac{M^2}{\mu_R^2},\as(\mu_R^2)\)
\nonumber\\
&\quad=-\gamma_i^{\rm AP}(N_1,\as( \mu_1^2))
C_{ij}\(N_1,N_2,\frac{M^2}{\mu_1^2},\frac{M^2}{\mu_2^2},\frac{M^2}{\mu_R^2},\as(\mu_R^2)\)
\end{align}
and an analogous equation with $\mu_1^2\to\mu_2^2$ and $\gamma_i^{\rm AP}(N_1,\as( \mu_1^2))\to\gamma_j^{\rm AP}(N_2,\as( \mu_2^2))$.

In the remainder of this section we omit the parton indices $i,j$ and denote with $C$ a generic
eigenstate coefficient function, which is the quantity that undergoes
resummation through renormalization group improvement. We will then
restore parton indices in the next Section when considering the explicit
construction of the resummed coefficient function in various limits.

Solving Eq.~(\ref{eq:ap}), the coefficient function
is found to have the form
\begin{align}\label{eq:cs}
&C\(N_1,N_2,\frac{M^2}{\mu_1^2},\frac{M^2}{\mu_2^2},\frac{M^2}{\mu_R^2},\as(\mu_R^2)\)
\nonumber\\
&\qquad=
C\(N_1,N_2,1,\frac{M^2}{\mu_2^2},\frac{M^2}{\mu_R^2},\as(\mu_R^2)\)
\exp\int_{\mu_1^2}^{M^2}\frac{dk^2}{k^2}\gamma^{\rm AP}(N_1,\as(k^2)).
\end{align}
Resummation is performed  in terms of a physical anomalous dimension
\begin{align}\label{eq:gamp}
\gamma\(N_1,N_2, \frac{M^2}{\mu^2}, \as(\mu^2)\)
= \frac{d}{d\ln M^2} \ln
C\(N_1,N_2, \frac{M^2}{\mu^2}, \frac{M^2}{\mu^2},\frac{M^2}{\mu^2},\as(\mu^2)\),
\end{align}
where in the $\overline{\rm MS}$ scheme there is a single common renormalization and factorization
scale $\mu^2$ identified with the scale arising when performing
dimensional regularization in    $4-2\epsilon$ space-time
dimensions. 

The argument then reproduces that of
Refs.~\cite{Forte:2002ni,Forte:2021wxe}, the only difference being
that now the coefficient function depends on two Mellin variables $N_1$, $N_2$:
the coefficient function is written in terms of bare coefficient
function and coupling $C^{(0)}$ and $\alpha_0$
\begin{equation}\label{eq:barecf}
 C\(N_1,N_2,\frac{M^2}{\mu^2},\frac{M^2}{\mu^2},\frac{M^2}{\mu^2},\as(\mu^2)\)=\lim_{\epsilon\to 0}Z^C(N_1,N_2,\as(\mu^2),\epsilon)C^{(0)}(N_1,N_2,M^2,\alpha_0,\epsilon).
\end{equation}
The  physical anomalous dimension (in mass-independent factorization schemes) is determined by dimensional
analysis from the bare coefficient function~\cite{Forte:2002ni} and it is finite as  $\epsilon\to0$ so that
\begin{align}\label{eq:gampbare}
\gamma\(N_1, N_2,\frac{M^2}{\mu^2},\as(\mu^2)\)=-\lim_{\epsilon\to 0}\epsilon \frac{d}{d\ln \alpha_0} \ln
C^{(0)}(N_1,N_2, M^2,\alpha_0,\epsilon).
\end{align}

\subsection{Double-soft resummation}
\label{eq:rgds}

In the double-soft limit, the kinematic analysis of the phase space presented in
Sect.~\ref{sec:kin}, together with the argument of
Ref.~\cite{Forte:2002ni},  implies that the leading contribution to
the hard coefficient function   as
$x_1\to1$ and $x_2\to1$ depends on $x_1$ and $x_2$ only through
the fixed combination of the scale and the scaling
variables
\begin{equation}\label{eq:gen}
\Lambda^2_{\rm ds}=M^2(1-x_1)(1-x_2),
\end{equation}
up to terms of relative order $(1-x_1)$ or $(1-x_2)$.
It is then easy to prove (see Appendix~\ref{app:dmel}) that  the Mellin-space coefficient function
only depends on $N_1$, $N_2$ through the dimensionful variable
\begin{equation}\label{eq:genn}
\tilde\Lambda^2_{\rm ds}=\frac{M^2}{N_1N_2}, 
\end{equation}
up to corrections of order $\frac{1}{N_1}$ or $\frac{1}{N_2}$.
The bare coefficient function can be consequently expanded as
\begin{align}
    \label{eq:cbarefact}
    C^{(0)}(N_1N_2,
    M^2,\alpha_0,\epsilon) & =C^{(0,c)}(M^2,\alpha_0,\epsilon)
    C^{(0,\,l)}(\tilde\Lambda_{\rm ds}^2,\alpha_0,\epsilon)
    \\
    \label{eq:cbarec}
    C^{(0,\,c)}(M^2,\alpha_0,\epsilon) & = \sum_n 
    C_n^{(0,\,c)}(\epsilon) M^{- 2n\epsilon}\alpha_0^n
    \\
    \label{eq:cbarel}
    C^{(0,\,l)}(\tilde\Lambda_{\rm ds}^2,\alpha_0,\epsilon) & =\sum_n 
    C_n^{(0,\,l)}(\epsilon) {\tilde\Lambda_{\rm ds}}^{- 2n\epsilon}\alpha_0^n,
\end{align}
where $C^{(0,\,l)}$ collects contributions due to real emission, which
have a nontrivial dependence on $N_1$, $N_2$, $C^{(0,\,c)}$ collects virtual contributions, that
have Born kinematics, and  it is assumed~\cite{Contopanagos:1996nh}
that virtual and real soft-emission contributions fully
factorize. 

The factorization Eq.~(\ref{eq:cbarefact}) implies that in the double-soft limit the
physical anomalous dimension is the sum of two contributions:
\begin{align}\label{eq:gampbard}
\gamma\(N_1N_2,\frac{M^2}{\mu^2},\as(\mu^2),\epsilon\)
=\bar\gamma^{(c)}\(\frac{M^2}{\mu^2},\as(\mu^2),\epsilon\)+\bar\gamma^{(l)}\(\frac{\tilde\Lambda_{\rm ds}^2(N_1N_2,M^2)}{\mu^2},\as(\mu^2),\epsilon\),
\end{align}
that are not separately finite and $\mu$-independent.
Note that  the dependence of $\bar\gamma^{(l)}$ on $N_1,N_2$ is only
through the scale $\tilde\Lambda_{\rm
    ds}^2$.
Renormalization-group invariance, i.e. $\mu$ independence of
the full anomalous dimension, then implies
\begin{equation}\label{eq:sudrge}
    -\frac{d}{d\ln\mu^2}
    \lim_{\epsilon\to0}\bar\gamma^{(l)}\(\frac{\tilde\Lambda_{\rm ds}^2}{\mu^2},\as(\mu^2),\epsilon\)
    = \frac{d}{d\ln\mu^2}\lim_{\epsilon\to0}\bar\gamma^{(c)}\(\frac{M^2}{\mu^2},\as(\mu^2),\epsilon\)=\bar g^{\rm ds}(\as(\mu^2))
\end{equation}
where $\bar g^{\rm ds}(\as)$ has an expansion in powers of $\as$ with finite coefficients.
Solving Eqs.~(\ref{eq:sudrge}) 
\begin{align}
&\bar\gamma^{(c)}\(\frac{M^2}{\mu^2},\as(\mu^2),\epsilon\)
=\int_{M^2}^{\mu^2}\frac{dk^2}{k^2}\,\bar g^{\rm ds}(\as(k^2))
+\bar g^{(c)}_0(\as(M^2),\epsilon)
\\
&\bar\gamma^{(l)}\(\frac{\tilde\Lambda_{\rm ds}^2(N_1N_2,M^2)}{\mu^2},\as(\mu^2),\epsilon\)
=-\int_{\tilde\Lambda_{ds}^2}^{\mu^2}\frac{dk^2}{k^2}\,\bar g^{\rm ds}(\as(k^2))
+\bar g^{(l)}_0(\as(\tilde\Lambda_{\rm ds}^2),\epsilon)
\end{align}
and adding up the solutions we get
\begin{align}\label{eq:gampbardgs}
  \gamma\(N_1N_2,\frac{M^2}{\mu^2},
\as(\mu^2)\)&=\lim_{\epsilon\to0}\left[
\bar g^{(c)}_0(\as(M^2),\epsilon)+\bar g^{(l)}_0(\as(\tilde\Lambda_{\rm ds}^2),\epsilon)\right]+\int_{M^2}^{\tilde\Lambda_{ds}^2} \frac{\dmu^2}{\mu^2}\bar g^{\rm ds}(\as(\mu^2))
\\
\label{eq:gampbardg}
&=g_0^{\rm ds} (\as(M^2))+\int_{M^2}^{\tilde\Lambda_{ds}^2} \frac{\dmu^2}{\mu^2} g^{\rm ds}(\as(\mu^2)),
\end{align}
where we have made use of the fact that the singularity must cancel
between $\bar g^{(l)}_0$  and $\bar g^{(c)}_0$, and
\begin{align}
  \label{eq:g0def}
  g_0^{\rm ds} (\as(M^2))&=\bar g^{(c)}_0(\as(M^2),0)+\bar g^{(l)}_0(\as(M^2),0),
  \\
  \label{eq:gdef}
g^{\rm ds}(\as)&=\bar
g^{\rm ds}(\as)+\beta(\as)\frac{d\bar g^{(l)}_0}{d\as}.
\end{align}
In Eq.~(\ref{eq:gampbardg}) both $g_0^{\rm ds} (\as(M^2)$ and  $g^{\rm  ds}(\as(\mu^2))$ are power series in $\as$ with finite coefficients.

The physical anomalous dimension Eq.~(\ref{eq:gampbardgs}) is now
written as the sum of two contributions that are separately finite and
$\mu$ independent:
\begin{align}
  \label{eq:paddec}
&  \gamma\(N_1N_2,\as(M^2)\)= \gamma^{(c)}\(\as(M^2)\)+
  \gamma^{(l)}\(N_1N_2,\as(M^2)\)\\   \label{eq:padpc}
  &\qquad \gamma^{(c)}\(\as(M^2)\)=g_0^{\rm ds} (\as(M^2))\\
  \label{eq:padpl}
  &\qquad \gamma^{(l)}\(N_1N_2,\as(M^2)\)=\int_{M^2}^{M^2/(N_1N_2)} \frac{dk^2}{k^2} g^{\rm ds}(\as(k^2)).
\end{align}
We conclude that in the double-soft
limit the finite renormalized coefficient function  
Eq.~(\ref{eq:cs}) can be factorized as
  \begin{align}\label{eq:cffact}
    C\(N_1,N_2,\frac{M^2}{\mu^2},\frac{M^2}{\mu^2},\frac{M^2}{\mu^2},\as(\mu^2)\)&=C^{(c)}\(\frac{M^2}{\mu^2},\as(\mu^2)\)C^{(l)}\(\frac{M^2/(N_1N_2)}{\mu^2},\as(\mu^2)\)
    \nonumber\\
    &+\mathcal{O}\(\frac{1}{N_1},\frac{1}{N_2}\)
  \end{align}
  where
  \begin{align}
\label{eq:pancc}
    M^2\frac{d}{dM^2} \ln C^{(c)}\(\frac{M^2}{\mu^2},\as(\mu^2)\)&= \gamma^{(c)}\(\as(M)\) 
    \\
\label{eq:pancl}
    M^2\frac{d}{dM^2} \ln C^{(l)}\(\frac{M^2/(N_1N_2)}{\mu^2},\as(\mu^2)\)&=
    \gamma^{(l)}\(N_1N_2,\as(M^2)\).
    \end{align}

  Logarithmically enhanced terms are contained in
$C^{(l)}(M^2/(N_1N_2\mu^2),\as(\mu^2))$, for which the expression
Eq.~(\ref{eq:padpl}) of   $ \gamma^{(l)}$ provides the resummed
prediction, while  $C^{(c)}$ contains the constants.
Note that the separation between $C^{(l)}$
and $C^{(c)}$ is ambiguous, reflecting the freedom in fixing the
finite part after the cancellation of the divergence: a constant term
can always be reabsorbed in a redefinition of  $C^{(l)}$.

\subsection{Single-soft resummation}\label{eq:rgss}
In the single soft limit the argument proceeds in the same way, except
that now, assuming for definiteness that the soft variable is $x_1$,  the dimensional dependence is through the scale
\begin{equation}\label{eq:genss}
\Lambda^2_{\rm ss}=M^2(1-x_1),
\end{equation}
up to corrections suppressed by powers of $(1-x_1)$.
The Mellin-space coefficient function then 
depends on $N_1$ through the dimensional variable
\begin{equation}\label{eq:gennss}
\tilde\Lambda^2_{\rm ss}=\frac{M^2}{N_1}. 
\end{equation}
The dependence on $N_2$ is  parametric: the argument runs through
unchanged, except that now $N_1N_2$ is everywhere replaced by $N_1$,
and we end up with the physical anomalous dimension
\begin{equation}\label{eq:gampbardgss}
\gamma(N_1,N_2, 1,\as(M^2),\epsilon)=
 g_0^{\rm ss}(\as(M^2), N_2)+\int_{M^2}^{\tilde\Lambda_{\rm ss}^2(M^2, N_1)} \frac{\dmu^2}{\mu^2}
   g^{\rm ss}(\as(\mu^2),N_2),
\end{equation}
where $g_0^{\rm ss}(\alpha,N_2)$ and  $g^{\rm ss}(\alpha,N_2)$ are power series in $\alpha$ with
$N_2$-dependent coefficients.

The resummed coefficient function now factorizes as
\begin{align}\label{eq:cffactss}
    C\(N_1,N_2,\frac{M^2}{\mu^2},\frac{M^2}{\mu^2},\frac{M^2}{\mu^2},\as(\mu^2)\)&=C^{(c)}\(N_2,\frac{M^2}{\mu^2},\as(\mu^2)\)C^{(l)}\(\frac{M^2/N_1}{\mu^2},N_2,\as(\mu^2)\)
    \nonumber\\
    +\mathcal{O}\(\frac{1}{N_1}\)
  \end{align}
  where
  \begin{align}
\label{eq:panccss}
    M^2\frac{d}{dM^2} \ln C^{(c)}\left(\frac{M^2}{\mu^2},N_2,\as(\mu^2)\right)&=  g_0^{\rm ss}(\as(M^2), N_2)\\
\label{eq:panclss}
    M^2\frac{d}{dM^2}\ln C^{(l)} \(\frac{M^2/N_1}{\mu^2},N_2,\as(\mu^2)\)&=\int_{M^2}^{M^2/N_1} \frac{\dmu^2}{\mu^2}
  g^{\rm ss}(\as(\mu^2),N_2).
    \end{align}
Terms that are logarithmically enhanced in $N_1$ are contained in $C^{(l)} \(\frac{M^2/N_1}{\mu^2},N_2,\as(\mu^2)\)$
 for which the expression
Eq.~(\ref{eq:panclss}) of   $ \gamma^{(l)}$ provides the resummed
prediction, while  $C^{(c)}(\frac{M^2}{\mu^2},N_2,\as(\mu^2))$ contain all
terms that do not vanish as $N_1\to\infty$ but remain finite in the
limit, i.e. constant in $N_1$ in Mellin space, thus proportional to a
$\delta(1-x_1)$ in physical space. All these terms depend
parametrically on $N_2$. Again, the separation between $C^{(l)}$
and $C^{(c)}$ is ambiguous.

\sect{The resummed coefficient function}
\label{sec:xsect}

The renormalization group argument leads to a resummed expression for
the logarithmic terms contained in
the physical anomalous dimension, which in turn determines the ratio of
the resummed coefficient function at two different scales. In order to
obtain a resummed expression for the coefficient function itself it is
necessary to separate off the dependence of the coefficient function
on the factorization scale from that on the soft scale that is being
resummed.  This
requires some care because the physical anomalous
dimension Eq.~(\ref{eq:gamp}) is a function of the pair of Mellin variables $N_1$ and
$N_2$, while  the dependence on the factorization scale is given by an
anomalous dimension that is a function of a single  Mellin variable, and
indeed, according to Eq.~(\ref{eq:ap}),  it provides the dependence on a single scale of the coefficient
when the factorization scales of the incoming partons are taken to be independent.
 To this
purpose, we first summarize how this is done in the inclusive
case~\cite{Forte:2002ni} in which only one scale and one Mellin variable is
present,
and then consider the double-soft limit (two Mellin variables and one
scale) and the 
single-soft limit (two Mellin variables and two scales).

\subsection{The inclusive coefficient function}\label{eq:cinf}
In the inclusive case, the resummed physical anomalous dimension is
given by Eq.~(\ref{eq:gampbardg}) but with $\Lambda^2_{\rm ds}$ replaced
by the soft scale $\Lambda^2_{\rm s}=\frac{M^2}{N^2}$. Logarithms of
$\frac{1}{N}$ are then resummed in the anomalous dimension $\gamma^{(l)}$, given by Eq.~(\ref{eq:padpl}) but with $N_1N_2$
replaced by $N^2$, still related to the coefficient function (now only
dependent on $N$) through Eq.~(\ref{eq:pancl}).
Integrating  Eq.~(\ref{eq:pancl}) the resummed coefficient function 
is given by~\cite{Forte:2002ni}
\begin{align}
E^{\rm res}(N;M_0^2,M^2)&\equiv\ln
C^{(l)}\(\frac{M^2/N^2}{\mu^2},\as(\mu^2)\)-\ln
C^{(l)}\(\frac{M_0^2/N^2}{\mu^2},\as(\mu^2)\)\nonumber\\\label{eq:eres}
&= -\int_1^{N^2} \frac{dn}{n}\, 
\int_{M_0^2}^{M^2}\frac{dk^2}{k^2}\,{g}\left(\as(k^2/n)\right).
\end{align}

Comparing to the solution to the renormalization group equation, which
has of course the form Eq.~(\ref{eq:cs}) but now with a single Mellin variable, it is clear that $\ln
C$ is the sum of a term that depends on the factorization scale
$\mu^2$ 
and a term that only depends on $\alpha_s$ evaluated at the scale of
the process. In order to rewrite the resummed result in this form we
let~\cite{Forte:2002ni}
\beq
\label{eq:gAB}
g(\as)=A(\as)-\frac{\partial D(\as(k^2))}{\partial\ln k^2}.
\eeq
where $A$ and $D$ are power series in $\as$, so that 
\begin{equation}\label{eq:eresab}
E^{\rm res}(N;M_0^2,M^2)=\int_{1}^{N^2} \frac{dn}{n} \left[\left(-\int_{M_0^2}^{M^2}
\frac{dk^2}{k^2} A(\as(k^2/n))\right)
+ D(\as(M^2/n))- D(\as(M_0^2/n))\right].
\end{equation}

We then conclude that the resummed coefficient function has the
general structure
\begin{equation}\label{eq:rescf}
 \ln C^{(l)}\(\frac{M^2/N^2}{\mu^2},\as(\mu^2)\)=\int_{1}^{N^2}
 \frac{dn}{n}\left[\left(-\int_{\bar \mu^2(\mu^2)}^{M^2}\frac{dk^2}{k^2}
A(\as(k^2/n))\right)+ D(\as(M^2/n))\right].
\end{equation}
Both the lower integration limit  $\bar\mu^2$ and the function
$A(\as(k^2/n))$ in Eq.~(\ref{eq:rescf}) are fully
determined by the renormalization group equation satisfied by the
coefficient  function and the known properties of the  $\overline{\rm MS}$ anomalous dimensions in the soft limit.

Indeed, differentiating Eq.~(\ref{eq:rescf}) we get
\begin{equation}\label{eq:diffcl}
\mu^2\frac{d}{d\mu^2} \ln C^{(l)}\(\frac{M^2/N^2}{\mu^2},\as(\mu^2)\)=\int_{1}^{N^2}
\frac{dn}{n}A(\as(\bar\mu^2/n))\frac{d\ln\bar\mu^2}{d\ln\mu^2},
\end{equation} 
but on the other hand, Eq.~(\ref{eq:ap}) implies that the right-hand
side of Eq.~(\ref{eq:diffcl})  must be matched to  the standard
anomalous dimension $\gamma^{\rm  AP}(N,\as(\mu^2))$ in the soft
limit. Because in this inclusive case there is a single scale, the
relevant anomalous dimension is the sum of the two anomalous dimensions
$\gamma_i$, $\gamma_j$ that correspond to the incoming evolution eigenstates.

Noting that by construction  only  $\gamma^{(l)}$ and $ C^{(l)}$
contain logarithmically enhanced contributions (constant
contributions having been included in $\gamma^{(c)}$),  the right-hand
side of Eq.~(\ref{eq:diffcl})  must be given by the logarithmically
enhanced contributions to the anomalous dimension  $\gamma^{\rm
  AP}$. Now, in the $\overline{\rm MS}$ scheme it can be
proven~\cite{Albino:2000cp}
that to all perturbative orders
\begin{equation}\label{eq:cusp}
\gamma_i^{\rm  AP}(N,\as(\mu^2))=-\ln N \sum_k \left(\frac{\as}{\pi}\right)^k A^{(i)}_k+O(N^0)+O\(\frac{1}{N}\),
\end{equation}
where $A^{(k)}_i$ are numerical coefficients and
only the diagonal splitting functions $P_{qq}$ and $P_{gg}$  are
logarithmically enhanced, so in the large-$N$ limit the anomalous
dimension eigenvectors are the 
quark singlet and gluon and  $i=q,g$.  The quantity on the right-hand
side of Eq.~(\ref{eq:cusp}) is
usually referred to as the cusp anomalous dimension.

Demanding that the right-hand side of  Eq.~(\ref{eq:diffcl}) coincides with (minus) the cusp anomalous
 dimension, with an extra factor of 2 to account for the presence of two partons of the same species in the initial state, we immediately find 
    \begin{align}\label{eq:lowint}
      \bar\mu^2&= n \mu^2\\\label{eq:Acusp}
      A^{(i)}(\as)&= \sum_k \left(\frac{\as}{\pi}\right)^k A^{(i)}_k,
    \end{align}
We therefore conclude that there are two independent resummed coefficient
functions, in the gluon-gluon and the quark-quark channel.
    
The full coefficient function in the soft limit is finally found by multiplying the
resummed expression for  $C^{(l)}\(\frac{M^2/N^2}{\mu^2},\as(\mu^2)\)$
by $C^{(c)}\(\frac{M^2}{\mu^2},\as(\mu^2)\)$, which in Mellin space is just a series in
$\as(M^2)$ with constant coefficients:
\begin{align}
&C(N,\frac{M^2}{\mu^2},\as(\mu^2))=C^{(c)}\(1,\as(M^2)\)
\nonumber\\
&\times\exp\int_{1}^{N^2}
 \frac{dn}{n}\left[\left(-\int_{\mu^2}^{M^2/n}\frac{dk^2}{k^2}
A(\as(k^2))\right)+ D(\as(M^2/n))\right]\(1+\mathcal{O}\(\frac{1}{N}\)\).
\label{eq:finresincalt}
\end{align}
 Specifically choosing $\mu^2=M^2$ and changing integration variables we get
\begin{align}
    &C(N,1,\as(M^2))=C^{(c)}\(1,\as(M^2)\)
    \nonumber\\
    &\times\exp\int_{M^2}^{M^2/N^2}
 \frac{dk^2}{k^2}\left[
A(\as(k^2))\ln\frac{M^2/N^2}{k^2}
+  D(\as(k^2))\right]\(1+\mathcal{O}\(\frac{1}{N}\)\).
\label{eq:finresm}
\end{align}

As well known (see e.g.~\cite{Forte:2025tli}), N$^k$LL
resummation is obtained by including both $C^{(c)}$ and $g(\as)$ up to
order $\alpha_s^{k+1}$, i.e.  $A(\as)$ up to order $\alpha_s^{k+1}$
and  $D(\as)$ up to order $\alpha_s^{k}$. Assuming knowledge of the
cusp anomalous dimension, all the remaining coefficients in $D$ and
$C^{(c)}$ are determined by comparing to a fixed N$^k$LO
computation. Note that in the dQCD literature (see e.g.~\cite{Moch:2005ba}) the
function $D$ is usually further divided into two separate functions
$B$ and $D$ that originate from different kinds of soft radiation,
though of course if they are determined by matching to fixed order no
such separation is possible. Note also that
in the SCET literature this is referred to as N$^k$LL', while N$^k$LL refers to including $C^{(c)}$ only up to order  $\alpha_s^{k}$
(see {\it e.g.} \cite{Becher:2007ty}).

\subsection{The coefficient function in the double-soft limit}
\label{sec:ds}
Coming now to the rapidity distribution in the soft limit, we can use
the expression Eq.~(\ref{eq:padpl}) of the resummed physical anomalous
dimension: we get
\begin{align}
E^{\rm res,\,ds}(N_1,N_2;M_0^2,M^2)&\equiv\ln
C^{(l),{\rm ds}}\(\frac{M^2/(N_1N_2)}{\mu^2},\as(\mu^2)\)-\ln
C^{(l),{\rm ds}}\(\frac{M_0^2/(N_1N_2)}{\mu^2},\as(\mu^2)\)\nonumber\\
\label{eq:eresds}
&= -\int_1^{N_1N_2}\frac{dn}{n}\int_{M_0^2}^{M^2} \frac{dk^2}{k^2} g^{\rm ds}(\as(k^2/n)).
\end{align}
The right-hand side coincides with the inclusive result
Eq.~(\ref{eq:eres}) with the substitution $N^2\to N_1N_2$. Given the
form  Eq.~(\ref{eq:cs}) of the solution to the renormalization group
equation, by simply making a common scale choice $\mu^2_1=\mu^2_2=\mu^2$  we thus
arrive through the same steps at the resummed coefficient function
\begin{equation}\label{eq:rescfds}
 \ln C^{(l),\rm ds}\(\frac{M^2/(N_1N_2)}{\mu^2},\as(\mu^2)\)=\int_{1}^{N_1N_2}
 \frac{dn}{n}\left[\left(-\int_{\bar \mu^2(\mu^2)}^{M^2}\frac{dk^2}{k^2}
A(\as(k^2/n))\right)
+ D^{\rm ds}(\as(M^2/n))\right].
\end{equation}

Matching to the Callan-Symanzik-Altarelli-Parisi
equation~(\ref{eq:ap}) is now performed based on the simple
observation that, because of Eqs.~(\ref{eq:N_1})
\begin{equation}\label{eq:N_1n2nsq}
  N_1N_2= N^2+\frac{b^2}{4}= N^2\( 1+ O\(\frac{b^2}{N^2}\)\).
\end{equation}
This  then immediately implies that consistency with the inclusive
case (which indeed corresponds to $b=0$) implies 
that 
Eqs.~(\ref{eq:lowint}-\ref{eq:Acusp}) still hold, and in particular
the functions $C^{(c)}$, $A$ and $D$ are the same as in the inclusive case,
so
\begin{align}
&C^{\rm ds}(N_1,N_2,\frac{M^2}{\mu^2},\frac{M^2}{\mu^2},\frac{M^2}{\mu^2},\as(\mu^2))
=C^{(c)}\(1,\as(M^2)\)
\nonumber\\
&\qquad\times\exp\int_{1}^{N_1N_2}
 \frac{dn}{n}\left[\left(-\int_{\mu^2}^{M^2/n}\frac{dk^2}{k^2}
A(\as(k^2))\right)+ D(\as(M^2/n))\right]\(1+\mathcal{O}\(\frac{1}{N_1},\frac{1}{N_2}\)\).
\label{eq:finresincaltds}
\end{align}

Hence double-soft resummation is simply obtained by performing the
substitution $N^2\to N_1N_2$ in the inclusive result:
\begin{equation}\label{eq:difftoinc}
C^{\rm ds}\(N_1,N_2,\frac{M^2}{\mu^2},\frac{M^2}{\mu^2},\frac{M^2}{\mu^2},\as(\mu^2)\)=C\(N,\frac{M^2}{\mu^2},\as(\mu^2)\)\( 1+\mathcal{O}\(\frac{b^2}{N^2}\)\)
\end{equation}
where $C(N,M^2/\mu^2,\as(\mu^2))$ is the coefficient function for the inclusive process, Eq.~\eqref{eq:finresincalt}, and $N^2=N_1N_2$ in the double-soft limit,
Eq.~\eqref{eq:N_1n2nsq}.
In particular, as in the
inclusive case, only the quark-quark and gluon-gluon  coefficient
survive the limit, and specifically for  the Drell-Yan process only the quark-quark channel
logarithmically enhanced.
As mentioned in the introduction, the result
Eq.~(\ref{eq:finresincaltds}) was first obtained to NLL accuracy in
Ref.~\cite{Catani:1989ne}, where it was used to derive resummation in
the double soft limit of the differential distribution with respect to
the longitudinal momentum of the $Z$ (sometimes called Feynman $x$). It
was then rewritten in Ref.~\cite{Westmark:2017uig} for the rapidity
distribution, and extended to NNLO and next-to-leading power in
Refs.~\cite{Banerjee:2018vvb,Ravindran:2022aqr}. In these references
an alternate form of the resummed results
Eq.~(\ref{eq:finresincaltds}-\ref{eq:finresmds}) is actually used;
its  equivalence to the resummed expression of this paper is proven in Appendix~\ref{app:equiv}.

Finally, choosing again $\mu^2=M^2$ we get
\begin{align}
    &C^{\rm ds}(N_1,N_2,1,1,1,\as(M^2))=C^{(c)}\(1,\as(M^2)\)
\nonumber\\
&\qquad\times\exp\int_{M^2}^{\tilde\Lambda^2_{\rm ds}}\left[ \frac{dk^2}{k^2}
A(\as(k^2))\ln\frac{\tilde\Lambda^2_{\rm ds}}{k^2}
+  D(\as(k^2))\right]\(1+\mathcal{O}\(\frac{1}{N_1},\frac{1}{N_2}\)\),
\label{eq:finresmds}
\end{align}
with $\tilde\Lambda^2_{\rm ds}$ the soft scale Eq.~(\ref{eq:genn}).
Of
course, we may equivalently choose any soft scale that is proportional
to  $\tilde\Lambda^2_{\rm ds}$. This is clear from the structure of the
renormalization group argument, since the latter only uses the fact
that the kinematic dependence on $N_1$ and $N_2$ goes through the
dimensionful variable   $\tilde\Lambda^2_{\rm ds}$. It is also manifest
from our final result Eq.~(\ref{eq:finresmds}), as a change of soft
scale by a constant factor can always reabsorbed by a change of the
coefficients of the expansion of the function $D^{\rm ds}$ in powers
of $\alpha_s$.
A common choice, suggested in Ref.~\cite{Catani:1989ne},
as it simplifies the expression of $D^{\rm ds}$ by reabsorbing some
subleading terms, is
\begin{equation}\label{eq:gennbar}
\bar\Lambda^2_{\rm ds}=\frac{M^2}{\bar N_1\bar N_2};\quad \bar
N_i\equiv e^{\gamma_E} N_i, 
\end{equation}
where $\gamma_E$ is the Euler constant.

\subsection{The coefficient function in the single-soft limit}\label{eq:ss}
The argument in the single-soft $N_1\to\infty$ limit requires treating
both the kinematic  variables  asymmetrically.
The scale dependence of  the coefficient function is now
given by 
\begin{align}
E^{\rm res,\,ss}(N_1,N_2;M_0^2,M^2)&\equiv\ln
C^{(l),\rm ss}\(\frac{M^2/N_1}{\mu^2},N_2,\as(\mu^2)\)-\ln
C^{(l),\rm ss}\(\frac{M^2_0/N_1}{\mu^2},N_2,\as(\mu^2)\)\nonumber\\\label{eq:eresss}
&= -\int_1^{N_1}\frac{dn}{n}\int_{M_0^2}^{M^2} \frac{d\mu^2}{\mu^2} g^{\rm ss}(\as(\mu^2/n),N_2),
\end{align}
leading to
\begin{align}\label{eq:rescfdsfin}
 &\ln C^{(l),\rm ss}\(\frac{M^2/N_1}{\mu^2},N_2,\as(\mu^2)\)
 \nonumber\\
 &=\int_{1}^{N_1}
 \frac{dn}{n}\left[\left(-\int_{\bar \mu^2(\mu^2)}^{M^2}\frac{dk^2}{k^2}
A^{\rm ss}(\as(k^2/n,N_2))\right)
+ D^{\rm ss}(\as(M^2/n),N_2)\right].
\end{align}

We now note that, because the PDF $f_2(N_2,\mu_2^2)$ in
Eq.~(\ref{eq:doublerg}) only depends on $N_2$, changing its
scale cannot introduce any large logs of the soft variable, which is
$M^2/N_1$. It follows that
\begin{equation}\label{eq:asymscale}
C^{\rm ss}\(N_1,N_2,\frac{M^2}{\mu_1^2},\frac{M^2}{\mu_2^2},\frac{M^2}{\mu_R^2},\as(\mu_R^2)\)
=C^{\rm ss}\(N_1,N_2,\frac{M^2}{\mu_1^2},\frac{M^2}{\mu_1^2},\frac{M^2}{\mu_R^2},\as(\mu_R^2)\)(1+\mathcal{O}(N_1^0)).
\end{equation}
We thus consider the coefficient function with the choice
$\mu_1^2=\mu^2_R=\mu^2$ but $\mu_2^2=M^2$, which
satisfies 
\begin{equation}\label{eq:diffclss}
  \mu^2\frac{d}{d\mu^2} \ln C^{\rm ss}_{ki}\(N_1,N_2,\frac{M^2}{\mu^2},1,\frac{M^2}{\mu^2},\as(\mu^2)\)= \ln N_1\sum_n \as^n(\mu^2)
A^{(n)}_k+\mathcal{O}\(N_1^0\),
\end{equation}
where we have restored the parton channel indices, and now $k$
corresponds to the diagonal quark or gluon logarithmically enhanced
channel, while $i$ runs over all eigenstates of perturbative evolution.
On the other hand, because of Eq.~(\ref{eq:asymscale}),
\begin{equation}\label{eq:diffmatch}
\mu^2\frac{d}{d\mu^2} \ln
C_{ki}^{(l),\rm ss}\(\frac{M^2/N_1}{\mu^2},N_2,\as(\mu^2)\)=
\mu^2\frac{d}{d\mu^2} \ln
C^{\rm ss}_{ki}\(N_1,N_2,\frac{M^2}{\mu^2},1,\frac{M^2}{\mu^2},\as(\mu^2)\)+\mathcal{O}(N_1^0),
\end{equation}
where we have restored the parton indices $ki$ also on the function $C_{ki}^{(l),\rm ss}$
that contains the logarithmically enhanced terms.

Hence the $\mu_1^2$ factorization scale dependence of the
coefficient function in the single-soft limit must match the cusp
anomalous dimension, and thus
cannot depend on $N_2$. We consequently
arrive at a resummed expression for the coefficient function with
asymmetric scale choice in the single soft limit of the form 
\begin{align}
&C^{\rm ss}_{ki}\(N_1,N_2,\frac{M^2}{\mu^2},1,\frac{M^2}{\mu^2},\as(\mu^2)\)=C_{ki}^{(c),{\rm ss}}\(1,\as(M^2),N_2\)
\nonumber\\
&\quad \times\exp\int_{1}^{N_1}\frac{dn}{n}\left[\left(-\int_{\mu^2}^{M^2/n}\frac{dk^2}{k^2}
A_k(\as(k^2))\right)+ D^{\rm ss}_{ki}(\as(M^2/n),N_2)\right]\(1+\mathcal{O}\(\frac{1}{N_1}\)\).
\label{eq:finresincaltss}
\end{align}
For the Drell-Yan process only quark channels are logarithmically
enhanced, so $k=q$ or $k=\bar q$, while $i$ runs over all eigenstates
of perturbative evolution, with $C^{\rm ss}_{qi}=C^{\rm ss}_{\bar qi}$. We will denote henceforth the single-soft
Drell-Yan coefficient functions as $C^{\rm ss}_{qi}$ for definiteness.

Finally restoring a symmetric scale choice  $\mu^2=M^2$ we get
\begin{align}
    &C^{\rm ss}_{qi}(N_1,N_2,1,1,1,\as(M^2))=C_{ki}^{(c),{\rm ss}}\(1,\as(M^2),N_2\)
\nonumber\\
&\quad\times
    \exp\int_{M^2}^{\tilde\Lambda^2_{\rm ss}}
 \frac{dk^2}{k^2}\left[
A_k(\as(k^2))\ln\frac{\tilde\Lambda^2_{\rm ss}}{k^2}+  D^{\rm ss}_{ki}(\as(k^2),N_2)\right]\(1+\mathcal{O}\(\frac{1}{N}\)\),
\label{eq:finresmss}
\end{align}
with the soft scale now given by $\tilde\Lambda^2_{\rm ss}$
Eq.~(\ref{eq:gennss}).
Again, we are free to redefine the soft scale by a constant factor, so
it might be useful to take instead 
\begin{equation}\label{eq:gennssbar}
\bar\Lambda^2_{\rm ss}=\frac{M^2}{\bar N_1};\quad \bar
N_1\equiv e^\gamma_E N_1. 
\end{equation}
However, in the single-soft limit, the non-soft variable $N_2$ is just a fixed parameter. Hence, we may also use  $\tilde\Lambda^2_{\rm ds}$ Eq.~(\ref{eq:genn})
or  $\bar\Lambda^2_{\rm ds}$ Eq.~(\ref{eq:gennbar}) instead of
$\tilde\Lambda^2_{\rm ss}$ in the resummation formula
Eq.~(\ref{eq:finresmss}), with the only effect of changing the
coefficients of the expansion of  the function $D^{\rm
  ss}_{ki}(\as(k^2),N_2)$ in powers of $\alpha_s$.

\section{Matching with fixed order calculations}
\label{sec:fixed-order}

The explicit values of the coefficients in the expansions of
$C^{(c)}$, $A$, and $D$ in powers of the strong coupling are obtained
by expanding the resummed  coefficient function at fixed order in
$\alpha_s$ and requiring the resummed result to match the Mellin
transform of the fixed-order computation. We perform
the matching procedure explicitly up to NNLO, which allows us to obtain a
resummed expression with NNLL accuracy, both in the double-soft and
single-soft limit. In the single-soft case in this paper we will only
consider the $C_{q\barq}$ coefficient function, and only resum the
quark nonsinglet channel.

To this purpose, we first compute explicitly the integrals in the
resummed expressions
Eqs.~(\ref{eq:finresmds},\ref{eq:finresmss}) and
expand out the result up to order $\alpha^2_s(M^2)$. We then provide
the fixed-order NNLO expressions, which were computed in
Ref.~\cite{Anastasiou:2003yy,Anastasiou:2003ds} and are available as a public {\tt
  FORTRAN} code~\cite{vrap}, but must be transformed into  the
variables $x_1$, $x_2$ in terms of whose Mellin conjugates $N_1$,
$N_2$ is expressed. We finally determine up to $\mathcal{O}(\alpha^2_s)$ the
functions $C^{(c)}$ and $D$ that, along with the cusp anomalous
dimension, control the resummed result.

\subsection{Explicit resummed results}
\label{sec:explicit}

We start by noting that the resummed expressions we are interested in,
namely the double-soft result Eq.~(\ref{eq:finresmds}) and the single-soft
result Eq.~(\ref{eq:finresmss})  have the same form if
expressed in terms of a variable
\begin{equation}\label{eq:lamdef}
 \lambda=\beta_0\alpha_s(M^2) \mathcal{L},
\end{equation}
where $\mathcal{L}$ is a large log, such as $\ln N_1N_2$, and of the
coefficients of the expansion of the function $D$, only with different
values of the coefficients of this expansion.

Specifically, we can write the resummed coefficient function as
\begin{equation}
    \label{eq:gexp}
    C_{qi}\left(N_1,N_2,\as(M^2)\right) = g_0(\as(M^2))\exp\left(\frac{1}{\as(M^2)}g_1(\lambda)+g_2(\lambda)+\as g_3(\lambda)+\ldots\right).
\end{equation}
Note that this defines unambiguously the separation between
logarithmic and soft contribution, i.e. it provides an unambiguous
definition of $C^{(l)}$
and $C^{(c)}$, which are otherwise not uniquely defined, as one can
always reshuffle a constant between them. 
As mentioned,  we provide results for the
Drell-Yan $C_{q\barq}$ coefficient function, with only the quark
nonsinglet channel included in the single-soft case. We will
consequently henceforth omit the parton indices on the coefficient
function.
Also, throughout the
section we choose the renormalization and
factorization scales
\begin{equation}\label{eq:scales}
  \mu_1^2=\mu_2^2=\mu_R^2=M^2,
\end{equation}
and in order to lighten the notation we omit the corresponding
arguments in the coefficient function.

Performing the integrals in
Eqs.~(\ref{eq:finresmds},\ref{eq:finresmss}) we
then get
\begin{align}\label{eq:gone}
g_1(\lambda)&=\frac{A_1 [\lambda +(1 -\lambda)\ln(1 - \lambda)]}{\pi\beta_0^2}
\\\label{eq:gtwo}
g_2(\lambda)&=\frac{A_1 \beta_1}{2 \pi \beta_0^3} 
\left( 2\lambda + \ln^2(1 - \lambda) + 2 \ln(1 - \lambda) \right) 
- \frac{A_2}{\pi^2 \beta_0^2} \left( \lambda + \ln(1 - \lambda) \right) 
+ \frac{D_1 \ln(1 - \lambda)}{\pi \beta_0}
\\\label{eq:gthree}
g_3(\lambda) &=
\frac{A_1 \beta_1^2}{\pi \beta_0^4 (1 - \lambda)}
\left( \frac{\lambda^2}{2} + \frac{1}{2} \ln^2(1 - \lambda) + \lambda \ln(1 - \lambda) \right) \nonumber \\
&+ \frac{A_1 \beta_2}{\pi \beta_0^3 (1 - \lambda)}
\left( \ln(1 - \lambda) - \lambda \ln(1 - \lambda) - \frac{\lambda^2}{2} + \lambda \right) \nonumber \\
&- \frac{A_2 \beta_1}{\pi^2 \beta_0^3 (1 - \lambda)}
\left( \frac{\lambda^2}{2} + \lambda + \ln(1 - \lambda) \right)
+ \frac{A_3 \lambda^2}{2 \pi^3 \beta_0^2 (1 - \lambda)} \nonumber \\
&- \frac{D_2 \lambda}{\pi^2 \beta_0 (1 - \lambda)}
+ \frac{D_1 \beta_1}{\pi \beta_0^2 (1 - \lambda)}
\left( \lambda + \ln(1 - \lambda) \right),
\end{align}
where we have written the beta function as
\begin{align}
&\mu^2\frac{d\as(\mu^2)}{d\mu^2}=-\beta_0\as^2(\mu^2)-\beta_1\as^3(\mu^2)+O(\as^4)
\\
&\beta_0=\frac{11 C_A - 4 T_F N_f}{12 \pi} \label{eq_beta_0}
\\
&\beta_1=\frac{17 C_A^2 - 10 C_A T_F N_f - 6 C_F T_FN_f}{24\pi^2}. \label{eq_beta_1}
\end{align}
$A_k$ are the coefficients in the expansion of the quark cusp
anomalous dimension Eq.~(\ref{eq:Acusp})  and $D_i$ are the
coefficients of the corresponding expansion
\begin{align}\label{eq:dexp}
D(\as)=D_1\frac{\as}{\pi}+D_2\(\frac{\as}{\pi}\)^2+\dots
\end{align}
of the $D$ function. Note that $D$ is a function of $N_2$ in the
single-soft limit, with different values according to the specific
soft scale choice, and just a constant in the double-soft limit.
We also define the $g_0$ function as an expansion of the form 
\begin{align}\label{eq:g0exp}
  g_0(\as)=1+g_{01}\frac{\as}{\pi}+g_{02}\(\frac{\as}{\pi}\)^2+g_{03}\(\frac{\as}{\pi}\)^3+\dots.
\end{align}

We finally give the explicit expression of the coefficient function
Eq.~(\ref{eq:gexp}), expanded out up to NNLO:
\begin{align}\label{eq:expres}
    C\left(N_1,N_2,\as(M^2)\right) &=  1+\frac{\as}{\pi}\left(\frac{A_1}{2} \mathcal{L}^2 - D_1\mathcal{L}+g_{01}\right) \nonumber \\
    &+\left(\frac{\as}{\pi}\right)^2 \Bigg[ \frac{A_1^2}{8}       
        \mathcal{L}^4+ \mathcal{L}^3\left(\frac{1}{6}A_1\pi \beta_0
        - \frac{1}{2} A_1 D_1\right)
\nonumber \\
&
+\mathcal{L}^2 \left(\frac{1}{2}A_1 g_{01} +\frac{1}{2} A_2 - \frac{1}{2}\pi \beta_0 D_1 +\frac{1}{2}D_1^2 \right)\nonumber \\
    &+\mathcal{L}\left(- D_1 g_{01}- D_2\right)+g_{02} \Bigg]+\mathcal{O}(\as^3).
\end{align}
The coefficients $A_i$, $D_i$ and $g_{0i}$ corresponding to the
various limits will be determined by matching Eq.~(\ref{eq:expres}) to
the fixed-order expression.

Note that all results given in this section are entirely generic, and
hold for any resummed formula of the form of
Eqs.~(\ref{eq:finresmds},\ref{eq:finresmss}) regardless of the
particular limit (single soft or double soft), of the choice of soft
scale, and of the specific coefficient function and parton channel
that is being resummed. What will change in each case is the form (and
functional dependence) of the coefficients  $D_i$ and $g_{0i}$.

\subsection{The NNLO Drell-Yan differential distribution}
\label{sec:nnlody}
The NLO Drell-Yan rapidity distribution was first computed in
Ref.~\cite{Altarelli:1979ub} in terms of the variable $\tau$
Eq.~(\ref{eq:taudef}) and the scaling variable $x_F$, which is related
to the difference of the variables $x_1$ and $x_2$
Eq.~(\ref{eq:variables}). The NNLO result was first obtained in
Refs.~\cite{Anastasiou:2003yy,Anastasiou:2003ds}, where it is expressed in terms of
$\tau$ and a variable $u$ which in terms of $\tau$ and the rapidity
$y$ Eq.~(\ref{eq:p_H}) is given by
\begin{equation}  \label{eq:udef}
u=\frac{1}{1-\tau}\frac{1-\tau e^{2y}}{1+e^{2y}}.
\end{equation}
The variable $u$ is simply related to the scattering angle of the $Z$
in the partonic center-of-mass frame in the single and double soft limits, as shown in
Appendix~\ref{app:u}.

In
Ref.~\cite{Anastasiou:2003yy,Anastasiou:2003ds} the NLO cross section is also
explicitly provided in terms of these variables, while the NNLO is
available through the {\tt Vrap}~\cite{vrap} code.
In order to compare and match to the resummation, however, the fixed
order result must be expressed in terms of the scaling variables
$x_i$. At NLO this was first done in
Ref.~\cite{Westmark:2017uig,Westmarkphd}, and then more recently, correcting
 some errors, in Ref.~\cite{Bonvini:2023mfj}. However, the NNLO expression
 in terms of $x_i$ is not available.

 The transformation between the variables $x_1$, $x_2$  and the variables $\tau$ and $u$ is 
\begin{align}\label{eq:uttox1}
  x_1&=\sqrt{\tau}\sqrt{\frac{1-u(1-\tau)}{\tau+u(1-\tau)}},\\\label{eq:uttox2}
  x_2&=\sqrt{\tau}\sqrt{\frac{\tau+u(1-\tau)}{1-u(1-\tau)}},
\end{align}
with inverse
\begin{align}\label{eq:x1x2tot}
  \tau&=x_1x_2,\\
  \label{eq:x1x2tou}
 u&=\frac{x_2(1-x_1^2)}{(x_1+x_2)(1-x_1x_2)}.
\end{align}
The coefficient function $C(x_1,x_2,M^2)$ defined in
Eq.~(\ref{eq:cfdeflo}) enters the computation of hadronic cross sections
through a convolution with a pair of parton distributions that
respectively depend on $x_1$ and $x_2$. We can thus view it as a
distribution acting on a test function $T(x_1,x_2)$. The results of
Ref.~\cite{Anastasiou:2003yy,Anastasiou:2003ds} provide an expression
for the coefficient function $\bar C(\tau, u,M^2)$ that satisfies
\begin{equation}\label{eq:cbardef}
  \int_0^1d\tau\int_0^1 du\, \bar C(\tau,u,M^2) t(\tau,u)=  \int_0^1dx_1
  \int_0^1 dx_2 \, C(x_1,x_2,M^2) T(x_1,x_2)
\end{equation}
where
\begin{equation}\label{eq:testfun}
  t(\tau,u)=T(x_1(\tau,u),x_2(\tau,u)),
  \end{equation}
so that
\begin{equation}\label{eq:ctocbar}
 C(x_1,x_2,M^2)= j(x_1,x_2)\bar C(\tau(x_1,x_2),u(x_1,x_2),M^2)
\end{equation}
with
\begin{equation}\label{eq:invjac}
j(x_1,x_2)=\left|\frac{\partial(\tau,u)}{\partial(x_1,x_2)}\right|.
\end{equation}

Equations~(\ref{eq:uttox1},\ref{eq:x1x2tou})  show that the single-soft limit  $x_1\to 1$ corresponds to $u\to0$, and
$x_2\to 1$ corresponds to $u\to1$: indeed, as mentioned 
(see  Appendix~\ref{app:u}) $u$ is related to the scattering angle and $u=0$ and $u=1$ are the collinear limits corresponding to the scattering angle of the final-state $Z$ being along  the directions of either of the
incoming partons.  The double-soft limit corresponds to
$\tau\to1$, and the transformation becomes singular in this
limit, i.e. the jacobian $j(x_1,x_2)$ Eq.~(\ref{eq:invjac}) diverges,
because when $\tau=1$ a regular function of $\tau$ and $u$
becomes $u$-independent.

The transformation of the coefficient function from $(\tau,u)$ to $(x_1,x_2)$
space is accordingly nontrivial in the soft limit, because it requires
mapping distributions in $\tau$ and $u$ localized at $\tau=1$ and at
$u=0$ or $u=1$ in terms of distributions in $x_1$ and $x_2$
localized at $x_1=1$ and $x_2=1$ through a singular jacobian.
A general expression of the transformations for all possible
distributions of $u$ only, $\tau$ only, or all products of
distributions in $u$ and $\tau$ is quite cumbersome. However, our task
is facilitated by the fact that only specific combinations of
distributions appear in the resummed result.

Indeed, in the double-soft limit we know from the discussion in
Sect.~\ref{sec:dsoft} that the coefficient function
only depends on $M^2(1-x_1)(1-x_2)$: hence, only logs of
$(1-x_1)(1-x_2)$ can appear, multiplied by  associate double
distributions localized at $x_1=1$ and $x_2=1$. Hence, we only need to
work  out the $(\tau,u)$ expression of these distributions,  which
corresponds to specific combinations of double and single distributions
in $\tau$ and $u$ which we can then recognize in the NNLO coefficient
function.
In the
single-soft limit $\tau$ cannot reach the value $\tau=1$, hence only
single distributions in $u$ localized at either $u=0$ or $u=1$, and
we only need to work out the expressions of these.

\subsubsection{The NNLO coefficient function in $(\tau,u)$ space}\label{sec:ctauu}
We write the expansion of the coefficient function $\bar
C(\tau,u,M^2)$ Eqs.~(\ref{eq:cbardef}-\ref{eq:ctocbar})
\begin{equation}
\bar C(\tau,u,\as) =\bar C^{(0)}(\tau,u)+ \bar
C^{(1)}(\tau,u)\frac{\as}{\pi} + \bar C^{(2)}(\tau,u)\left(\frac{\as}{\pi}\right)^2+O(\as^3).
\end{equation}
The LO term fixes the normalization of the coefficient function: as we
shall show explicitly in Sect.~\ref{sec:x1x2} below,
\begin{equation}\label{eq:tulo}
  \bar C^{(0)}(\tau,u)  = \delta(1-\tau)\frac{\delta(u)+\delta(1-u)}{2}
  \end{equation}
is consistent with the choice of normalization of Eq.~(\ref{eq:cfdeflo}).

At higher orders, following Ref.~\cite{Anastasiou:2003ds}, we write
the coefficient function as the sum of three contributions: $\bar
C^{(i)}_\text{Born}$, which is purely virtual and has the same
kinematics of the leading order;  $\bar
C^{(i)}_\text{Boost}$, which is proportional to
$\delta(u)+\delta(1-u)$ and thus contains real radiation contributions
that are  collinear to either of the incoming partons; and  $\bar
C^{(i)}_\text{Real}$ that contains the rest of the coefficient
function, corresponding to generic real emission contributions:
\begin{equation}\label{eq:fctudecomp}
\bar C^{(n)}(\tau,u)=\delta(1-\tau) \frac{\delta(u) +
  \delta(1-u)}{2}\bar C^{(n)}_\text{Born} 
+\frac{\delta(u)+\delta(1-u)}{2}\bar C^{(n)}_\text{Boost} (\tau)
+\bar C^{(n)}_\text{Real} (\tau,u).
\end{equation}
At
NLO we provide the full expression of the quark-quark coefficient
function. At NNLO we  provide the
expression for the contributions that are distributional in at least one
variable to $C_{q\bar q}$ in the quark-quark
nonsinglet channel, i.e. such that the incoming quark or antiquark leg
is the same that couples to the gauge boson. These have been extracted
by us
from the  {\tt
  FORTRAN} code~\cite{vrap}, following the decomposition into
individual contributions given in Ref.~\cite{Anastasiou:2003ds}, Eqs.~(5.1-5.3).

At NLO we have
\begin{align}
&\bar C^{(1)}_\text{Born} = C_F\left(\frac{\pi ^2}{3}-4\right) 
\\
&\bar C^{(1)}_\text{Boost}(\tau) =\frac{C_F}{\tau}\left(4\left[\frac{\ln(1-\tau)}{1-\tau}\right]_+-\tau(1+\tau^2)\frac{\ln\tau}{1-\tau}-2(2+\tau+\tau^2)\ln(1-\tau)+\tau(1-\tau)\right) 
\\
&\bar C^{(1)}_\text{Real}(\tau,u)=C_F\left(\left[\frac{1}{u}\right]_+ + \left[\frac{1}{1-u}\right]_+ \right)\left(\left[\frac{1}{1 - \tau}\right]_+-\frac{1+\tau}{2}\right)-(1-\tau).
\end{align}

At NNLO we find
\begin{align}\label{eq:cborn2}
&\bar C^{(2)}_\text{Born} = - \frac{2561}{144}  + \frac{13 }{9}\pi^2+ \frac{1}3\zeta_3  - \frac{19 }{1620}\pi^4 +  \left( \frac{127}{72}- \frac{14 }{81}\pi^2+\frac{2 }3\zeta_3   \right)N_f,
\\
    &\bar C^{(2)}_\text{Boost} (\tau) \nonumber \\\label{eq:cboost2}
    &= \frac{128}{9} \left[ \frac{\ln^3(1-\tau)}{1-\tau} \right]_+ + \left(-\frac{44}3+\frac{8}{9}N_f\right) \left[ \frac{\ln^2(1-\tau)}{1-\tau} \right]_++ \left( \frac43 - \frac{40 }{27}N_f - \frac{4 }{27}\pi^2 \right) \left[ \frac{\ln(1-\tau)}{1-\tau} \right]_+ \nonumber \\
    &+ \left( -\frac{404}{27} + \frac{11 \pi^2}{9} + N_f \left(
\frac{56}{81} - \frac{2\pi^2}{27} \right) + \frac{190}{9} \zeta_3
\right)\left[ \frac{1}{1-\tau} \right]_+ + 2\bar \kappa(\tau),\\\label{eq:creal2}
    &\bar C^{(2)}_{\text{Real}} (\tau,u) = \frac{16}3 \Bigg( \frac{1}{2} \left[ \frac{1}{1-\tau} \right]_+ \left[ \frac{\ln^2u}{u} \right]_+ + 2 \left[ \frac{1}{u} \right]_+ \left[ \frac{\ln^2(1-\tau)}{1-\tau} \right]_+ + 2 \left[ \frac{\ln u }{u} \right]_+ \left[ \frac{\ln(1-\tau)}{1-\tau} \right]_+ \nonumber \\
    &+2\frac{\ln u }{1-u} \left[ \frac{\ln(1-\tau)}{1-\tau} \right]_++ \left(\frac{\ln(1-u)\ln u}{1-u} + \frac{1}{2} \frac{\ln^2 u}{1-u} \right) \left[ \frac{1}{1-\tau} \right]_+\Bigg)  \nonumber \\
    &+ \left( \frac{2}{9}N_f -\frac{11}3  \right) \left(\left[ \frac{1}{1-\tau} \right]_+ \left[ \frac{\ln u}{u} \right]_+ + 2\left[ \frac{1}{u} \right]_+ \left[ \frac{\ln(1-\tau)}{1-\tau} \right]_+ + \frac{\ln u}{1-u} \left[\frac{1}{1-\tau}\right]_+ \right) \nonumber \\
    &+\left( \frac13 - \frac1{27}\pi^2 - \frac{10}{27}N_f \right) \left[ \frac{1}{u} \right]_+ \left[ \frac{1}{1-\tau} \right]_+ - \frac{4}{3}(1+\tau) \left[ \frac{\ln^2 u }{u} \right]_+ \nonumber \\ 
    &+ \Bigg(\left(\frac{11}{6}-\frac{1}{9}N_f\right)(1+\tau)-\frac{16}{3}(1+\tau)\ln(1-\tau) \nonumber \\
    &-\frac{4}{9} (7+9\tau^2) \frac{\ln\tau}{1-\tau} + \frac{16}{9} (1+\tau^2) \frac{\ln\left(\frac{1+\tau}{2}\right)}{1-\tau}\Bigg) \left[ \frac{\ln u }{u} \right]_+ \nonumber \\
    &+\Bigg( -\frac{16}{3}\ln^2(1-\tau) + \left(\frac{11}{3} - \frac{2}{9}N_f\right)(1+\tau)\ln(1-\tau) +\frac{1}{18}(37+61\tau^2)\frac{\ln^2\tau}{1-\tau} \nonumber \\
    &+\frac{4}{9} \left( 8(1+\tau^2)\ln(1-\tau) - (3+5\tau^2)\ln \tau \right)\frac{\ln\frac{1+\tau}2}{1-\tau} 
\nonumber \\
    &+\frac{1}{9} \Bigg( 26-8\tau+36\tau^2 - 2N_f(1+\tau^2) + 4(1+\tau)(1-\tau)\ln 2 - 4 (17+19\tau^2) \ln(1-\tau)\Bigg) \frac{\ln(\tau)}{1-\tau} \nonumber \\
    & + \frac{4}{9}(1+\tau) \left(\text{Li}_2(-\tau)-\text{Li}_2(\tau)\right) + \frac{1}{9}(14-17 \tau)+\frac{7}{54} (1+\tau)\pi^2 + \frac{2}{27}(1+4\tau)N_f \Bigg) \left[\frac{1}{u}\right]_+ \nonumber \\
    &+ (u \mapsto 1-u).
\end{align}
The function $\bar\kappa(\tau)$ is given explicitly in Appendix~\ref{app:xconst}.

\subsubsection{The NNLO coefficient function in $(x_1,x_2)$ space}
\label{sec:x1x2}
We now turn to the expressions of the expansion coefficients of the
coefficient function $C(x_1,x_2, M^2)$ Eq.~(\ref{eq:cfdeflo}). In order
to obtain it we have transformed all the combinations of distributions
that appear in the NLO and NNLO coefficient functions from 
$(\tau,u)$ space to   $(x_1,x_2)$ space. We have performed the
transformation by obtaining the relevant distributions from a
generating function, following the approach used  in
Ref.~\cite{Forte:2002ni} for the inclusive coefficient function, and
performing the transformation at the level of generating
functions. The procedure is discussed in Appendix~\ref{app:distr}, and
specifically all the distributional identities needed in order to
perform the transformation are collected in Appendix~\ref{sec:distresults}.

At leading order we get
\begin{equation}\label{eq:locf}
  C^{(0)}(x_1,x_2)=\delta(1-x_1) \delta(1-x_2),
\end{equation}  
consistently with Eq.~(\ref{eq:cfdeflo}), as already mentioned.
Beyond leading order we separate again terms with Born kinematics,
which, like the leading order, are proportional to a double
delta distribution; terms that are proportional to a single Dirac delta
in $x_1$ or $x_2$, and which therefore, based on the analysis of
Sect.~\ref{sec:kinssl}, include contributions from radiation that is  collinear to either of the incoming
partons; and terms that correspond to generic real emission:
\begin{equation}\label{eq:cfxdecomp}
    C^{(n)} (x_1,x_2) = \delta(1-x_1)\delta(1-x_2)
    C^{(n)}_\text{Born}+ \delta(1-x_1)C^{(n)}_\text{Boost} (x_2) +
    \delta(1-x_2) C^{(n)}_\text{Boost} (x_1) +C^{(n)}_\text{Real} (x_1,x_2).
\end{equation}
Note that, as it is clear from the relations between distributions
from Appendix~\ref{sec:distresults}, the three contributions to
Eq.~(\ref{eq:cfxdecomp}) are not respectively obtained by transforming
the corresponding three contributions in
Eq.~(\ref{eq:fctudecomp}). This is due to the fact that it is
generally possible to redefine a plus distribution by addition term
proportional to a Dirac delta, corresponding to the fact that in
Mellin space it is generally possible to redefine a log by addition of
a constant. Hence the three terms can get reshuffled according to how
one defines the plus distribution in each case.

At NLO we get
\begin{align}
&C^{(1)}_\text{Born}=\frac{C_F}{2} \left(\pi ^2-8\right)
\\
&C^{(1)}_\text{Boost} (x_2) = C_F \left(\left[\frac{\ln(1-x_2)}{1-x_2}\right]_+-\frac{1+x_2^2}{2} \frac{\ln\frac{1+x_2}{2}}{1-x_2}-\frac{1+x_2}{2}\ln(1-x_2)+\frac{1-x_2}{2}\right)
\\
&C^{(1)}_\text{Real}=C_F\Bigg(\left[\frac{1}{1-x_1}\right]_+\left[\frac{1}{1-x_2}\right]_+- \frac{1+x_1}{2} \left[\frac{1}{1 - x_2}\right]_+-\frac{1+x_2}{2} \left[\frac{1}{1 - x_1}\right]_+ 
\nonumber \\
 &+\frac{(x_1^2+x_2^2)((1+x_1^2)+(1+x_2^2)+2x_1 x_2(3+x_1 + x_2 +x_1 x_2))}{2(1+x_1)(1+x_2)(x_1+x_2)^2}\Bigg).
\end{align}

At NNLO the individual terms are given by
\begin{align}\label{eq:cborn2x}
&C^{(2)}_\text{Born} =-\frac{2561}{144} + \frac{13}{9}\pi^2 + \frac{\zeta_3}{3} - \frac{19}{1620}\pi^4 +\left( \frac{127}{72} - \frac{14}{81}\pi^2 + \frac{2}{3}\zeta_3 \right) N_f
\\\label{eq:cboost2x}
& C^{(2)}_\text{Boost} (x_2) = \frac{64}{9}\left[\frac{\ln^3(1-x_2)}{1-x_2}\right]_{+} + \left( \frac{4}{9} N_{f}-\frac{22}{3}\right) \left[\frac{\ln^2(1-x_2)}{1-x_2}\right]_{+} \nonumber \\
    &+ \left(-\frac23 - \frac2{27}\pi^2 - \frac{20}{27} N_{f}\right) \left[\frac{\ln(1-x_2)}{1-x_2}\right]_{+} \nonumber \\
    & + \left(-\frac{202}{27} + \frac{11}{18}\pi^2 + \frac{95}{9}\zeta_3 + N_{f}\left(\frac{28}{81} - \frac{\pi^2}{27}\right)\right)\left[\frac{1}{1-x_2}\right]_{+} + \frac{1}{2} \kappa(x_2),\\\label{eq:real2x}
    & C^{(2)}_\text{Real} (x_1, x_2) = \left\{\frac{8}{3}\left(\left[\frac{1}{1 - x_{2}}\right]_{+}-\frac{1+x_2}{2} \right) \left[\frac{\ln^{2}(1 - x_{1})}{1 - x_{1}}\right]_{+} + (x_1 \leftrightarrow x_2) \right\} \nonumber \\
    & + \frac{16}{3}\left[\frac{\ln(1 - x_{1})}{1 - x_{1}}\right]_{+}\left[\frac{\ln(1 - x_{2})}{1 - x_{2}}\right]_{+} + \Bigg\{ \left[\frac{\ln(1 - x_{1})}{1 - x_{1}}\right]_{+} \Bigg(\left(-\frac{11}{3} + \frac{2}{9} N_{f}\right) \left[\frac{1}{1 - x_{2}}\right]_{+} \nonumber \\
    &-\frac{8}{9} (1 + x_{2}^{2}) \frac{\ln\frac{1 + x_{2}}{2}}{1 - x_{2}} -\frac{4}{9}\left(1 + 3 x_{2}^{2}\right) \frac{\ln x_{2}}{1 - x_{2}} -\frac{8}{3}\left(1+x_2 \right)\ln(1 - x_{2}) \nonumber \\
    &+\frac{11}{6} (1+x_2)-\frac{1}{9}(1+x_2) N_{f} \Bigg) +(x_1 \leftrightarrow x_2)\Bigg\} \nonumber \\
    &+ \left(\frac{1}{3} - \frac{10}{27} N_{f} - \frac{1}{27} \pi^{2}\right) \left[\frac{1}{1 - x_{1}}\right]_{+}\left[\frac{1}{1 - x_{2}}\right]_{+} \nonumber \\
    &\Bigg\{\left[\frac{1}{1 - x_{1}}\right]_{+}\Bigg(-\frac{4}{9} (1 + x_{2}) \left(\text{Li}_{2}(x_{2}) - \text{Li}_{2}(-x_{2})\right) -\frac{4}{9}  (1 + x_2^2)  \frac{\ln^2\frac{1+x_2}{2}}{1 - x_2} \nonumber \\
    &+\Big(\frac{11}{6}(1+x_2^2)-\frac{1}{9}(1+x_2^2)N_f - \frac{8}{9} (1+x_2^2)\ln(1-x_2)+\frac{1}{9}(11+5x_2^2)\ln(x_2)\Big) \frac{\ln\left(\frac{1+x_2}{2}\right)}{1 - x_2} \nonumber \\
    &-\frac{1 }{18 }(5+13x_2^2) \frac{\ln^2(x_2)}{1-x_2}+\Bigg(-\frac{1}{9}(1+x_2^2) N_f + \frac{7}{9}(1-x_2^2) \ln(2)- \frac{16}{9} (1 + x_2^2) \ln(1 - x_2) \nonumber \\
    & \frac{19-16x_2+39x_2^2}{18}-\frac{1}{3}(1-x_2^2)\ln(1+x_2)\Bigg)\frac{ \ln(x_2)}{1 - x_2} -\frac{4}{3} (1+x_2)\ln^2(1-x_2) \nonumber \\
    &+\left(-\frac{1}{9}N_f+\frac{11}{6}\right)(1+x_2)\ln(1-x_2) + \frac{2}{27}(1+4x_2) N_f + \frac{7}{54} (1+x_2) \pi^2 + \frac{1}{9} (14-17x_2)\nonumber \Bigg) \nonumber \\
    &+ (x_1 \leftrightarrow x_2) \Bigg\},
\end{align}
where again the explicit expression of $\kappa(x_2)$ is given in Appendix~\ref{app:xconst}.

\subsection{Resummation coefficients up to NNLO}\label{sec:nnlores}

The resummation coefficients can now be determined by performing a
double Mellin transform of the $(x_1,x_2)$-space  expression of the
coefficient function  $C(x_1,x_2, M^2)$ and comparing to the expanded resummed result
Eq.~(\ref{eq:expres}).
The Mellin transform is
\begin{align}\label{eq:melcf}
C(N_1,N_2,M^2)&=\int_0^1 d x_1\int_0^1 dx_2   x_1^{N_1-1}   x_2^{N_2-1}
C(x_1,x_2,M^2)\\
\label{eq:melcfexp} &= 1+\frac{\alpha_s(M^2)}{\pi}
C^{(1)}(N_1,N_2)+\left(\frac{\alpha_s(M^2)}{\pi}\right)^2
C^{(1)}(N_1,N_2),
\end{align}
where $C(x_1,x_2,M^2)$ is given in Eqs.~(\ref{eq:cfxdecomp}-\ref{eq:cboost2x}) .
Most of the Mellin transforms  are all elementary and
can be performed using well-known identities as collected for instance
in Appendix~B of Ref.~\cite{Bonvini:2012sh}.
In practice, we have used th ${\tt MT}$ code~\cite{Hoschele:2013pvt},
and simplified the result using  ${\tt HarmonicSums}$~\cite{ablinger2010computeralgebratoolboxharmonic}.

We will choose as a soft scale both in the single-soft and the
double-soft case $\bar\Lambda^2_{\rm ds}$ Eq.~(\ref{eq:gennbar}). This
means that the resummed result and its expansion will have an
identical expression in the single-soft and double soft case, with the
large log $\mathcal{L}$ everywhere given by
\begin{equation}\label{eq:ln1n2}
  \mathcal{L}=\ln\bar N_1\bar N_2.
  \end{equation}  
We will denote with $D_i^{\rm ds}$, $g_{0i}^{\rm ds}$ and $D_i^{\rm
  ss}$, $g_{0i}^{\rm ss}$ the expansion coefficients Eq.~(\ref{eq:dexp}) and
Eq.~(\ref{eq:g0exp}) determined in the double-soft and single-soft
limit respectively. The double-soft coefficients, as discussed in
Sect.~\ref{sec:ds}, are expected to coincide with those found in the
inclusive case, as we will explicitly check.  Furthermore, with this
choice of scale the single-soft result reduces to the double-soft one
in the double-soft limit.

\subsubsection{Double-soft limit}\label{sec:resds}

We expand the 
coefficients $C^{(n)}(N_1,N_2)$ of Eq.~(\ref{eq:melcfexp})  in powers of $\mathcal{L}$ 
according to
\begin{equation}\label{eq:cellexp}
	C^{(n)} (N_1,N_2)= \sum_{m=0}^{2n} C^{(n)}_{m}\mathcal{L}^m,
\end{equation}
where all coefficients $C^{(n)}_{m}$ are evaluated in the limit
$N_1\to\infty$, $N_2\to\infty$ and are consequently
constant. Comparing the coefficients of the expansion
Eq.~(\ref{eq:cellexp}) of the fixed order expression to the expansion
of the resummed result Eq.~(\ref{eq:expres} we get the matching
relations
\begin{align}\label{eq:nlodsmatch}
&\hbox{NLO}:\quad 	C^{(1)}_{2} = \frac{A_1}{2}, \qquad C^{(1)}_{1}= -D^\text{ds}_1, \qquad
       C^{(1)}_{0} = g^\text{ds}_{01},\\
&\hbox{NNLO:}\quad        \label{eq:nnlodsmatch}
	C^{(2)}_{4} = \frac{A_1^2}{8}, \quad
	C^{(2)}_{3} = \frac{4}{3} \frac{\pi \beta_0 A_1}{8}, \quad 
	C^{(2)}_{2} = \frac{A_2 + g_{01}^\text{ds} A_1}{2}, \quad	C^{(2)}_{1} = -D_2^\text{ds}, \quad C^{(2)}_{0} = g_{02}^\text{ds},
\end{align}

Comparing to the explicit expression  of the fixed-order coefficient
function we get
\begin{align}\label{eq:dsnlo}
&\hbox{NLO:}\qquad\quad	A_1=4/3, \qquad D_1^\text{ds}=0, \qquad g^\text{ds}_{01}= \frac{16}{3} \left(\frac{\pi^2}{6}-1\right); \\\label{eq:dsnnloa}
&\hbox{NNLO:}\qquad	A_2 = \frac{67}{9}-\frac{\pi ^2}{3}-\frac{10}{27}N_f, \\\label{eq:dsnnlod}
&\phantom{\hbox{NNLO}}\qquad	D_2^\text{ds} =-\frac{202}{27}+7 \zeta_3+ \frac{28}{81}N_f, \\\label{eq:dsnnlog}
&\phantom{\hbox{NNLO}}\qquad	g^\text{ds}_{02} = -\frac{2561}{144}+\frac{14 }{9}\pi ^2 + \frac{91 }{9}\zeta_3+\frac{23 }{108}\pi ^4 + N_f \left(\frac{2 }{27}\zeta_3-\frac{8 }{27}\pi ^2+\frac{127}{72}\right).
\end{align}
The coefficients $A_i$ of the expansion of the cusp
anomalous dimension are well known (see e.g. Ref.~\cite{Moch:2005ba}), and in fact the NNLO cusp
anomalous dimension $A_3$, which only enters at fixed N$^3$LO is
needed for NNLL resummation.

As mentioned in the introduction and
proven in Sect.~\ref{sec:ds} the functions $D^\text{ds}(\as)$ and
$g^\text{ds}_{0}(\as)$ coincide with their inclusive counterparts, and
thus the coefficients in Eqs.~(\ref{eq:dsnlo}-\ref{eq:dsnnlog}) must
coincide with the NLO~\cite{Catani:1989ne} and
NNLO~\cite{Vogt:2000ci} inclusive Drell-Yan threshold resummation
coefficients. We have explicitly checked that the resummation $D$
function agrees with the
inclusive case by using
the expressions given in Ref.~\cite{Moch:2005ba}. Note that in
this reference the perturbative expansion is performed in powers of
$\frac{\alpha_s}{4\pi}$, and furthermore that the alternate form of
the resummation given in Appendix~\ref{app:equiv} Eq.~(\ref{Rav}) is
adopted, hence  they must be compared using the relations given in the
Appendix, in particular
Eqs.~(\ref{eq:cmatch}-\ref{eq:dmatch}). Finally, we have checked that
the constants collected in $g^\text{ds}_{0}(\as)$ also agree with the
NNLO result, first obtained in Ref.~\cite{Matsuura:1988sm}, by checking against the
form given in Ref.~\cite{Bonvini:2023mfj}, Eq.~(A.5). This provides a
nontrivial check of the consistency of the double-soft resummation.

\subsubsection{Single-soft limit}
\label{sec:resss}

We now expand the 
coefficients $C^{(n)}(N_1,N_2)$ of Eq.~(\ref{eq:melcfexp})  in powers
of $\ln \bar N_1$ 
according to
\begin{equation}\label{eq:cellexpss}
	C^{(n)} (N_1,N_2)= \sum_{m=0}^{2n} \bar C^{(n)}_{m}(N_2) (\ln
        \bar N_1) ^m,
\end{equation}
where the  coefficient $\bar C^{(n)}_{m}$ are evaluated in the limit
$N_1\to\infty$, and are consequently a function of $N_2$
only. Furthermore, we substitute $\mathcal{L}=\ln \bar N_1+\ln \bar N_2$ in the
expanded resummed result Eq.~(\ref{eq:expres}) and similarly expand in
powers of $\ln \bar N_1$. Comparing the two expansions we get the
matching relations at NLO
\begin{align}
\bar C^{(1)}_{2} =\frac{A_1}{2}, \qquad \bar C^{(1)}_{1}=-B^\text{ss}_1-A_1 \ln\bar{N}_2,
\qquad \bar C^{(1)}_{10}=g^\text{ss}_{01} - D_1^\text{ss} \ln
(\bar{N}_2)+\frac{1}{2} A_1 \ln^2(\bar{N}_2)
\end{align}
and at NNLO
\begin{align}
\bar C^{(2)}_{4} &=\frac{A_1^2}{8}
\\
\bar C^{(2)}_{3} &=\frac{\pi}{6} \beta_0 A_1 - \frac{1}{2} A_1 D_1^\text{ss} + \frac{1}{2} A_1^2 \ln\bar{N}_2,
\\
\bar C^{(2)}_{2} &=\frac{1}{2} A_2 + \frac{1}{2} (D_1^\text{ss})^2 + \frac{1}{2} A_1 g^\text{ss}_{01} - \frac{\pi}{2} \beta_0 D_1^\text{ss} + \left(\frac{1}{2} A_1 \pi \beta_0- \frac{3}{2} A_1 D_1^\text{ss} \right)  \ln\bar{N}_2 + \frac{3}{4} A_1^2 \ln^2\bar{N}_2,
\\
\bar C^{(2)}_{1} &=-D^\text{ss}_2-B^\text{ss}_1 g_{01}^\text{ss} + \left(A_2+ (D_1^\text{ss})^2+ A_1 g_{01}^\text{ss} - \pi \beta_0 D_1^\text{ss} \right) \ln\bar{N}_2 \nonumber \\ 
	& + \left(\frac{\pi}{2} \beta_0 A_1- \frac{3}{2} A_1 D_1^\text{ss} \right) \ln^2 \bar{N}_2 + \frac{2}{3} A_1 \ln^3\bar{N}_2
\\
\bar C^{(2)}_{0} &= g_{02}^\text{ss} - (D_2^\text{ss}+ D_1^\text{ss} g_{01}^\text{ss}) \ln\bar N_2 + \frac{1}{2} \left(A_2 + (D_1^\text{ss})^2 + A_1 g_{01}^\text{ss} - \pi \beta_0 D_1^\text{ss}\right) \ln^2\bar{N}_2 \nonumber \\
&+ \frac{A_1}{2} \left(\frac{\pi}{3} \beta_0 - D_1^\text{ss}\right) \ln^3\bar{N}_2 + \frac{1}{8} A_1^2 \ln^4\bar{N}_2.
\end{align}

The resummation coefficients are again obtained by substituting the explicit expressions of the fixed-order result
coefficients $\bar C_{ij}$. Because we have adopted the same soft
scale choice as in the double-soft limits, all coefficients must
reduce to their double-soft counterparts in the $N_2\to\infty$
limit. We get
\begin{align}\label{eq:nlosscoef}
&\hbox{NLO}:\qquad	g^\text{ss}_{01} (N_2) = g^\text{ds}_{01} + F(N_2),
  \qquad D_1^\text{ss} (N_2)= \hat \gamma_{qq}^{(0)}(N_2), \\\label{eq:nnlosscoef}
&\hbox{NNLO}:\qquad 	D_2^\text{ss} (N_2) = D_2^\text{ds} - \pi \beta_0 F(N_2) + \hat \gamma_{qq}^{(1)}(N_2),
\end{align}
where   $g^\text{ds}_{01}$ and $D_2^\text{ds}$ were respectively given
in Eqs.~(\ref{eq:dsnlo}) and~(\ref{eq:dsnnlod}).

The explicit expressions of $F(N_2)$ and $\hat P^{(i)}(N_2)$ are
\begin{align}\label{eq:fexp}
    F(N_2) 
    & = \Cf \bigg[ 
    \ln\bar N_2 \bigg(\frac{1}{2N_2(N_2+1)} - S_1(N_2)\bigg)  
    + \bigg( \frac{(-1)^{N_2} (2N_2+1)}{2N_2(N_2+1)} - \ln{2} \bigg) S_{-1}(N_2) 
    \notag \\
    & + \frac12 \Big[S_1^{\, 2}(N_2) - S_{-1}^{\, 2}(N_2) + \ln^2\bar N_2 - \ln^2{2} \Big] 
    + \frac{1}{2N_2(N_2+1)} \Big[1 - S_1(N_2)\Big] 
    \nonumber\\
    & + \frac{(-1)^{N_2} (2N_2+1)}{2N_2(N_2+1)} \ln{2} \bigg]
    \\
\hat \gamma_{qq}^{(0)}(N_2)  &  = \Cf \(\frac{1}{2N_2(N_2+1)} + \ln\bar N_2 - S_1(N_2)\) 
\label{eq:gammahatzero}
\\
\label{eq:gammahatone}
	\hat \gamma_{qq}^{(1)}(N_2)  & = \Ca \Cf \bigg[\frac{N_2 (N_2+1) \big[N_2 (151 N_2+85)+3 \big] + 18}{72 N_2^3 (N_2+1)^3}+\left(\frac{\pi ^2}{12}-\frac{67}{36}\right) \Big(S_1(N_2) - \ln\bar N_2\Big) 
\notag \\
& + \frac{11 S_2(N_2)}{12} - \frac{S_3(N_2)}{2} + \frac{\pi^2}{72} \left(\frac{3}{N_2+1}-\frac{3}{N_2}-11\right) + \frac{\zeta_3}{2}\bigg]
+ \Cf^2 \bigg\{S_1(N_2) \bigg[\frac{1}{2(N_2+1)^2}-\frac{1}{2N_2^2} 
\notag \\
& + S_2(N_2) - \frac{\pi^2}{6}\bigg] - \left(\frac{1}{2N_2(N_2+1)} + \frac{3}{4}\right) S_2(N_2) + \frac{\pi^2}{4} \left(\frac{1}{3N_2(N_2+1)} + \frac{1}{2}\right)+\frac{3 N_2^3+N_2^2-1}{4 N_2^3 (N_2+1)^3} 
\notag \\
& +S_3(N_2)-\zeta_3 \bigg\}
    + \Cf \nf \TR \[\frac{3-N_2 (11 N_2+5)}{18 N_2^2
      (N_2+1)^2}+\frac{5}{9} \(S_1(N_2) - \ln\bar N_2\)
    -\frac{S_2(N_2)}{3} - \frac{\pi^2}{18} \].
\end{align}
The lengthy expression of the NNLO constant $g_{02}(N_2)$ is given in
Appendix~\ref{app:nconst}. 
We have explicitly checked that 
\begin{equation}\label{eq:psubtract}
\lim_{N\to\infty}F(N)= \lim_{N\to\infty} \hat \gamma_{qq}^{(0)}(N)=
\lim_{N\to\infty} \hat \gamma_{qq}^{(1)}(N)=0.
 \end{equation} 

The functions $\hat \gamma_{qq}^{(i)}(N)$ are easily identified in
terms of the
coefficients in the expansion in powers of $\frac{\alpha_s}{\pi}$ of
the leading and next-to-leading order nonsinglet
quark-quark anomalous dimensions
\begin{equation}\label{eq:apaasexp}
  \gamma_{qq}(N)= \frac{\alpha_s}{\pi}\gamma_{qq}^{(0)}(N)+\left(\frac{\alpha_s}{\pi}\right)^2\gamma_{qq}^{(1)}(N),
\end{equation}
by separating off the contributions to the anomalous dimension that
survive in the soft limit. Namely, by writing
\begin{equation}\label{eq:fullap}
 \gamma_{qq}^{(i)}(N)=  \hat\gamma_{qq}^{(i)}(N) + \gamma_{qq}^{(i, 0)} -\ln N A^{(i)}_q,
\end{equation}
where $\gamma_{qq}^{(i, 0)}$ are constants, and we recalled the
definition Eq.~(\ref{eq:cusp}) of the cusp anomalous dimension (with
$\gamma_{qq}=\gamma_q^{\rm  AP}$).

The physical meaning of this single-soft resummation contribution is
understood recalling that, as discussed in Sect.~\ref{sec:kinssl},
 the single-soft limit is purely
collinear. These terms then 
resum collinear evolution in the non-soft variable from the hard scale
up to the soft
scale. The cusp contribution is soft-collinear, and thus not included
in $D$ because it is already
included in the double logarithmic $A$ term: indeed, it already
contributes to the double soft limit. The constants  $\gamma_{qq}^{(i, 0)}$
are not included either, because we choose to define the constant
according to Eq.~(\ref{eq:gexp}), i.e. not to exponentiate any
constant. They could of course be equivalently  included in $D$,
reflecting the arbitrariness in defining the separation between
constants and logs already discussed above.
Finally, the remaining  contribution to $D$ at NNLO is proportional to the
contribution $F(N_2)$ to the single soft constant $g^\text{ss}_{01}
(N_2)$ that vanishes in the double-soft limit:  it is recognized to
originate from evolving up to the soft scale the argument of
$\alpha_s$ in this contribution.

The single-soft resummation of the nonsinglet coefficient function up to NNLL
accuracy in dQCD found using the coefficients
Eqs.~(\ref{eq:nlosscoef}-\ref{eq:gammahatone}) and
Eq.~(\ref{eq:g02ss}) in the general expression Eq.~(\ref{eq:gexp})
with the form Eqs.~(\ref{eq:gone}-\ref{eq:gthree},\ref{eq:g0exp}) of
the $g_i$ functions is a new result of this paper. The result
was also implicitly given in
Refs.~\cite{Lustermans:2019cau,Mistlberger:2025lee} using a SCET
formalism. The translation of the SCET result into explicit dQCD expressions
will be given in the next section.

\sect{Comparison to SCET}
\label{sec_dQCD_vs_SCET}

As mentioned in the introduction, resummation of the rapidity
distribution for Drell-Yan and Higgs production in both the
double-soft and single-soft limits was derived using SCET methods in
Ref.~\cite{Lustermans:2019cau}. The SCET resummation in the
single-soft limit was
recently re-derived and extended to N$^3$LL in
Ref.~\cite{Mistlberger:2025lee}. Also as mentioned, the comparison of
SCET and dQCD results is not immediate. In this Section we will show
that the SCET result agrees with the dQCD resummation Eqs.~(\ref{eq:finresmds}, \ref{eq:finresmss})
in the double- and single-soft limits, and it predicts the respective
$B$ and $D$ functions, specifically agreeing with the form of the
double-soft $D$
function (which starts at NNLL) of Ref.~\cite{Banerjee:2018vvb}.

Resummed results in SCET are obtained as the solution of suitable
momentum-space renormalization group equations depending on a number
of soft and hard scales. Comparison to dQCD thus requires solving these
equations in closed form, transforming to Mellin space, and making a
suitable choice of the various scales. We will follow through these
steps, also exploiting the previous results on the SCET-dQCD
comparison of  inclusive resummed results performed by some of us in
Refs.~\cite{Bonvini:2012az,Bonvini:2014qga}. 
We first solve the relevant evolution equations; next we transform to
Mellin space, and finally, following our previous
comparison~\cite{Bonvini:2012az,Bonvini:2014qga}  of the
inclusive SCET result of  Ref.~\cite{Becher:2007ty} to the dQCD
language, we make the appropriate choice of soft, hard and
factorization scales, substitute explicit expressions of the relevant
SCET anomalous dimensions and prove the equivalence of the SECT result
of Ref.~\cite{Lustermans:2019cau} to our result in both the
double-soft and single-soft limit.

In this Section, we mostly follow the notation originally used in the SCET
literature, specifically
Refs.~\cite{Lustermans:2019cau,Becher:2007ty}, in order to ease
comparison with these references, and we only depart from it when needed to
make contact with our results presented in the previous Sections.
 
\subsection{SCET resummation from evolution}
\label{sec:scetevol}

We start considering  the single-soft limit, as the
double-soft   limit can be obtained from it as a particular
case. The single soft limit is referred to as
generalized threshold limit in Ref.~\cite{Lustermans:2019cau}, and
collinear approximation in Ref.~\cite{Mistlberger:2025lee}. 
We specifically discuss results up to NNLL accuracy. Resummation in SCET is generally performed at the level of hadronic
cross sections, because the partonic cross section is sometimes chosen
to depend on a hadronic scale (see
Ref.~\cite{Bonvini:2013td}).

The
expression for the resummed hadronic differential distribution  is given in
Ref.~\cite{Lustermans:2019cau} in the
single-soft $\xa \to 1$ limit in the form
\begin{equation}
    \frac{d \sigma(z_1,z_2,M^2)}{d\tau_H dY}
    = \sum_{ij}
    H_{ij}(M^2, \muF^2) \int_{0}^{M^2(1-z_1)}\!\! \dt \, f_i\left(z_1\left(1 + \frac{t}{M^2}\right), \muF^2\right) \tildeB_j (t, z_2, \muF^2).
    \label{eq_LMT_main_result}
    \end{equation}
Here $i,j$ are parton indices, and
\begin{align} 
\label{eq:variableshada}
 \tau_H&=\frac{M^2}{S}
\\
\label{eq:variableshadb}
z_1&=\sqrt{\tau_H} \, e^Y\\
z_2&=\sqrt{\tau_H} \, e^{-Y}.
\end{align}
In \eq\eqref{eq_LMT_main_result}, $f_i(z,\mu^2)$ is a 
PDF,  $H_{ij}(\Q^2, \muF^2)$ is a hard function, and
$\tildeB_j$ a modified beam function, all of which will be defined below.
In the SCET formalism, resummation is performed by evolving the hard function $H_{ij}$ from the factorization scale $\muF^2$ to a hard scale $\muH^2$, and the beam function $\tildeB_j$ from the factorization scale $\muF^2$ to a soft scale $\muS^2$.


The dependence of $H_{ij}(M^2,\mu^2)$ on $\mu^2$ is governed by an evolution factor $U_H$:~\cite{Stewart:2010qs, Berger:2010xi}
\begin{equation}
    H_{ij}(M^2, \mu^2) = U^j_H(M^2, \muH^2, \mu^2) \, H_{ij}(M^2, \muH^2),
    \label{eq_H_qqb_evolution}
\end{equation}
where
\begin{equation}
    U^j_H(M^2, \muH^2, \mu^2)
    =
    \bigg(\frac{M^2}{\muH^2}\bigg)^{\!\!2\eta_\Gamma^j(\muH^2,\mu^2)} \exp{-4K_\Gamma^j(\muH^2, \mu^2) + 2K_{\gamma_H^j}(\muH^2, \mu^2)},
\label{eq_UH_as_BN}
\end{equation}
and we have assumed that partons $i$ and $j$ evolve with the same anomalous dimension 
\begin{equation}
    \gamma_H^j(\as) = \sum_{n=0}^{\infty} \gamma_{H,n}^j \left(\frac{\as}{4\pi}\right)^{\! n+1} .
\end{equation}
The coefficients of the expansion of $\gamma_H^j$ up to order $\as^2$ are given in Eqs. (\ref{eq_gamma_H_q} -- \ref{eq_gamma_H_q1}).

In \eq\eqref{eq_UH_as_BN}, the evolution kernels $K_\Gamma^j$ and $\eta_\Gamma^j$ are defined as integrals of the cusp anomalous dimension $\Gamma_\cusp^j$ as
\begin{align}
\label{eq:kGamma}
    K_\Gamma^j(\mu_0^2, \mu^2) &\equiv  \int_{\as(\mu_0^2)}^{\as(\mu^2)} \frac{\dalpha}{\tildebeta(\alpha)} \, \Gamma_\cusp^j(\alpha) \int_{\as(\mu_0^2)}^{\alpha} \frac{\dalpha'}{\tildebeta(\alpha')}
    \\
    \label{eq:etaGamma}    
    \eta_\Gamma^j(\mu_0^2, \mu^2) &\equiv  \int_{\as(\mu_0^2)}^{\as(\mu^2)} \frac{\dalpha}{\tildebeta(\alpha)} \, \Gamma_\cusp^j(\alpha),
\end{align}
while, for a generic anomalous dimension $\gamma$,
\begin{equation}
    \label{eq:kgammah}    
    K_{\gamma}(\mu_0^2, \mu^2) \equiv \int_{\as(\mu_0^2)}^{\as(\mu^2)} \frac{\dalpha}{\tildebeta(\alpha)} \, \gamma(\alpha).
\end{equation}
In our case, $\gamma= \gamma_H^j$.

The scale dependence of the beam function $\tildeB_j(t,z_2,\mu^2)$ is
given by~\cite{Stewart:2010qs, Lustermans:2019cau}  
\begin{equation}
    \tildeB_j(t,z_2,\mu^2) = \int_{0}^{t} \dt'\, U_B^j(t - t',\muS^2, \mu^2) \, \tildeB_j(t',z_2,\muS^2),
    \label{eq_TK_tilde_B_solution_of_RGE}
\end{equation}
where 
\begin{equation}
    U^j_B(t, \muS^2, \mu^2) 
    = 
    \frac{\exp{K_B^j(\muS^2,\mu^2) - \gammaE \, \eta_B^j(\muS^2,\mu^2)}}{\Gamma\big(1 + \eta_B^j(\muS^2,\mu^2)\big)} \left[ \frac{\eta_B^j(\muS^2,\mu^2)}{\muS^2} \calL^{\eta_B^j}\left( \frac{t}{\muS^2}\right) + \delta(t)\right] \,.
    \label{eq_UB_def}
\end{equation}
The evolution kernels now are
\begin{align}
    \label{eq_KB_def}
    K_B^j(\muS^2, \mu^2) & = 4K_\Gamma^j(\muS^2, \mu^2) + K_{\gamma_B^j}(\muS^2, \mu^2) \\
    \label{eq_etaB_def}
    \eta_B^j(\muS^2, \mu^2) & = -2\eta_\Gamma^j(\muS^2, \mu^2),
\end{align}
where $K_\Gamma^j$ and $\eta_\Gamma^j$ are given in Eqs.~\eqref{eq:kGamma} and \eqref{eq:etaGamma}, respectively. The relevant coefficients of the $\gamma_B^j$ anomalous dimension are given in Eqs. (\ref{eq_gamma_B_q} -- \ref{eq_gamma_B_q1}).
Finally, $\hat\calL^\eta(t,\mu^2)$ is a generating function for plus distributions, defined in Eq.~(\ref{eq:lscal}).

The resummed result is obtained by substituting in
\eq\eqref{eq_LMT_main_result}  the
expressions Eqs.~\eqref{eq_H_qqb_evolution} and
\eqref{eq_TK_tilde_B_solution_of_RGE} of  the hard and beam functions, respectively evolved to the hard scale $\muH^2$ and the soft scale $\muS^2$. We get
\begin{align}
\frac{d \sigma(z_1,z_2,\Q^2)}{d\tau_H d Y} 
    & = \sum_{ij} H_{ij}(\Q^2, \muH^2) \, U^j_H(\Q^2,\muH^2,\mu^2)
    \, \frac{\exp{K_B^j(\muS^2,\mu^2) - \gammaE \, \eta_B^j(\muS^2,\mu^2)}}{\Gamma\big(1 + \eta_B^j(\muS^2,\mu^2)\big)}  
\nonumber\\
& \times \int_{0}^{\Q^2(1-z_1)}d t \, f_i\left(\xa\left(1 + \frac{t}{\Q^2}\right), \mu^2\right)
\nonumber\\
&\times
    \int_{0}^{t} \dt' \bigg[\delta(t -t')+\frac{\eta_B^j(\muS^2,\mu^2)}{\muS^2} \calL^{\eta_B^j} \bigg( \frac{t -t'}{\muS^2}\bigg) \bigg] \tildeB_j(t',z_2, \muS^2).
\label{eq_LMT_main_result_first_step}
\end{align}

This resummed result can be simplified by changing the scale
integration variable $w = t'/t$, and exploiting the
scale transformation property of the distribution $\hat\calL^{\eta}(t,
\mu^2)$ given in \eq\eqref{eq:letaresc}:  
\begin{align}
    & \int_0^t d t'\, \left[\delta(t -t')+\frac{\eta_B^j}{\muS^2} \calL^{\eta_B^j} \bigg( \frac{t -t'}{\muS^2}\bigg) \right] \tildeB_j(t',z_2, \muS^2) 
    \nonumber\\
   &=\tildeB_j(t,z_2, \muS^2)+\eta_B^j \int_{0}^{1} d w \, \frac{t}{\muS^2} \, \calL^{\eta_B^j} \( \frac{t}{\muS^2}(1-w)\) \, \tildeB_j(t w, z_2, \muS^2) 
    \nonumber\\
     &= \eta_B^j \int_{0}^{1} \dw \, \(\frac{t}{\muS^2}\)^{\eta_B^j} \, \frac{\tildeB_j(t w, z_2, \muS^2) - \tildeB_j(t, z_2, \muS^2)}{(1-w)^{1-\eta_B^j}}
    + \(\frac{t}{\muS^2}\)^{\eta_B^j} \tildeB_j(t, z_2, \muS^2) 
    \nonumber\\
   &= \eta_B^j \(\frac{t}{\muS^2}\)^{\eta_B^j} \int_{0}^{1} \dw \, \frac{\tildeB_j(t w,z_2, \muS^2)}{(1-w)^{1-\eta_B^j}}.
   \label{eq:intBt}
\end{align}
Substituting this simplified expression in Eq.~\eqref{eq:intBt} in \eq\eqref{eq_LMT_main_result_first_step}, the resummed
result now takes the form
\begin{align}
    \frac{d \sigma(z_1,z_2,\Q^2)}{d\tau_H dY} 
    & = \sum_{ij} H_{ij}(\Q^2, \muH^2) \, U^j_H(\Q^2,\muH^2,\mu^2)
    \, \frac{\eta_B^j \, \exp{K_B^j - \gammaE \, \eta_B^j}}{\Gamma(1 + \eta_B^j)}  
    \nonumber\\
    & \times \int_0^{\Q^2(1-\xa)}dt \, f_i\left(\xa\left(1 + \frac{t}{\Q^2}\right), \mu^2\right)
    \(\frac{t}{\muS^2}\)^{\eta_B^j}
    \int_0^1 dw \, \frac{\tildeB_j(t w, z_2, \muS^2)}{(1-w)^{1-\eta_B^j}},
\label{eq_LMT_main_result_second_step}
\end{align}
where we have omitted the arguments $(\muS^2,\mu^2)$ of $K_B^j$ and $\eta_B^j$.

The explicit expression of the beam function $\tildeB_j(t,x,\mu^2)$ is given by~\cite{Lustermans:2019cau}
\begin{equation}
\tildeB_j(t,x,\mu^2) = \sum_k \int_{x}^{1} \frac{\dz}{z} \, f_k\bigg(\frac{x}{z},\muS^2 \bigg) \, \I_{jk}(t,z,\mu^2),
\end{equation}
where
\begin{equation}
\I_{jk}(t,z,\mu^2) = \sum_{n=0}^{\infty} \left[\amu{\mu^2}\right]^n \I_{jk}^{(n)}(t,z,\mu^2)
    \label{eq_tildeB_I_PDF_relation}
\end{equation}
and NNLL accuracy (in the sense of SCET\footnote{In
Ref.~\cite{Lustermans:2019cau} the $n=2$ term
was also given, as needed in order to achieve NNLL' accuracy in the
SCET nomenclature, which corresponds to the NNLL accuracy of our  dQCD
result. A comparison also including this term is left for  future work.}) is achieved by including terms up to $n=1$ in
Eq.~(\ref{eq_tildeB_I_PDF_relation}). The relevant coefficients are
\begin{align}     \label{eq_I0_defs}
    \I_{jk}^{(0)} (t,z,\mu^2) & = \delta(t) \delta_{jk} \delta(1-z) \,, \\
    \I_{jk}^{(1)} (t,z,\mu^2) & = \hat\calL_1(t,\mu^2) \, \Gamma_0^j \, \delta_{jk} \delta(1-z) 
    + \hat\calL_0(t,\mu^2) \bigg[4\PAP_{jk}(z) - \frac{\gamma_{B,0}^j}{2}\delta_{jk} \delta(1-z)\bigg] 
    + \delta(t) \, \Ileft_{jk}(z),
    \label{eq_I1_defs}
\end{align}
where 
\begin{align}
    & \Gamma_0^q = 4\Cf \,, \qquad \gamma_{B,0}^j = 6 \Cf  \\
    & \Gamma_0^g = 4\Ca \,, \qquad \gamma_{B,0}^j = 8\pi\beta_0,
    \label{eq:Gamma0coeff}
\end{align}
$\PAP_{jk}(z)$ are the leading order coefficients in the expansion in
powers of $\frac{\as}{\pi}$ of the splitting functions\footnote{Note that the definition of $\PAP_{jk}(z)$ in Ref.~\cite{Lustermans:2019cau} brings a factor of 4 with respect to the convention assumed in this paper. This is the reason why a factor a 4 appears in front of $\PAP_{jk}(z)$ in \eq\eqref{eq_I1_defs}, which is not present in Ref.~\cite{Lustermans:2019cau}.}, and the explicit expressions of the coefficients $\Ileft_{jk}(z)$ are given in \eq(S51) of Ref.~\cite{Lustermans:2019cau}.

The integral over the beam function in
Eq.~(\ref{eq_LMT_main_result_second_step}) can now be performed
explicitly, with the result
\begin{equation}
    \int_{0}^{1} dw \, \frac{\tildeB_j(t w,z_2, \muS^2)}{(1-w)^{1-\eta_B^j}}
    = \sum_k \frac{1}{t} \int_{z_2}^{1} \frac{dx_2}{x_2} \, f_k\bigg(\frac{z_2}{x_2},\muS^2 \bigg) \, \F_{jk}(t,x_2,\muS^2) + \order{\as^2(\muS^2)}\,,
\label{eq_integrate_tildeB_over_w}
\end{equation}
where
\begin{align}
    \F_{jk}(t,x_2,\muS^2) 
    & = \delta_{jk}\delta(1-x_2) 
    + \amu{\muS^2} \bigg\{ \Gamma_0^j \, \delta_{jk} \delta(1-x_2) \(V_1 + V_0 \ln\frac{t}{\muS^2}+ \frac{1}{2}\ln^2\frac{t}{\muS^2}\)
    \nonumber\\
    & + \bigg[4\PAP_{jk}(x_2) - \frac{\gamma_{B,0}^j}{2}\delta_{jk} \delta(1-x_2)\bigg] \(V_0 + \ln\frac{t}{\muS^2}\) + \Ileft_{jk}(x_2) \bigg\}.
\label{eq_F_jk_def}
\end{align}
In \eq\eqref{eq_F_jk_def}, 
\begin{align}\label{eq_V0_def}
    V_0 & 
    = \int_{0}^{1} dw \, \frac{\calL_0(w)}{(1-w)^{1 - \eta_B^j}} 
    = - \psi_0(\eta_B^j) - \gammaE \,, \\
\label{eq_V1_def}
    V_1 & 
    = \int_{0}^{1} dw \, \frac{\calL_1(w)}{(1-w)^{1 - \eta_B^j}} 
    = \frac{1}{2} \Big[-\psi_1(\eta_B^j) + \psi_0^2(\eta_B^j) + 2\gammaE\,\psi_0(\eta_B^j) + \gammaE^2 + \zeta_2 \Big].
\end{align}

Substituting the expression
\eq\eqref{eq_integrate_tildeB_over_w} of the beam function into the
resummed expression
\eq\eqref{eq_LMT_main_result_second_step}, changing the integration
variable from $t$ to $x_1$ according to $t = \Q^2(1-x_1)$, so that 
$x_1 \in[0,z_1]$, and noting that
\begin{equation}
f_i\left(z_1\left(1 + \frac{t}{\Q^2}\right), \mu^2\right)
    =
    \frac{1}{x_1} f_i\bigg(\frac{z_1}{x_1},\mu^2 \bigg) \big(1+ \order{1-x_1} \big) ,
\end{equation}
 we obtain the final resummed expression, up to NNLO accuracy in the
 single soft $x_1\to 1$ limit
\begin{align}
    \frac{d \sigma(z_1,z_2,M^2)}{d\tau_H dY} 
    & = \sum_{ijk} H_{ij}(\Q^2, \muH^2) \, U^j_H(\Q^2,\muH^2,\mu^2)
    \, \frac{\exp{K_B^j - \gammaE \, \eta_B^j}}{\Gamma(\eta_B^j)}  
    \nonumber\\
    & \times
    \int_{\xa}^{1} \frac{dx_1}{x_1} \, f_i\bigg(\frac{z_1}{x_1},\mu^2 \bigg) 
    \int_{\xb}^{1} \frac{dx_2}{x_2} \, f_k\bigg(\frac{z_2}{x_2},\muS^2 \bigg)
    \T_{jk}\big(x_1,x_2,\Q^2,\muS^2\big) \,, 
\label{eq_LMT_main_result_third_step}
\end{align}
where
\begin{equation}\label{eq:scetpartxsect}
    \T_{jk}\big(x_1,x_2,M^2,\muS^2\big)
    \equiv
    \bigg(\frac{M^2}{\muS^2}\bigg)^{\!\! \eta_B^j} (1-x_1)^{\eta_B^j - 1} \, \F_{jk}\big(\Q^2(1-x_1),x_2,\muS^2\big) + \order{\as^2(\muS^2)} \,.
\end{equation}
In this resummed expression, the hard and soft contributions are
evaluated at the scales $\muH^2$ and $\muS^2$, respectively.

We further observe that if the hard and soft scales are chosen to be 
independent of the hadronic scaling variables $z_1,z_2$, then
Eq.~(\ref{eq_LMT_main_result_third_step}) expresses the hadronic
cross section as a convolution of the PDFs $f_i,f_k$ with a coefficient function
\begin{align}
    &C^{\rm scet}_{ik}\(x_1,x_2,\frac{M^2}{\mu^2},\frac{M^2}{\muS^2},\frac{M^2}{\muH^2},\as(\muH^2)\)
    \nonumber\\
    & = \frac{1}{M^2\sigma^0_{ik}}\sum_j H_{ij}(\Q^2, \muH^2) U^j_H(\Q^2,\muH^2,\mu^2)
    \, \frac{\exp{K_B^j - \gammaE  \eta_B^j}}{\Gamma(\eta_B^j)}  
    \T_{jk}\big(x_1,x_2,\Q^2,\muS^2\big).
\label{eq:SCETcf}
\end{align}
Because of the definition Eqs.~(\ref{eq:variableshada}, \ref{eq:variableshadb}) of the
hadronic scaling variables, it then follows that the partonic
variables $x_1, x_2$ can be identified with the variables defined in \eq\eqref{eq:variables}.


\subsection{SCET resummation in Mellin space}
\label{sect:SCETmellin}

The Mellin transform of the coefficient function is given by
\begin{align}
&C^{\rm scet}_{ik}\(N_1,N_2,\frac{M^2}{\mu^2},\frac{M^2}{\muS^2},\frac{M^2}{\muH^2},\as(\muH^2)\)
\nonumber\\
& =    \int_{0}^{1}dx_1 \, x_1^{N_1-1} \int_{0}^{1}dx_2\,x_2^{N_2-1} \, 
C^{\rm scet}_{ik}\(x_1,x_2,\frac{M^2}{\mu^2},\frac{M^2}{\muS^2},\frac{M^2}{\muH^2},\as(\muH^2)\).
\label{eq:CscetN1N2}
\end{align}
We are specifically interested in evaluating the Mellin transform in the
single-soft limit $N_1\to\infty$, by only retaining terms that do not
vanish in this limit.

The dependence of the coefficient function on $x_1,x_2$ is contained in the
function $\T_{jk}$, which can be expanded in powers of $\as(\muS^2)$:
\begin{equation}
    \T_{jk} = \T_{jk}^{(0)} + \amu{\muS^2} \T_{jk}^{(1)} + \order{\as^2(\muS^2)}.
\end{equation}
Using \eq\eqref{eq_F_jk_def}, the coefficients $\T_{jk}^{(0,1)}$ read 
\begin{align}
\T_{jk}^{(0)}\big(x_1,x_2,\Q^2,\muS^2\big) &= \bigg(\frac{\Q^2}{\muS^2}\bigg)^{\!\! \eta_B^j} (1-x_1)^{\eta_B^j - 1} \delta_{jk} \delta(1-x_2)
    \nonumber\\
 \T_{jk}^{(1)}\big(x_1,x_2,\Q^2,\muS^2\big) &= \bigg(\frac{\Q^2}{\muS^2}\bigg)^{\eta_B^j} (1-x_1)^{\eta_B^j - 1} \bigg\{\Gamma_0^j \delta_{jk} 
 \delta(1-x_2)\bigg[\bigg(V_1 + V_0 \ln\frac{M^2}{\muS^2}+ \frac{1}{2} \ln^2\frac{M^2}{\muS^2} \bigg) 
     \nonumber\\
        & + \(V_0 + \ln\frac{M^2}{\muS^2}\) \ln(1-x_1) + \frac{1}{2}\ln^2(1-x_1)\bigg] 
        \nonumber\\
        &+ \bigg[4\PAP_{jk}(x_2)  - \frac{\gamma_{B,0}^j}{2}\delta_{jk} \delta(1-x_2)\bigg] \(V_0 + \ln\frac{M^2}{\muS^2}+ \ln(1-x_1)\) + \Ileft_{jk}(x_2) \bigg\},
\end{align}
where $\eta_B^j$ is defined in Eq.~\eqref{eq_etaB_def}, $V_0$ and $V_1$ are
given in Eqs.~(\ref{eq_V0_def}, \ref{eq_V1_def}), and $\Gamma_0^j$ are given in Eq.~\eqref{eq:Gamma0coeff}.
We can now perform the Mellin transform in the soft limit using
Eq.~(\ref{eq:scetmel1}) and its derivatives with respect to $\eta$. We find
\begin{align}
& \T_{jk}^{(0)}(N_1,N_2,M^2,\muS^2) = \delta_{jk} \bigg(\frac{M^2}{\muS^2 N_1}\bigg)^{\eta_B^j}
\label{eq_Tjk_0_Mellin_space} 
\\
& \begin{aligned}
        \T_{jk}^{(1)}(N_1,N_2,\Q^2,\muS^2) & = \bigg(\frac{\Q^2}{\muS^2 N_1}\bigg)^{\eta_B^j} \bigg\{\Gamma_0^j \delta_{jk} 
        \(\frac12 \ln^2\frac{M^2}{\muS^2 \bar N_1}+ \frac{\pi^2}{12} \) \\
        & + \bigg[4\gamma^{(0)}_{jk}(N_2) - \frac{\gamma_{B,0}^j}{2}\delta_{jk}\bigg] \ln\frac{M^2}{\muS^2 \bar N_1} + \Ileft_{jk}(N_2) \bigg\},
    \end{aligned}
    \label{eq_Tjk_1_Mellin_space}
\end{align}
where $\bar N_i = e^{\gammaE} N_i$, and $\gamma^{(0)}_{jk}(N_2)$ is the Mellin transform of $P^{(0)}_{jk}(z)$.

The Mellin-transformed  resummed coefficient function in
\eq\eqref{eq:CscetN1N2} in the single-soft limit thus takes the form
\begin{align}
&C^{\rm scet}_{ik}\(N_1,N_2,\frac{M^2}{\mu^2},\frac{M^2}{\muS^2},\frac{M^2}{\muH^2},\as(\muH^2)\)
\nonumber\\
& =\frac{1}{M^2\sigma^0_{ik}} \sum_j H_{ij}(\Q^2, \muH^2) \, U^j_H(\Q^2,\muH^2,\mu^2) 
\exp\big[4K_\Gamma^j(\muS^2, \mu^2) + K_{\gamma_B^j}(\muS^2, \mu^2)\big] 
\nonumber\\
&\times\bigg(\frac{\Q^2}{\muS^2 \bar N_1}\bigg)^{\eta_B^j} \, \sproc^{jk}(N_1,N_2,\muS^2)+\mathcal{O}\(\frac{1}{N_1}\),
\label{eq_LMT_main_result_fifth_step}
\end{align}
where we have used the explicit expression Eq.~\eqref{eq_KB_def} for the evolution kernels $K_B^j$.
The $N_2$ dependence is collected in the function
\begin{align}
    \sproc^{jk}(N_1,N_2,\muS^2) & = \delta_{jk} + \amu{\muS^2} \bigg\{\Gamma_0^j \delta_{jk} \bigg(\frac12 \ln^2\frac{M^2}{\muS^2 \bar N_1 \bar N_2} + \frac{\pi^2}{12} \bigg) 
    \nonumber\\
    &+ \bigg[4\gamma^{(0)}_{jk}(N_2) - \frac{\gamma_{B,0}^j}{2}\delta_{jk}
     + 4C_j \delta_{jk} \ln\bar N_2\bigg] \ln\frac{M^2}{\muS^2 \bar N_1 \bar N_2}
    + 4 \Fscet_{jk}(N_2) \bigg\} + \order{\as^2(\muS^2)},
\label{eq_sDY_def}
\end{align}
where
\begin{equation}
    4 \Fscet_{jk}(N_2)=
    \Ileft_{jk}(N_2) - 2 C_j\delta_{jk} \ln^2\bar N_2
    + \bigg[4\gamma^{(0)}_{jk}(N_2) - \frac{\gamma_{B,0}^j}{2}\delta_{jk} + 4C_j \delta_{jk} \ln\bar N_2 \bigg] \ln\bar N_2.
\label{eq_Fscet_def}
\end{equation} 
Note that from the definition of $\Ileft_{jk}(N_2)$, given in
Eq.~(S51) of Ref.~\cite{Lustermans:2019cau}, it follows that
\begin{equation}
\lim_{N_2\to\infty}\[\Ileft_{jk}(N_2) - 2 C_j\delta_{jk} \ln^2\bar N_2\]=0.
\end{equation}

We get to the final Mellin-space resummed expression by trading the
beam anomalous dimension for a contribution to
the anomalous dimension that governs the
evolution of the PDF, and also expressing the result in terms of PDFs
that are all
evaluated at the factorization scale $\mu^2$.
The first step can be accomplished by noting that (see Appendix \ref{app_subsec_anom_dim})
\begin{equation}
    \gamma_B^j(\as) = - 2\gamma_H^j(\as) - 2\gamma_\phi^j(\as) \,,
\end{equation}
where $\gamma_\phi^j$ is defined in Eqs.~(\ref{eq_gamma_phi_q} -- \ref{eq_gamma_phi_q_1}). The evolution kernels $K_\Gamma^j$ (Sudakov exponents) that originate
from the hard and beam function can then be combined. 
Specifically, using the relation  
\begin{equation}
    K_\Gamma^j(\muS^2, \mu^2) - K_\Gamma^j(\muH^2, \mu^2) 
    = 
    - K_\Gamma^j(\muH^2, \muS^2) + \eta_\Gamma^j(\muS^2,\mu^2) \ln\frac{\muH^2}{\muS^2},
\end{equation}
we find
\begin{align}
&U^j_H(\Q^2,\muH^2,\mu^2) \, \exp{4K_\Gamma^j(\muS^2, \mu^2) + K_{\gamma_B^j}(\muS^2, \mu^2)}
 \(\frac{\Q^2}{\muS^2}\)^{\eta_B^j} 
\nonumber\\
& = \exp{2K_{\gamma_\phi^j}(\muS^2,\mu^2)} \, U_j(\Q^2,\muH^2,\muS^2,\mu^2),
\label{eq_U_hard_plus_soft_appo}
\end{align}
where  
\begin{equation}
    U_j(\Q^2,\muH^2,\muS^2,\mu^2) 
    =
    \bigg(\frac{\Q^2}{\muH^2}\bigg)^{-\eta_B^j}   
    \exp{- 4K_\Gamma^j(\muH^2, \muS^2) + 2K_{\gamma_H^j}(\muH^2, \muS^2) - 4K_{\gamma_\phi^j}(\muS^2, \mu^2)},
    \label{eq_U_hard_plus_soft_def}
\end{equation}
and $K_{\gamma_\phi^j}$ is defined as in \eq\eqref{eq:kgammah} with $\gamma= \gamma_\phi^j$.

Substituting \eq\eqref{eq_U_hard_plus_soft_appo} in  \eq\eqref{eq_LMT_main_result_fifth_step}, we obtain
\begin{align}
&C^{\rm scet}_{ik}\(N_1,N_2,\frac{M^2}{\mu^2},\frac{M^2}{\muS^2},\frac{M^2}{\muH^2},\as(\muH^2)\)
\nonumber\\
&=\frac{1}{M^2\sigma^0_{ik}}\sum_j H_{ij}(M^2, \muH^2)  \exp{2K_{\gamma_\phi^j}(\muS^2,\mu^2)} 
    U_j(\Q^2,\muH^2,\muS^2,\mu^2) \, \bar N_1^{- \eta_B^j} \sproc^{jk}(N_1,N_2,\muS^2).
\label{eq_LMT_main_result_sixth_step}
\end{align}

Finally, the PDF $f_k(N_2,\muS^2)$ can be evolved to the factorization
scale $\mu^2$ through standard perturbative evolution
\begin{equation}\label{eq:pdfevol}
    f_k(N_2,\muS^2) = \sum_l U^{\pdf}_{kl}(N_2,\mu^2,\muS^2) \, f_l(N_2,\mu^2),
\end{equation}
thereby leading to an expression of the SCET coefficient function which is especially suited for the comparison with the dQCD result:
\begin{align}
&\; C^{\rm scet}_{ik}\(N_1,N_2,\frac{M^2}{\mu^2},\frac{M^2}{\muS^2},\frac{M^2}{\muH^2},\as(\muH^2)\)
=\frac{1}{M^2\sigma^0_{ik}}
\sum_{jl} H_{ij}(\Q^2, \muH^2) U^{\pdf}_{lk}(N_2,\mu^2,\muS^2) 
    \nonumber\\
    & \times \exp{2K_{\gamma_\phi^j}(\muS^2,\mu^2)} U_j(M^2,\muH^2,\muS^2,\mu^2) \, \bar N_1^{- \eta_B^j} \sproc^{jl}(N_1,N_2,\muS^2).
\label{eq_LMT_main_result_final_step}
\end{align}

\subsection{Comparison to the dQCD result}

We finally turn to the comparison of the SCET resummed result
\eq\eqref{eq_LMT_main_result_final_step} with the dQCD result first
derived in this paper. 
As discussed in Refs.~\cite{Bonvini:2012az,Bonvini:2014qga},
agreement between SCET and dQCD resummation is obtained by setting the
SCET soft scale $\muS^2$ equal to the soft scale of QCD
resummation. In order to facilitate the comparison, we choose the scale  $\bar\Lambda^2_{\rm ds}$
Eq.~(\ref{eq:gennbar}) that we had chosen as a common scale for both
the single-soft and double soft result in Sect.~\ref{sec:nnlores}:
\begin{equation}
    \muS^2 =  \bar\Lambda^2_{\rm ds}\equiv    \frac{\Q^2}{\bar N_1 \bar N_2} \,.
\label{eq:muschoice}
\end{equation}  
Furthermore, we fix the hard scale and the factorization scale as
\begin{equation}
  \label{eq:mufix}
  \muH^2 = \mu^2 =M^2.
\end{equation}
With this choice of the hard scale, the
evolution factor $U_H$ \eq\eqref{eq_UH_as_BN} equals 1, and 
\eq\eqref{eq_LMT_main_result_final_step} becomes
\begin{align}
C^{\rm scet}_{ik}\(N_1,N_2,1,\frac{M^2}{\muS^2},1,\as(M^2)\)
& = \frac{1}{M^2\sigma^0_{ik}}\sum_{jl} H_{ij}(M^2,M^2) U^{\pdf}_{lk}(N_2,M^2,\muS^2)
\nonumber\\
    & \times \exp{2K_{\gamma_\phi^j}(\muS^2,M^2)} \, U_j(M^2,\muS) \, \bar N_1^{\,- \eta_B^j} \sDY^{jl}(N_1,N_2,\muS^2),
\label{eq_LMT_main_result_seventh_step}
\end{align}
where
\begin{equation}
    U_j(\Q^2,\muS^2) 
    =  
    \exp{-4K_\Gamma^j(\Q^2,\muS^2) + 2K_{\gamma_H^j}(\Q^2,\muS^2) + 4K_{\gamma_\phi^j}(\Q^2,\muS^2)} \,.
    \label{eq_U_hard_plus_soft_def_muH_fixed}
\end{equation}

We now note that $K_\Gamma^j(\Q^2, \muS^2)$ \eq\eqref{eq:kGamma}
is defined in terms of the cusp anomalous dimension $\Gamma_\cusp^j(\alpha)$, which coincides with the function $A^j(\as)$ which appears in the dQCD resummation formula \eq\eqref{eq:finresmss}.
Furthermore, the combination of anomalous
dimensions that enters the kernels $K_{\gamma_H^j}$ and
$K_{\gamma_\phi^j}$ can be arranged as (see Appendix \ref{app_subsec_anom_dim})
\begin{equation}
    \gamma_W^j(\as) = \gamma_H^j(\as) + 2\gamma_\phi^j(\as)
    =
    \sum_{n=0}^{\infty} \gamma_{W,n}^j \left(\frac{\as}{4\pi}\right)^{\! n+1} \,.
\label{eq_gamma_W_j_def}
\end{equation}
We note that the coefficient $\gamma_{W,0}^j$ vanishes, so that the expansion of $\gamma_W^j$ starts at order $\as^2$. The coefficient $\gamma_{W,1}^j$ is given in \eq\eqref{eq_gamma_W_j}. Thus
\begin{equation}
    U_j(\Q^2,\muS^2)
    =
    \exp{ \int_{\Q^2}^{\muS^2} \frac{\dk^2}{k^2} \bigg[A^j(\as(k^2)) \, \ln\frac{\Q^2}{k^2}
    + \frac{\gamma_{W,1}^j}{16} \frac{\as^2(k^2)}{\pi^2} \bigg] } .
\label{eq_minus_4_K_gamma_in_dQCD}
\end{equation}  
The remaining evolution factors in
\eq\eqref{eq_LMT_main_result_seventh_step}, using the definitions \eq\eqref{eq:kgammah} and \eq\eqref{eq_etaB_def}, are 
\begin{align}\label{eq:nevolfac}
    \bar N_1^{- \eta_B^j}&= 
    \exp{ - \int_{M^2}^{\muS^2} \frac{\dk^2}{k^2} \, A^j(\as(k^2)) \ln{\bar N_1} }
    \\
    \label{eq:kevolfac}
    \exp{2K_{\gamma_\phi^j}(\muS^2,\mu^2)}
    &=
    \exp{ - \int_{M^2}^{\muS^2} \frac{\dk^2}{k^2} \, \gamma_\phi^j(\as(k^2)) }.
\end{align}
Combining these three evolution factors  Eqs.~(\ref{eq_minus_4_K_gamma_in_dQCD},\ref{eq:nevolfac},\ref{eq:kevolfac}) we obtain the full soft evolution function
\begin{align}
&\exp{2K_{\gamma_\phi^j}(\muS^2,\mu^2)} \, U_j(\Q^2,\muS^2) \, \bar N_1^{- \eta_B^j} 
\nonumber\\
& = \exp{ \int_{M^2}^{\muS^2} \frac{\dk^2}{k^2} \bigg[A^j(\as(k^2)) \, \ln\frac{M^2/\bar N_1}{k^2}
    - \gamma_\phi^j(\as(k^2)) + \frac{\gamma_{W,1}^j}{16} \frac{\as^2(k^2)}{\pi^2} \bigg]} 
    .
\label{eq_soft_contribution_scet}
\end{align}

Furthermore,
with the soft scale choice \eq\eqref{eq:muschoice} the matching
function is 
\begin{align}
    \sDY^{jl}(N_1,N_2,\muS^2)
    & = \delta_{jl} + \frac{\as(\muS^2)}{4\pi} \bigg[\delta_{jl} \Gamma_0^j \frac{\pi^2}{12} + 4 \Fscet_{jl}(N_2) \bigg] 
 \nonumber\\
    & = \exp{\frac{\as(\muS^2)}{\pi} \bigg[\delta_{jl} \frac{C_j \pi^2}{12} + \Fscet_{jl}(N_2) \bigg] }  ,
\label{eq_sDY_muS_fixed}
\end{align}
where we have used $\Gamma_0^j = 4C_j$. Expanding $\as(\muS^2)$ in powers of
$\as(M^2)$ through 
\begin{equation}
    \as(\muS^2) = \as(M^2) - \beta_0 \int_{M^2}^{\muS^2} \frac{\dk^2}{k^2}  \as^2(k^2)   
\end{equation}
this can be rewritten as 
\begin{align}
    &\sDY^{jl}(N_1,N_2,M^2/\bar N_1)
    = \left[\delta_{jl} + \frac{\as(M^2)}{\pi} \left(\delta_{jl} \frac{C_j \pi^2}{12} + \Fscet_{jl}(N_2) \right)\right] 
    \nonumber \\
    & \times \exp{ - \int_{M^2}^{\muS^2} \frac{\dk^2}{k^2} \bigg(\delta_{jl} \frac{C_j\beta_0\pi^3}{12} + \pi\beta_0 \Fscet_{jl}(N_2) \bigg) \frac{\as^2(k^2)}{\pi^2}} .
\label{eq_sDY_muS_fixed_final}
\end{align}

Collecting the above results, we finally obtain
\begin{align}
C^{\rm scet}_{ik}\(N_1,N_2,1,\frac{M^2}{\muS^2},1,\as(M^2)\)
&=\frac{1}{M^2\sigma^0_{ik}}
 \sum_{jl} \hatH_{ij,l}(\Q^2)U^{\pdf}_{lk}(N_2, M^2,\muS^2)
\nonumber\\
& \times \exp\Bigg\{\int_{\Q^2}^{\muS^2} \frac{\dk^2}{k^2} \bigg\{\delta_{jl} A^j(\as(k^2)) \, \ln\frac{M^2/\bar N_1}{k^2} - \delta_{jl} \gamma_\phi^j(\as(k^2)) 
\nonumber\\
 & + \bigg[\delta_{jl} \bigg(\frac{\gamma_{W,1}^j}{16} - \frac{C_j\beta_0\pi^3}{12} \bigg) - \pi\beta_0 \Fscet_{jl}(N_2) \bigg] \frac{\as^2(k^2)}{\pi^2} \bigg\} \Bigg\}
\label{eq_LMT_main_result_eighth_step}
\end{align}
where
\begin{equation}
    \hatH_{ij,l}(\Q^2) = H_{ij}(M^2,M^2) \left[\delta_{jl} + \frac{\as(\Q^2)}{\pi} \left(\delta_{jl} \frac{C_j\pi^2}{12} + \Fscet_{jl}(N_2) \right)\right]. 
\label{eq_hatH_ijk_def}
\end{equation}

Equation~\eqref{eq_LMT_main_result_eighth_step} can be compared with the dQCD result \eq\eqref{eq:finresmss} in any parton channel.
In Sect.~\ref{sec:nnlores} we have presented results for the
resummation of the quark nonsinglet channel of the  $C_{q\barq}$
coefficient function for the Drell-Yan process, hence we now specialize to this case.
To this purpose we make the following observations.  
\begin{enumerate}[label=\roman*)]
    \item For the Drell-Yan process  $C_{q\barq}$ coefficient function  \eq\eqref{eq_hatH_ijk_def} becomes \cite{Lustermans:2019cau, Becher:2007ty} 
    \begin{equation}
        \hatH_{ij,l}(\Q^2) 
        = \delta_{jl} 
        (\delta_{iq}\delta_{j\qb} + \delta_{i\qb}\delta_{jq}) \sigma^0_{ij}
        \left[1 + \frac{\as(\Q^2)}{\pi} \left(H_1 + \frac{\Cf\pi^2}{12} + \Fscet_{jl}(N_2) \right)\right] ,
    \end{equation}
    where
    \begin{equation}
        H_1 = \Cf \bigg(\frac{7 \zeta_2}{2} - 4 \bigg).     
    \label{eq_H_1_scet_def}
    \end{equation}  
  \item The exponential in the second and third lines of \eq\eqref{eq_LMT_main_result_eighth_step} is diagonal in the indices $j,l$ except for the term $\Fscet_{jl}(N_2)$.  
    However, parton $j$ is fixed to be a quark or an antiquark from
    point i), and in the non-singlet channel $l$ is also a quark or
    antiquark, with  
    \begin{align}
        \Ileft_{q_j q_l}(N_2) 
        & = \Ileft_{\qb_j \qb_l}(N_2) 
        = \delta_{jk} \Ileft_{qq}(N_2) 
        \\
        \gamma^{(0)}_{q_j q_l}(N_2) 
        & = \gamma^{(0)}_{\qb_j \qb_l}(N_2) 
        = \delta_{jl} \gamma^{(0)}_{qq}(N_2).
    \end{align}
      \item In the non-singlet channel, the evolution function $U^{\pdf}_{lk}$ in \eq\eqref{eq_LMT_main_result_eighth_step} is diagonal in the indices $l,k$. 
    Therefore, combining observations i) and ii), $j = k = l$ and
   $U^{\pdf}_{lk}$ becomes
    \begin{equation}
        U^{\pdf}_{lk}(N_2,\Q^2,\muS^2)= \delta_{lk} \exp\Bigg\{\int_{\Q^2}^{\muS^2} \frac{\dk^2}{k^2} 
        \bigg(\frac{\as(k^2)}{\pi} \gamma^{(0)}_{qq}(N_2) + \frac{\as^2(k^2)}{\pi^2} \, \gamma^{(1)}_{qq}(N_2) \bigg)  \Bigg\},
    \label{eq_U_pdf_Nb_general_solution}
    \end{equation}
    where $\gamma^{(0)}_{qq}(N_2)$ and $\gamma^{(1)}_{qq}(N_2)$ were introduced in \eq\eqref{eq:apaasexp}.
    \item 
The anomalous dimension  $\gamma_\phi^q(\as)$ appearing in
\eq\eqref{eq_LMT_main_result_eighth_step} and whose coefficients are
given in Eqs.~(\ref{eq_gamma_phi_q}, \ref{eq_gamma_phi_q_1}) is
defined by decomposing the quark-quark anomalous dimension as in
Eq.~(\ref{eq:fullap}) and identifying 
      \begin{equation}
        \gamma_\phi^q(\as)=  \frac{\alpha_s}{\pi} \gamma_{qq}^{(0,
            0)}+\left( \frac{\alpha_s}{\pi} \right)^2 \gamma_{qq}^{(1, 0)}.
    \label{eq_pdf_fq_DGLAP_Nb_infinity}
    \end{equation}
\end{enumerate}

Combining these observations, we can write \eq\eqref{eq_LMT_main_result_eighth_step} as
\begin{align}\label{eq:scetcfin}
    &C_{q\bar q}\(N_1,N_2,1,\frac{M^2}{\muS^2},1,\as(M^2)\)
    =  \gZeroscet(\as(M^2), N_2) 
    \nonumber\\
    & \times 
    \exp\Bigg\{\int_{\Q^2}^{\Q^2/(\bar N_1\bar N_2)} \frac{\dk^2}{k^2} \bigg\{A^q(\as(k^2)) \, \ln\frac{\Q^2/(\bar N_1\bar N_2)}{k^2} 
    + \frac{\as(k^2)}{\pi}  D^{\rm scet}_1
    + \frac{\as^2(k^2)}{\pi^2} D^{\rm scet}_2 \Bigg\},
\end{align}
where
\begin{align}
    &\gZeroscet(\as(\Q^2), N_2) = 1 + \frac{\as(\Q^2)}{\pi} \bigg(H_1 + \frac{\Cf\pi^2}{12} + \Fscet_{qq}(N_2) \bigg) \,, 
     \label{eq_gZeroscet_def} \\
    & D^{\rm scet}_1 = \hat\gamma^{(0)}_{qq}(N_2), 
    \label{eq_Bscet_1_def} \\
    & D^{\rm scet}_2 = \bigg(\frac{\gamma_{W,1}^q}{16} - \frac{\Cf\beta_0\pi^3}{12} \bigg) - \pi\beta_0 \Fscet_{qq}(N_2) + \hat\gamma^{(1)}_{qq}(N_2).
    \label{eq_Bscet_2_def} 
\end{align}

In \eq\eqref{eq_gZeroscet_def}, $H_1$ is given in
\eq\eqref{eq_H_1_scet_def}, while $\Fscet_{qq}(N_2)$ is defined
Eq.~\eqref{eq_Fscet_def} in terms of the function  $\Ileft_{jk}(N_2)$
of Ref.~\cite{Lustermans:2019cau} and the beam and standard anomalous
dimensions. Combining these definitions we get
\begin{equation}
\Fscet_{qq}(N_2)= F(N_2),
\end{equation}
where $= F(N_2)$ was given in Eq.~(\ref{eq:fexp}), and using the
explicit expression of $H_1$ we also verify that
\begin{equation}\label{eq:g0dsok}
H_1 + \frac{\Cf\pi^2}{12} =g^\text{ds}_{01}
\end{equation}
which then implies
\begin{equation}\label{eq:g0ssok}
\gZeroscet(\as(\Q^2), N_2)=1+\frac{\as(\Q^2)}{\pi}g^\text{ds}_{01}.
\end{equation}
Equations~(\ref{eq:g0dsok}) and ~(\ref{eq:g0ssok}) verify that the
constants agree between SCET and dQCD respectively in the
double-soft and single-soft  limit, up to $O(\alpha_s)$; the
$O(\alpha^2_s)$ single-soft constant (see Eq.~(\ref{eq:g02ss}) is not given in Ref.~\cite{Lustermans:2019cau}.

In Eq.~(\ref{eq_Bscet_1_def}) $\hat\gamma^{(0)}_{qq}(N_2)$ is the leading-order
subtracted anomalous dimension Eq.~(\ref{eq:gammahatzero}). Comparing
to the dQCD expression Eq.~(\ref{eq:nlosscoef})  verifies the agreement of the
single-soft $D$ function up to $O(\alpha_s)$. Finally, using the
expression of $\gamma_{W,1}^q$ of \eq\eqref{eq_gamma_W_j} it is easy
to check that
\begin{equation}\label{eq:d1dsok}
  \frac{\gamma_{W,1}^q}{16} - \frac{\Cf\beta_0\pi^3}{12}=D_2^\text{ds},
\end{equation}
where $D_2^\text{ds}$ was given in Eq.~(\ref{eq:dsnnlod}). The
contribution  $\hat\gamma^{(0)}_{qq}(N_2)$ to
Eq.~(\ref{eq_Bscet_2_def})  is the next-leading-order
subtracted anomalous dimension Eq.~(\ref{eq:gammahatone}), so it
follows that the  $O(\alpha_s^2)$ contribution to the SCET
$D$-function also agrees with its dQCD counterpart of
Eq.~(\ref{eq:nnlosscoef}). We conclude that we find full agreement
between the SCET result Eq.~(\ref{eq:scetcfin}) and our result in the
single soft limit Eqs.~(\ref{eq:nlosscoef}-\ref{eq:gammahatone}).

\sect{Conclusion}
\label{sec:conc}

In this paper we have derived a resummed result for rapidity
distributions of  a colorless final state object (such as a gauge 
or Higgs boson) in the limit in which the
center-of-mass energy tends to the minimum value which is needed in
order to impart to the final state gauge boson  a fixed value of the
rapidity in the center-of-mass frame of the partonic collision. This
result is complementary to  our previous~\cite{Forte:2021wxe} resummation of transverse
momentum distributions, in which instead the gauge boson transverse
momentum was kept fixed in the limit.

In comparison to that case,
rapidity distributions are  simpler in that the resummation is
characterized by a single soft and a single hard scale while for
transverse momentum distributions there are two. On the other hand,
it is more subtle because the parton distributions do not depend on
transverse momentum but they do depend on rapidity, hence the
resummation dynamics gets tangled with parton evolution.
In fact, resummation of rapidity distributions  in this limit turns out to be collinear, and it
is consequently characterized by the conventional anomalous
dimensions, now however evolving parton distributions up to a suitable
soft scale.

A byproduct of the work presented here is the derivation of a set of
relations between partonic cross-sections expressed in  the variables
that are most naturally used for phase space parametrization, and thus
for fixed order calculations, and those that are needed for QCD
factorization and resummation, namely the scaling variables on which
PDFs depend. Another significant byproduct is the relation of dQCD and
SCET approaches to resummation, which we have pursued in the
past~\cite{Bonvini:2012az,Bonvini:2014qga}, and that was needed here in
order to compare our results to previous approaches to the same
problem using SCET techniques.

Our results are a first proof of concept, in that we
have only presented full explicit expressions  for one particular partonic subchannel and
evolution eigenstate.
The extension to the complete set of parton states using the methods
presented here, which is required for full phenomenology, is in
principle straghtforward, but in practice laborious, and will be left
for future work. A natural spinoff of our results is also the
application of these techniques to the resummation of semi-inclusive
deep-inelastic  scattering (SIDIS), a process that has attracted
considerable attention in recent
years~\cite{Goyal:2023zdi,Goyal:2024emo,Bonino:2024qbh,Bonino:2025qta}
and whose resummation can be attacked with closely related
methods~\cite{Sterman:2006hu,Anderle:2012rq,Abele:2022wuy,Abele:2021nyo}.
This will also be the focus of forthcoming work.
\vskip 1cm

{\bf Acknowledgments:} We thank Johannes Michel for several illuminating exchanges  spread over many years on
the subject matter of this paper and specifically on
his work~\cite{Lustermans:2019cau}, and Luca Rottoli for pointing out
the collinear nature of the single-soft limit, as well as Gherardo
Vita,  Bernhard Mistlberger, Marco
Bonvini and Cesare Carlo Mella for discussions.  
The work of SF and GR has received funding from the European Union
NextGeneration EU program – NRP Mission 4 Component 2 Investment 1.1 –
MUR PRIN 2022 – CUP G53D23001100006 through the Italian Ministry of
University and Research (MUR). This research was
started at the Munich Institute for Astro-, Particle and BioPhysics
(MIAPbP) which is funded by the Deutsche Forschungsgemeinschaft (DFG,
German Research Foundation) under Germany's Excellence Strategy –
EXC-2094 – 390783311.  The work of SF  was 
performed in part at Aspen Center for Physics, which is supported by
National Science Foundation grant PHY-2210452.
SF also thans the Institute for Nuclear Theory
at the University of Washington for its 
hospitality and the Department of Energy for partial support during
the completion of this work.

\appendix
\sect{Mellin transforms in the soft limit}
\label{app:dmel}

The argument presented in Sect.~\ref{sec:rg} implies that in the soft
limit the coefficient function only depends on the scale $M^2$ and the
scaling variable $\Lambda^2_{\rm ds}=M^2(1-x_1)(1-x_2)$
Eq.~(\ref{eq:gen}), where in the single-soft limit $1-x_2$ is a fixed
finite value, while in the double-soft limit both $x_1$ and $x_2$ are close to 1.
This implies that in the soft limit the bare coefficient function
has the form
\begin{equation}
C^{(0)}(x_1,x_2,M^2,\alpha_0,\epsilon)=\sum_{n=0}^\infty\alpha_0^n C^{(0)}_n(x_1,x_2,M^2,\epsilon)
\label{CF}
\end{equation}
where
\begin{equation}
C^{(0)}_n(x_1,x_2,M^2,\epsilon)
=(M^2)^{-n\epsilon}\left[\delta(1-x_1)\delta(1-x_2)+\sum_{p=1}^nC^{(0)}_{np}(\epsilon)(1-x_1)^{-1+p\epsilon}(1-x_2)^{-1+p\epsilon}\right].
\label{CFn}
\end{equation}
The Mellin transform of this quantity is divergent due to soft and
collinear singularities, and indeed the renormalization group argument
of Sect.~\ref{sec:rg} is presented for a dimensionally regularized coefficient
function.

We now show that the Mellin transform of $C^{(0)}_n(x_1,x_2,M^2,\epsilon)$ 
\begin{equation}
C^{(0)}_n(N_1,N_2,M^2,\epsilon)=\int_0^1\dx_1\,x_1^{N_1-1}\int_0^1\dx_2\, x_2^{N_2-1}C^{(0)}_n(x_1,x_2,M^2,\epsilon)
\end{equation}
is a function of the product $N_1N_2$ up to terms that vanish as $N_1,N_2\to\infty$. Indeed 
\begin{align}
&\int_0^1\dx_1\,x_1^{N_1-1}\int_0^1\dx_2\, x_2^{N_2-1}
    (1-x_1)^{-1+p\epsilon}(1-x_2)^{-1+p\epsilon}
\nonumber\\
&\qquad=
    \frac{\Gamma(N_1)\Gamma(p\epsilon)}{\Gamma(N_1+p\epsilon)}\frac{\Gamma(N_2)\Gamma(p\epsilon)}{\Gamma(N_2+p\epsilon)}
\nonumber\\
&\qquad=\Gamma^2(p\epsilon)(N_1N_2)^{-p\epsilon} +\order{\frac{1}{N_1}}+\order{\frac{1}{N_2}},
    \label{eq:ipnres}
\end{align}
where we have used the well-known identity
\begin{equation}
\label{eq:scetmel1}
\int_0^1 dx\, x^{N-1} (1-x)^{\eta-1} 
 = \frac{\Gamma(N)\Gamma(\eta)}{\Gamma(N+ \eta)} 
 =\frac{\Gamma(\eta)}{N^{\eta}} \Big[1 + \order{1/N} \Big].
 \end{equation} 
Hence, in the  limit $N_1\to\infty,N_2\to \infty$
\begin{equation}
\label{eq:iplargen}
C^{(0)}_n(N_1,N_2,M^2,\epsilon)
=(M^2)^{-n\epsilon}\left[1+\sum_{p=1}^nC^{(0)}_{np}(\epsilon)\Gamma^2(p\epsilon)(N_1N_2)^{-p\epsilon}\right]
+\mathcal{O}\left(\frac{1}{N_1}\right)+\mathcal{O}\left(\frac{1}{N_2}\right)
\end{equation}
which is a function of $N_1 N_2$, thus establishing
Eq.~(\ref{eq:cbarel}).

\sect{Alternate forms of the resummed coefficient function}
\label{app:equiv}

The  form of the resummed result in the inclusive case in the original
paper Ref.~\cite{Catani:1989ne} (and also e.g. in
Ref.~\cite{Moch:2005ba}) was given in a somewhat different form, whose
relation to the form given in Eq.~(\ref{eq:finresm}) is discussed
e.g. in Ref.~\cite{Forte:2002ni} Eq.~(3.19) (see also
Ref.~\cite{Forte:2025tli} Eqs.~(110-111)), namely
\begin{align}\label{eq:catanires}
  C (N,1,\as(M^2))=&C^{(c)}\left(\as(M^2)\right)\nonumber\\\times&\exp\left[\int_0^1 \! dx
  \frac{x^{N-1}-1}{1-x}\left(\int_{M^2}^{\left(1-x\right)^2 M^2}
  \frac{dk^2}{k^2} A(\as(k^2)\right)+\bar D(\as((1-x)^2 M^2))\right],
\end{align}
where the function $\bar  D$ is a series in $\alpha_s$ whose
coefficients are determined order by order 
in terms of the coefficients of the expansions of $A$ and $D$.

The double-soft resummed
result, also given in Ref.~\cite{Catani:1989ne}, was also written similarly up to
NLL (where no $D$ function appears for Drell-Yan) and then extended to
higher logarithmic accuracy in Ref.~\cite{Banerjee:2018vvb}:
\begin{align}
&\ln C^{(l)}\left(\frac{M^2}{\mu^2 N_1N_2},\as(\mu^2)\right)=I_1+I_2+J_1+J_2
\label{Rav}
\\
&I_1=\int_0^1dx_1\int_0^1dx_2\frac{(x_1^{N_1-1}-1)(x_2^{N_2-1}-1)}{(1-x_1)(1-x_2)} A(\as(M^2(1-x_1)(1-x_2)))
\label{Rav1}
\\
&I_2=\int_0^1dx\,\frac{(x^{N_1-1}-1)+(x^{N_2-1}-1)}{1-x}\int_{\mu^2}^{M^2(1-x)}\frac{dk^2}{k^2}\,A(\as(k^2))
\label{Rav2}
\\
&J_1=\int_0^1dx_1\int_0^1dx_2\frac{(x_1^{N_1-1}-1)(x_2^{N_2-1}-1)}{(1-x_1)(1-x_2)}\frac{d}{d\ln M^2}D(\as(M^2(1-x_1)(1-x_2)))
\label{RavJ1}
\\
&J_2=\int_0^1dx\,\frac{(x^{N_1-1}-1)+(x^{N_2-1}-1)}{1-x}D(\as(M^2(1-x))).
\label{RavJ2}
\end{align}

Proving the equivalence of this  to the double-soft resummation as
given in  Eq.~(\ref{eq:finresmds}) is nontrivial because of the
double-plus distribution appearing in $I_1$ Eq.~(\ref{Rav1}), which
requires a double subtraction in order to be finite, as this
equation shows. However, the interference of the subtraction in the  Mellin
transform with respect to $x_1$ with the Mellin with respect to
$x_2$ and conversely generates extra contributions that are
compensated by the terms in Eq.~(\ref{Rav2}), as we will now show. 

Specifically, we prove that,  up to terms that are either constant or
power-suppressed in the large $N_1,N_2$ limit,
\be
I_1+I_2+J_1+J_2=\int_1^{N_1N_2}\frac{dn}{n}\left[\left(-\int_{\mu^2n}^{M^2}\frac{dk^2}{k^2}\,A(\as(k^2/n))\right)+D^{\rm ds}(\as(M^2/n))\right].
\label{lnWRG}
\ee
We also provide the relationship between the functions $D^{\rm
  ds}(\as)$ in this equation and $D(\as)$ in Eq.~\eqref{Rav}.

The proof relies on the  same identity used to prove the equivalence
of the inclusive resummation formulae Eq.~(\ref{eq:catanires}) and
Eq.~(\ref{eq:finresm}), namely
\be
\int_0^1dx\,\frac{x^{N-1}-1}{1-x}f(\ln(1-x))
=-\sum_{k=0}^\infty
\frac{\Gamma^{(k)}(1)}{k!}
\frac{d^k}{dL^k}\int_0^{1-1/N}\frac{dx}{1-x}f(\ln(1-x))+\order{\frac{1}{N}},
\label{idL}
\ee
where $L=\ln\frac{1}{N}$, for any function $f(z)$ with a Taylor expansion in its argument. The leading log, next-to-leading log, \ldots approximations are achieved by including terms up to $k=0,1,\ldots$ in the sum.
Using eq.~(\ref{idL}) we find, in the large $N_1,N_2$ limit,
\begin{align}
I_1&=\sum_{k=0}^\infty\frac{\Gamma^{(k)}(1)}{k!}\frac{d^k}{dL_1^k}\sum_{j=0}^\infty\frac{\Gamma^{(j)}(1)}{j!}
\frac{d^j}{dL_2^j}G(L_1,L_2)
\label{I1}
\\
I_2&=-\sum_{k=0}^\infty\frac{\Gamma^{(k)}(1)}{k!}\frac{d^k}{dL_1^k}H(L_1)
-\sum_{k=0}^\infty\frac{\Gamma^{(k)}(1)}{k!}\frac{d^k}{dL_2^k}H(L_2)
\label{I2}
\end{align}
where 
\be
L_1=\ln\frac{1}{N_1};\qquad L_2=\ln\frac{1}{N_2}
\ee
and
\begin{align}
G(L_1,L_2)&=\int_0^{1-\frac{1}{N_1}}\frac{dx_1}{1-x_1}
\int_0^{1-\frac{1}{N_2}}\frac{dx_2}{1-x_2}A(\as(M^2(1-x_1)(1-x_2)))
\\
H(L_1)&=\int_0^{1-\frac{1}{N_1}}\frac{dx}{1-x}
\int_{\mu^2}^{M^2(1-x)}\frac{dk^2}{k^2}A(\as(k^2)).
\end{align}

As a first step, we show that
\be
G(L_1,L_2)=H(L_1)+H(L_2)-H(L),
\ee
where
\be
L=L_1+L_2=\ln\frac{1}{N_1N_2}.
\ee
To this purpose, we trade the integration variable $x_2$ in $G(L_1,L_2)$ for
\be
k^2=M^2(1-x_1)(1-x_2);\qquad  \frac{M^2(1-x_1)}{N_2}\le k^2\le M^2(1-x_1).
\ee
We find
\be
G(L_1,L_2)=\int_0^{1-\frac{1}{N_1}}\frac{dx_1}{1-x_1}\int_{\frac{M^2(1-x_1)}{N_2}}^{M^2(1-x_1)}\frac{dk^2}{k^2}\,A(\as(k^2)).
\ee

We may now split the $k^2$ integration range at $k^2=\mu^2$,
in order to isolate a term which depends only on $N_1$ and coincides with $H(L_1)$:
\begin{align}
G(L_1,L_2)&=\int_0^{1-\frac{1}{N_1}}\frac{dx_1}{1-x_1}\int_{\frac{M^2(1-x_1)}{N_2}}^{\mu^2}\frac{dk^2}{k^2}\,A(\as(k^2))
+\int_0^{1-\frac{1}{N_1}}\frac{dx_1}{1-x_1}\int_{\mu^2}^{M^2(1-x_1)}\frac{dk^2}{k^2}\,A(\as(k^2))
\nonumber\\
&=-\int_0^{1-\frac{1}{N_1}}\frac{dx_1}{1-x_1}\int^{\frac{M^2(1-x_1)}{N_2}}_{\mu^2}\frac{dk^2}{k^2}\,A(\as(k^2))+H(L_1).
\end{align}
By rescaling $\frac{1-x_1}{N_2}=1-x$ in the first term, we can further isolate a $N_2$ dependent term, equal to $H(L_2)$,  and a $N_1N_2$ dependent one:
\begin{align}
G(L_1,L_2)
&=-\int_{1-\frac{1}{N_2}}^{1-\frac{1}{N_1N_2}}\frac{dx}{1-x}\int^{M^2(1-x)}_{\mu^2}\frac{dk^2}{k^2}\,A(\as(k^2))
+H(L_1)
\nonumber\\
&=-\int_0^{1-\frac{1}{N_1N_2}}\frac{dx}{1-x}\int_{\mu^2}^{M^2(1-x)}\frac{dk^2}{k^2}\,A(\as(k^2))
+\int_0^{1-\frac{1}{N_2}}\frac{dx}{1-x}\int_{\mu^2}^{M^2(1-x)}\frac{dk^2}{k^2}\,A(\as(k^2))
\nonumber\\
&+H(L_1)
\nonumber\\
&=-H(L)+H(L_2)+H(L_1),
\label{GHH}
\end{align}
as promised.

Next, we show that the contribution to $I_1$ from the term $H(L_1)+H(L_2)$ is exactly canceled by $I_2$. Indeed
\be
\sum_{k=0}^\infty\frac{\Gamma^{(k)}(1)}{k!}\frac{d^k}{dL_1^k}
\sum_{j=0}^\infty\frac{\Gamma^{(j)}(1)}{j!}\frac{d^j}{dL_2^j}
\left[H(L_1)+H(L_2)\right]
=\sum_{k=0}^\infty\frac{\Gamma^{(k)}(1)}{k!}\frac{d^kH(L_1)}{dL_1^k}+\sum_{j=0}^\infty\frac{\Gamma^{(j)}(1)}{j!}\frac{d^jH(L_2)}{dL_2^j}
\label{H}
\ee
which is precisely $-I_2$, as one can check by inspection of eq.~(\ref{I2}). Therefore
\be
I_1+I_2=-\sum_{k=0}^\infty\frac{\Gamma^{(k)}(1)}{k!}\sum_{j=0}^\infty\frac{\Gamma^{(j)}(1)}{j!}
\frac{d^k}{dL_1^k}\frac{d^j}{dL_2^j}H(L),
\label{I1plusI2}
\ee
where
\be
H(L)=\int_0^{1-\frac{1}{N_1N_2}}\frac{dx}{1-x}\int_{\mu^2}^{M^2(1-x)}\frac{dk^2}{k^2}\,A(\as(k^2))
=\int_1^{N_1N_2}\frac{dn}{n}\int_{\mu^2n}^{M^2}\frac{dk^2}{k^2}\,A(\as(k^2/n)).
\ee
Now, because $L=L_1+L_2$, Eq.~\eqref{I1plusI2} can be rewritten
\begin{equation}
I_1+I_2=-\sum_{k=0}^\infty\frac{\Gamma^{(k)}(1)}{k!}\sum_{j=0}^\infty\frac{\Gamma^{(j)}(1)}{j!}\frac{d^{k+j}}{dL^{k+j}}H(L)
=-\sum_{m=0}^\infty\frac{\gamma_m}{m!}\frac{d^mH(L)}{dL^m}
\label{lnWf}
\end{equation}
where
\be
\gamma_m=\sum_{k=0}^m\binom{m}{k}\Gamma^{(k)}(1)\Gamma^{(m-k)}(1)=\left.\frac{d^m\Gamma^2(z)}{dz^m}\right|_{z=1}.
\ee

In order to show that Eq.~\eqref{lnWf}
can be written in the form of Eq.~\eqref{lnWRG}
we first separate off the term $m=1$ in the sum, $-H(L)$, which is equal to the first term in Eq.~\eqref{lnWRG}:
\be
I_1+I_2=-\int_1^{N_1N_2}\frac{dn}{n}\int_{\mu^2n}^{M^2}\frac{dk^2}{k^2}\,A(\as(k^2/n))+R\left(\frac{M^2}{\mu^2 N_1N_2},\as(\mu^2)\right)
\ee
with
\be
R\left(\frac{M^2}{\mu^2 N_1N_2},\as(\mu^2)\right)=-\sum_{m=1}^\infty\frac{\gamma_m}{m!}\frac{d^m}{dL^m}\int_1^{N_1N_2}\frac{dn}{n}\int_{\mu^2n}^{M^2}\frac{dk^2}{k^2}\,A(\as(k^2/n)).
\ee

The function $R$ is in fact $\mu^2$-independent, up to constant terms in $N_1N_2$. Indeed
\be
\frac{d}{d\ln\mu^2}R\left(\frac{M^2}{\mu^2 N_1N_2},\as(\mu^2)\right)=\sum_{m=1}^\infty\frac{\gamma_m}{m!}\frac{d^m}{dL^m}\int_1^{N_1N_2}\frac{dn}{n}\,A(\as(\mu^2))=-2\gammaE A(\as(\mu^2))
\ee
which is a constant in $N_1,N_2$, and can be neglected. We may therefore compute $R$ for any choice of $\mu^2$, for example $\mu^2=M^2$:
\be
R\left(\frac{1}{N_1N_2},\as(M^2)\right)=-\sum_{m=1}^\infty\frac{\gamma_m}{m!}\frac{d^m}{dL^m}\int_1^{N_1N_2}\frac{dn}{n}\int_{M^2n}^{M^2}\frac{dk^2}{k^2}\,A(\as(k^2/n)).
\label{R}
\ee
It is now useful to define new integration variables
\be
t=\ln\frac{1}{n};\qquad t'=\ln\frac{k^2}{M^2n}
\ee
so that, for $\mu^2=M^2$,
\be
H(L)=\int_1^{N_1N_2}\frac{dn}{n}\int_{M^2n}^{M^2}\frac{dk^2}{k^2}\,A(\as(k^2/n))=-\int_0^Ldt\int_0^tdt'\,{\mathcal A}(t'),
\label{HM}
\ee
where 
\be
{\mathcal A}(t')=A(\as(k^2/n))=A(\as(M^2 e^{t'})).
\ee

We now show that for $m\ge 1$
\be
\frac{d^mH(L)}{dL^m}=-\int_0^Ldt\,{\mathcal A}^{(m-1)}(t)
\label{dmH}
\ee
up to terms that are either constant or vanishing in the large $N_1,N_2$ limit.
This is obviously true for $m=1$: differentiating Eq.~(\ref{HM}) with respect to $L$ we get
\be
\frac{dH(L)}{dL}=-\int_0^Ldt'\,{\mathcal A}(t').
\ee
Now assume that eq.~(\ref{dmH}) holds for a given value of $m$. Then
\be
\frac{d^{m+1}H(L)}{dL^{m+1}}=-{\mathcal A}^{(m-1)}(L)=-\int_0^Ldt\,{\mathcal A}^{(m)}(t)-{\mathcal A}^{(m-1)}(0)
\ee
and the last term can be dropped in the large $N_1,N_2$ limit.
This completes the proof of Eq.~(\ref{dmH}) for all $m\ge 1$.

We may now substitute Eq.~(\ref{dmH}) in Eq.~(\ref{R}), obtaining
\begin{align}
R\left(\frac{1}{N_1N_2},\as(M^2)\right)&=-\sum_{m=1}^\infty\frac{\gamma_m}{m!}\frac{d^mH(L)}{dL^m}
\nonumber\\
&=\sum_{m=1}^\infty\frac{\gamma_m}{m!}\int_0^Ldt\,{\mathcal A}^{(m-1)}(t)
\nonumber\\
&=
-\int_1^{N_1N_2}\frac{dn}{n}\,\sum_{m=1}^\infty\frac{\gamma_m}{m!}\left(\frac{d}{d\ln M^2}\right)^{(m-1)}A\left(\as(M^2/n)\right).
\end{align}
So
\be\label{eq:isum}
I_1+I_2=\int_1^{N_1N_2}\frac{dn}{n}\left[\left(-\int_{\mu^2n}^{M^2}\frac{dk^2}{k^2}\,A(\as(k^2/n))\right)+D_A(\as(M^2/n))\right]
\ee
with
\be
D_A(\as(M^2))=-\sum_{m=1}^\infty\frac{\gamma_m}{m!}\left(\frac{d}{d\ln M^2}\right)^{(m-1)}A\left(\as(M^2)\right).
\ee
We now note that
\be
J_1+J_2=\left.\frac{d}{d\ln M^2}(I_1+I_2)\right|_{A\to D},
\ee
as one can check by inspection of Eqs.~(\ref{Rav1}-\ref{RavJ2}). Hence
\be\label{eq:jsum}
J_1+J_2=\int_1^{N_1N_2}\frac{dn}{n}\left[-D(\as(M^2/n))+\frac{d}{d\ln M^2}D(\as(M^2/n))\right],
\ee
and therefore, combining Eqs.~(\ref{eq:isum}) and (\ref{eq:jsum}).
\be\label{eq:cmatch}
I_1+I_2+J_1+J_2
=\int_1^{N_1N_2}\frac{dn}{n}\left[\left(-\int_{\mu^2n}^{M^2}\frac{dk^2}{k^2}\,A(\as(k^2/n))\right)+D^{\rm ds}(\as(M^2/n))\right],
\ee
as we set out to prove. Also, as announced, we have found the relation
between the function $D$ appearing in the resummation Eq.~(\ref{Rav}) and
the function $D^{\rm ds}$ appearing in our resummed result
Eq.~(\ref{eq:finresmds}), namely
\be\label{eq:dmatch}
D^{\rm ds}(\as(M^2))=D_A(\as(M^2))-D(\as(M^2))+\frac{d}{d\ln M^2}D(\as(M^2)).
\ee
\sect{The variable $u$}\label{app:u}
Fixed-order differential cross sections are expressed in terms of
the scaling variables $\tau=x_1x_2$ and $u$, Eq.~(\ref{eq:udef}). We show
that $u$ is simply related to the scattering angle $\theta$ of the $Z$ with respect to the incident beam direction
in the partonic center-of-mass frame. In this frame
the $Z$ momentum can be parametrized as
\begin{equation}
p=(E,0,|\vec p|\sin\theta,|\vec p|\cos\theta);\qquad    E=\sqrt{|\vec p|^2+M^2}.
\end{equation}
Hence
\begin{equation}
e^{2y}=\frac{E+|\vec p|\cos\theta}{E-|\vec p|\cos\theta}.
\end{equation}
Using the definition
Eq.~(\ref{eq:udef}) of $u$ we get
\begin{align}
u&=\frac{1}{1-\tau}\frac{1-\tau e^{2y}}{1+e^{2y}}
\nonumber\\
&=\frac{1}{1-\tau}\frac{E(1-\tau)-|\vec p|(1+\tau)\cos\theta}{2E}
\nonumber\\
&=\frac{1}{2}\left[1-\frac{|\vec p|}{E}\frac{1+\tau}{1-\tau}\cos\theta\right].
\end{align}

In the limit of vanishing invariant mass $m_X^2$ of emitted radiation,
i.e.  when emitted partons are collinear, the energy conservation relation
\begin{equation}
\sqrt{s }=|\vec p|+\sqrt{|\vec p|^2+M^2}
\end{equation}
can be solved for $|\vec p|$ to give
\begin{equation}
|\vec p|=\frac{s -M^2}{2M};\qquad E=\frac{s +M^2}{2M};\qquad
\frac{|\vec p|}{E}=\frac{s -M^2}{s -M^2}=\frac{1-\tau}{1+\tau}.
 \end{equation}
Therefore
\begin{equation}
u=\frac{1-\cos\theta}{2}
\end{equation}
in this limit. This implieas that the collinear limits $\theta=0$ and
$\theta=\pi$ correspond respectively to $u=0$ and $u=1$.

\sect{Single-soft constant coefficients}
\label{app:Fixed-order_expressions}
We give here the explicit expression of
some lengthy coefficients that contribute to the soft limit as
constants, i.e. such that they do not vanish but are also
not logarithmically enhanced as 
$x_1\to1$ or $N_1\to\infty$ 

\subsection{Fixed-order direct space}
\label{app:xconst}
We give the expression of
$\bar\kappa(\tau)$ that enters the expression of the contribution
$\bar C^{(2)}_\text{Boost} (\tau)$ Eq.~(\ref{eq:cboost2}) to the NNLO
coefficient function in $(\tau,u)$ space and $\kappa(x_2)$ that enters the expression of the contribution
$C^{(2)}_\text{Boost} (x_2)$ Eq.~(\ref{eq:cboost2x}) to the NNLO
coefficient function in $(x_1,x_2)$ space.

We have
{\small \begin{align}
    &\bar\kappa(\tau)=\frac{(16 + 30 \tau^2) }{9}\frac{\text{Li}_3\left(\frac{2 \tau}{1+\tau}\right) - \zeta_3}{1-\tau} - \frac{2 (1 + \tau^2) }{9} \frac{ \text{Li}_3\left(\frac{1+\tau}2\right) - \zeta_3}{1-\tau} - \frac{4 (1 + \tau^2)}3 \frac{\text{Li}_3(\tau) - \zeta_3}{1-\tau} \nonumber \\
    &+ \frac{10 (1 + \tau^2)}{9 } \frac{\text{Li}_3(-\tau) + \frac3{4}\zeta_3}{1 - \tau} + \left( \frac{1 + \tau^2}{9}N_f - \frac{8 + 27 \tau - 22 \tau^2 + 22 \tau^3 - 2 \tau^4}{9 \tau} \right) \frac{\text{Li}_2\left(\frac{1+\tau}2\right)-\frac{\pi^2}{6} }{1 - \tau} \nonumber \\
    & -\left( \frac{2 - 14 \tau + 11 \tau^2 - 9 \tau^3 - 2 \tau^4}{9 \tau} + \frac{(19 - 9 \tau^2)}{9}\ln(1-\tau) \right) \frac{\text{Li}_2(\tau)-\frac{\pi^2}{6} }{1 - \tau} \nonumber \\
    &+\left( -\frac{16 + 14 \tau - 19 \tau^2 + \tau^3}{9 \tau} + \frac{14}{9} (1+\tau)(1-\tau)\ln(1-\tau) \right) \frac{ \text{Li}_2(-\tau)+\frac{\pi^2}{12}}{1 - \tau} +\frac{( 1 +9 \tau^2 ) }{9 } \frac{\text{Li}_3\left( \frac{1-\tau}{1+\tau} \right)}{1-\tau } \nonumber \\
    & + \frac{5 (1 + \tau^2) }{18} \frac{\text{Li}_3\left( 1 - \tau^2 \right)}{1-\tau} +\frac{1+\tau}{3} \text{Li}_3\left( \frac{2\tau}{1+\tau} \right) - \frac{8}{9} \frac{\text{Li}_3\left( \frac{1-\tau}2 \right)}{1-\tau}-\frac{25 \left(1+\tau ^2\right) }{9 }  \frac{\text{Li}_3(1-\tau )}{1-\tau}\nonumber \\
    & + \left(\frac{(1+\tau)(1-\tau)(16 + 5 \tau + 4 \tau^2)}{18 \tau} + \frac{8(1 + \tau^2)}{9}\ln\tau - (1+\tau^2)\ln\frac{1+\tau}2 \right) \frac{\text{Li}_2\left(\frac{1+\tau}2\right)}{1-\tau} \nonumber \\
    &+\left( \frac{2+3\tau+6\tau^2+2\tau^3}{9\tau} + \frac{2}{3}(1+\tau)\ln(1-\tau)+\frac{4}{9}(1-3\tau^2) \frac{\ln\tau}{1-\tau}+\frac{3+11\tau^2}{9} \frac{\ln\frac{1+\tau}{2}}{1-\tau}\right) \text{Li}_2(\tau) \nonumber \\
    &-\left( \frac{-12 - \frac{16}{\tau} - 3 \tau}{9} + \frac{2}{3}(1+\tau)\ln\left(1-\tau\right) + \frac{1-15\tau^2}{9}\frac{\ln\tau}{1-\tau} + \frac{3 + 11 \tau^2}{9}\frac{\ln\frac{1+\tau}2}{1-\tau} \right) \text{Li}_2(-\tau) \nonumber \\
    &-\frac{17 + 25 \tau^2}{27} \frac{\ln^3\frac{1+\tau}2}{1-\tau} \nonumber \\
    &+\left( -\frac{1+\tau^2}{18}N_f +\frac{11}{12}(1+\tau^2)-\frac{1-7\tau^2}{18}\ln\left(\frac{1-\tau}{2}\right) \right) \frac{\ln^2\frac{1+\tau}2}{1-\tau} \nonumber \\
    & \Bigg[\Bigg( -\frac{4(1+\tau^2)}{9}\ln 2  -\frac{40(1+\tau^2)}{9}\ln(1-\tau) +\frac{4}{9} \left(\tau ^2-2 \tau -2\right)+\frac{4}{3} \left(1+\tau ^2\right) \ln (1+\tau) \Bigg)\ln\tau \nonumber \\
    &+\frac{1}{9} \left(8 \tau ^2-31 \tau +13\right)+\frac{1}{108} \left(11 \tau ^2+19\right) \pi ^2+\frac{1}{18} \left(49 \tau ^2-32 \tau +49\right) \ln 2 \nonumber \\
    &+ \left(-\frac{2}{9} (1-\tau )-\frac{1}{9} \left(\tau ^2+1\right) \ln 2\right)N_f+\left(\frac{1}{9} \left(1+\tau ^2\right) N_f-\frac{1}{18} \left(49 \tau ^2-32 \tau +49\right)\right) \ln (1-\tau ) \nonumber \\
    &+ \frac{32(1+\tau^2)}{9}\ln^2(1-\tau) + \frac{17(1+\tau^2)}{18}\ln^2\tau\Bigg] \frac{\ln\left(\frac{1+\tau}2\right)}{1-\tau} \nonumber \\
    & -\frac{\left(107+211 \tau ^2\right) }{108} \frac{\ln ^3(\tau )}{1-\tau} + \Bigg( \frac{7(1+\tau^2)}{36}N_f + \frac{1+\tau^2}{18}\ln 2 + \frac{16+23\tau^2}3\ln(1-\tau) \nonumber \\
    &-\frac{1}{72} \left(267 \tau ^2-76 \tau +151\right)+\frac{1}{6} \left(43 \tau ^2+35\right) \ln (1-\tau )-\frac{2}{3} \left(1-\tau ^2\right) \ln (1+\tau ) \Bigg)\frac{\ln^2\tau}{1-\tau} \nonumber \\
    &+\Bigg(-\frac{1}{18} \left(69 \tau ^2-82 \tau +43\right)-\frac{1}{36} \left(1-15 \tau ^2\right) \pi ^2-\frac{2}{9} \left(2 \tau ^2-5 \tau +9\right) \ln 2 \nonumber \\
    & +\frac{1}{18} \left(7 \tau ^2-4 \tau +7\right) N_f- \frac{83}{9}(1+\tau^2)\ln^2(1-\tau) \nonumber \\
    &- \left(\frac{4}{9} \left(\tau ^2+1\right) N_f+\frac{1}{9} \left(-73 \tau ^2+16 \tau -63\right)-\frac{8}{9} \left(1-\tau^2\right) \ln (1+\tau)\right)\ln(1-\tau) \Bigg)\frac{\ln\tau}{1-\tau} \nonumber \\
    &-\frac{32}{9}(1+\tau)\ln^3(1-\tau)  +\left(\frac{11}{3}-\frac{2}{9} N_f \right)(1+\tau)\ln^2(1-\tau) \nonumber \\
    &+ \left( \frac{4}{27} (4 \tau +1)N_f-\frac{7 \tau }{2}-\frac{7}{54} (1+\tau ) \pi ^2+\frac{28}{9}\right)\ln(1-\tau) +\frac{2}{27} (82 \tau +19) \nonumber \\
    &+ \left(\frac{1+\tau}{54} \pi ^2+\frac{-19 \tau -37}{162} \right) N_f -\frac{ \left(8 \tau ^3+106 \tau ^2-7 \tau +4\right)}{108 \tau } \pi ^2-\frac{101}{18} (\tau +1) \zeta_3
\end{align}}

{\small \begin{align}
    & \kappa (x_2) = \frac{4 \left(8 + 15 x_2^2\right)}{9} \frac{\text{Li}_3 \left(\frac{2 x_2}{1+x_2}\right) - \zeta_3}{1-x_2} -\frac{4 \left(1 + x_2^2\right)}{9} \frac{\text{Li}_3\left(\frac{1+x_2}2\right) - \zeta_3}{1-x_2} \nonumber \\
    &-\frac{8(1+x_2^2)}{3}  \frac{ \text{Li}_{3}(x_{2}) - \zeta_3 }{1-x_2}+ \frac{20 \left(1 + x_2^2\right)}{9}\frac{\text{Li}_3(-x_2) + \frac3{4}\zeta_3}{1 - x_2}  \nonumber \\
    &+ \left(\frac{2 (1 + x_2^2)}{9}N_{f}- \frac{16 + 54 x_2 - 44 x_2^2 + 44 x_2^3 - 4 x_2^{4}}{9x_2}\right)\frac{\text{Li}_2\left(\frac{1 + x_2}2\right) - \frac{\pi^2}{6}}{1 - x_2} \nonumber \\
    &+\Bigg(\frac{2}{9} \left(9 x_2^2-19\right) \ln \left(1-x_2\right)+\frac{2 \left(2 x_2^4+9 x_2^3-11 x_2^2+14 x_2-2\right)}{9 x_2}\Bigg)\frac{\text{Li}_2(x_2) - \frac{\pi^2}{6}}{1 - x_2} \nonumber \\
    &+ \left(\frac{28}{9} \left(1-x_2\right) \left(x_2+1\right) \ln \left(1-x_2\right)-\frac{2 \left(x_2^3-19 x_2^2+14 x_2+16\right)}{9 x_2}\right)\frac{\text{Li}_2(-x_2) + \frac{\pi^2}{12}}{1 - x_2} \nonumber \\
    &-\frac{16}{9}\frac{\text{Li}_3\left(\frac{1-x_2}{2}\right)}{1-x_2}+\frac{2}{3} \left(1+x_2\right) \text{Li}_3\left(\frac{2 x_2}{1+x_2}\right)-\frac{50}{9} \left(1+x_2^2\right)\frac{ \text{Li}_3\left(1-x_2\right)}{1-x_2}\nonumber \\
    &+\frac{5}{9}\left(1+x_2^2\right)\frac{ \text{Li}_3\left(1-x_2^2\right)}{1- x_2} +\frac{2}{9}\frac{\left(1+9 x_2^2\right) \text{Li}_3\left(\frac{1-x_2}{1+x_2}\right)}{1- x_2} \nonumber \\
    &+\left(\frac{7}{3}+x_2+\frac{4 x_2^2}{9}+\frac{16}{9 x_2}-2 \left(1+x_2^2\right)\frac{ \ln \left(\frac{1+x_2}{2}\right)}{1-x_2}+\frac{16}{9}\left(1+x_2^2\right)\frac{ \ln \left(x_2\right)}{1-x_2} \right)  \text{Li}_{2}\left(\frac{1 + x_{2}}{2}\right) \nonumber \\
    &+ \left(\frac{2 \left(2 x_2^3+6 x_2^2+3 x_2+2\right)}{9 x_2}+\frac{28}{9}x_2^2\frac{ \ln \left(\frac{1+x_2}{2}\right)}{1-x_2}+\frac{2}{3} \left(1+x_2\right) \ln \left(1-x_2\right)+\frac{2}{9}(7-15x_2^2)\frac{\ln \left(x_2\right)}{1- x_2}\right) \text{Li}_2\left(x_2\right)\nonumber \\
    &+ \left(\frac{2}{9} \left(3 x_2+\frac{16}{x_2}+12\right)-\frac{28}{9}x_2^2\frac{ \ln \left(\frac{1+x_2}{2}\right)}{\left(1-x_2\right)}-\frac{2}{3} \left(1+x_2\right) \ln \left(1-x_2\right)+\frac{4}{9}\left(9 x_2^2-2\right)\frac{  \ln \left(x_2\right)}{1-x_2}\right) \text{Li}_2\left(-x_2\right)\nonumber \\
    & +\Bigg(-\frac{1}{9} N_f \left(x_2^2+1\right)+\frac{-2 x_2^4+11 x_2^3-15 x_2^2+21 x_2+18}{9 x_2}-\frac{1}{3} \left(1+x_2\right) \left(1-x_2\right) \ln \left(x_2\right) \nonumber \\
    &+\frac{1}{3} \left(x_2+1\right) \left(1-x_2\right) \ln \left(\frac{1-x_2}{2} \right)+\frac{2}{9} \left(5 x_2^2-2\right) \ln \left(1-x_2\right)+\frac{1}{9} \left(4-10 x_2^2\right) \ln (2)\Bigg) \frac{\ln ^2\left(\frac{1+x_2}{2}\right) }{1-x_2}  \nonumber \\
    &+ \Bigg(\frac{ \pi ^2}{27} \left(x_2^2+14\right)-\frac{-44 x_2^4+62 x_2^3+120 x_2^2+44 x_2-50}{27 x_2}+\frac{64}{9} \left(x_2^2+1\right) \ln ^2\left(1-x_2\right)+\frac{1}{9} \left(20 x_2^2+14\right) \ln ^2\left(x_2\right) \nonumber \\
    &+\frac{\left(147 x_2^3-96 x_2^2+147 x_2+80\right) }{27 x_2}\ln (2)+N_f \left(-\frac{4}{9} \left(1-x_2\right)-\frac{16}{27} \left(x_2^2+1\right) \ln (2)\right) \nonumber \\
    &+\left(\frac{2}{9} N_f \left(1+ x_2^2\right)+\frac{4 x_2^4-56 x_2^3+8 x_2^2-46 x_2+24}{9 x_2}\right) \ln \left(1-x_2\right)+ \Bigg(-\frac{8}{9}  \left(x_2^2+1\right) \ln (2) \nonumber \\
    &-\frac{80}{9} \left(x_2^2+1\right) \ln \left(1-x_2\right)-\frac{8 x_2^4-7 x_2^3-14 x_2^2+17 x_2-6 \left(3 x_2^2+5\right) x_2 \ln \left(x_2+1\right)+20}{9 x_2}\Bigg)\ln \left(x_2\right)\Bigg) \frac{\ln\left(\frac{1+x_2}{2}\right)}{1-x_2} \nonumber \\
    &- \frac{\left(247 x_2^2+71\right) }{54 } \frac{\ln ^3\left(x_2\right)}{1-x_2}\nonumber \\
    &+ \Bigg(-\frac{1}{18} \left(1-x_2\right) \left(7 N_f \left(x_2^2+1\right)\right)+\frac{40 x_2^4-241 x_2^3+16 x_2^2-165 x_2+8}{36 x_2}-2 \left(1-x_2\right) \left(1+x_2\right) \ln \left(1+x_2\right) \nonumber \\
    &+\frac{1}{9} \left(x_2^2+1\right) \ln (2)+\frac{1}{3} \left(38 x_2^2+40\right) \ln \left(1-x_2\right)\Bigg) \frac{\ln^2(x_2)}{1-x_2} \nonumber \\
    & + \Bigg(\frac{1}{9} N_f \left(7 x_2^2-4 x_2+7\right)-\frac{\pi ^2}{9}  \left(2-9 x_2^2\right)-\frac{44 x_2^4+97 x_2^3-180 x_2^2+7 x_2+50}{27 x_2}-\frac{1}{9} \left(169 x_2^2+163\right) \ln ^2\left(1-x_2\right) \nonumber \\
    &-\frac{4}{9} \left(2 x_2^2-5 x_2+9\right) \ln (2)+\ln \left(1-x_2\right) \Bigg(-\frac{8}{9} N_f \left(x_2^2+1\right)-\frac{16 x_2^4-141 x_2^3+8 x_2^2-131 x_2+8}{9 x_2} \nonumber \\
    &+\frac{22}{9} \left(1-x_2\right) \left(1+x_2\right) \ln \left(1+x_2\right)\Bigg)\Bigg) \frac{\ln(x_2)}{1-x_2} -\frac{64}{9} \left(1+x_2\right) \ln ^3\left(1-x_2\right) \nonumber \\
    &+\left(\frac{-4 x_2^3+41 x_2^2+47 x_2+4}{6 x_2}-\frac{4}{9} N_f \left(x_2+1\right)\right) \ln ^2\left(1-x_2\right) \nonumber \\
    &+\left(\frac{8}{27} N_f \left(4 x_2+1\right)-\frac{5}{54} \pi ^2 \left(1+x_2\right)+\frac{-44 x_2^3-123 x_2^2+168 x_2+50}{27 x_2}\right) \ln \left(1-x_2\right) \nonumber \\
    &+N_f \left(\frac{\pi ^2}{27}  \left(x_2+1\right)-\frac{19 x_2+37}{81} \right)-\frac{101}{9} \left(x_2+1\right) \zeta_3+\frac{4}{27} \left(82 x_2+19\right)-\frac{ \left(8 x_2^3+106 x_2^2-7 x_2+4\right)}{54 x_2}\pi ^2
\end{align}}

\subsection{Resummed Mellin space}
\label{app:nconst}

We give the expression of $g^\text{ss}_{02} (N_2)$, i.e.
the NNLO constant Eq.~(\ref{eq:g0exp}) in the resummed expression in
the single-soft limit.

{\small \begin{align}\label{eq:g02ss}
	g^\text{ss}_{02} (N_2) &=\frac{20 }{9}\text{Li}_4(1/2)+ \frac{17 }{54}\ln^4 2+\frac{341 }{3240}\pi ^4-\frac{17 }{54}\pi^2 \ln^2 2 + \frac{65 }{36} \zeta_3\ln 2  \nonumber \\
	&+\frac{\left(66 N_2^2+\left(66-34 (-1)^{N_2}\right) N_2-17 (-1)^{N_2}-7\right) }{54 N_2 (N_2+1)}\ln^3 2 \nonumber\\
	&-\frac{\left(26 N_2^4+\left(52+66 (-1)^{N_2}\right) N_2^3+\left(42+95 (-1)^{N_2}\right) N_2^2+\left(20+13 (-1)^{N_2}\right) N_2-10 (-1)^{N_2}+2\right)}{36 N_2^2 (N_2+1)^2} \ln^2 2 \nonumber \\
	&+\frac{(-1)^{N_2} \left(26 N_2^5+22 N_2^4+10 N_2^3+41 N_2^2+31 N_2+10\right) }{18 N_2^3 (N_2+1)^3}\ln 2+\frac{2 }{9}\ln^4(\bar{N}_2)-\frac{1}{108} S(-1,N_2)^4 \nonumber \\
	& +\frac{2}{9} S(1,N_2)^4 +\left(\frac{(-1)^{N_2} (2 N_2+1)}{54 N_2 (N_2+1)}-\frac{\ln 2}{27}\right) S(-1,N_2)^3 +\frac{\left(11 N_2^2+11 N_2-8\right) }{18 N_2 (N_2+1)} S(1,N_2)^3 \nonumber \\
	&+\left(\frac{\ln^2 2}{18}-\frac{5 (-1)^{N_2} (2 N_2+1) \ln 2}{9 N_2 (N_2+1)}+\frac{6 N_2^4+12 N_2^3-11 N_2^2+31 N_2+32}{36 N_2^2 (N_2+1)^2}+\frac{17 \pi^2}{36}\right) S(1,N_2)^2 \nonumber \\
	& -\frac{17}{9} S(-2,N_2)^2+\left(\frac{\ln 2}{9}-\frac{5 (-1)^{N_2} (2 N_2+1)}{9 N_2 (N_2+1)}\right) S(-1,N_2) S(1,N_2)^2 -\frac{3}{4} S(2,N_2)^2 -\frac{53}{18} S(-1,1,N_2)^2 \nonumber \\
	&+\Bigg(\frac{-9 N_2^2+\left(-9+74 (-1)^{N_2}\right) N_2+37 (-1)^{N_2}+9 }{18 N_2 (N_2+1)}\ln 2 \nonumber \\
	&+\frac{404 N_2^4+808 N_2^3+\left(493-2 (-1)^{N_2}\right) N_2^2+\left(97-6 (-1)^{N_2}\right) N_2-2 \left(-5+(-1)^{N_2}\right)}{36 N_2^2 (N_2+1)^2}\Bigg) \frac{\pi^2}{6} \nonumber \\
	&+\frac{766 N_2^2+\left(766+24 (-1)^{N_2}\right) N_2+12 (-1)^{N_2}+67}{72 N_2 (N_2+1)} \zeta_3 \nonumber \\
	&+\left(\frac{31 (-1)^{N_2} (2 N_2+1)}{27 N_2 (N_2+1)}-\frac{68 }{27}\ln 2\right) S(-3,N_2) \nonumber \\
	&+\left(-\frac{11 }{9}\ln^2 2+\frac{\left(7 N_2^2+7 N_2-3\right) }{3 N_2 (N_2+1)}\ln 2+\frac{\pi^2}{54}-\frac{(-1)^{N_2} \left(66 N_2^3+97 N_2^2+19 N_2-8\right)}{18 N_2^2 (N_2+1)^2}\right) S(-2,N_2) \nonumber \\
	&+\Bigg(\frac{17 }{27}\ln^3 2+\frac{\left(33 N_2^2+\left(33+2 (-1)^{N_2}\right) N_2+(-1)^{N_2}+2\right) }{18 N_2 (N_2+1)}\ln^2 2 \nonumber \\
	&-\frac{\left(13 N_2^4+26 N_2^3+21 N_2^2+10 N_2+1\right) }{9 N_2^2 (N_2+1)^2}\ln 2+\frac{(-1)^{N_2} \left(26 N_2^5+22 N_2^4+10 N_2^3+41 N_2^2+31 N_2+10\right)}{18 N_2^3 (N_2+1)^3} \nonumber \\
	&+\left(\frac{37 (-1)^{N_2} (2 N_2+1)}{18 N_2 (N_2+1)}-\frac{37 }{9}\ln 2\right) \frac{\pi^2}{6}-\frac{\zeta_3}{3}\Bigg) S(-1,N_2) -\frac{101}{27} S(-3,N_2) S(-1,N_2) \nonumber \\
	&+\frac{\left(11 N_2^2+11 N_2-3\right) }{6 N_2 (N_2+1)}S(-2,N_2) S(-1,N_2) +\Bigg(\frac{7 }{27}\ln^3 2-\frac{\left(36 (-1)^{N_2} N_2+18 (-1)^{N_2}+1\right) }{18 N_2 (N_2+1)}\ln^2 2 \nonumber \\
	&-\frac{7 (-1)^{N_2} \left(2 N_2^2+2 N_2+1\right) }{9 N_2^2 (N_2+1)^2}\ln 2 +\frac{404 N_2^5+793 N_2^4+227 N_2^3-66 N_2^2-27 N_2-30}{54 N_2^3 (N_2+1)^2} \nonumber \\
	&+\left(\frac{33 N_2^2+\left(33+4 (-1)^{N_2}\right) N_2+2 (-1)^{N_2}-51}{18 N_2 (N_2+1)}-\ln 2\right) \frac{\pi^2}{6} -\frac{67 }{36}\zeta_3\Bigg) S(1,N_2) \nonumber \\
	& +\left(2 \ln 2-\frac{16 (-1)^{N_2} (2 N_2+1)}{9 N_2 (N_2+1)}\right) S(-2,N_2) S(1,N_2)+S(-2,N_2) S(-1,N_2) S(1,N_2) \nonumber \\
	&+\left(-\frac{2 \ln^2 2}{9}-\frac{\ln 2}{9 N_2^2+9 N_2}-\frac{7 (-1)^{N_2} \left(2 N_2^2+2 N_2+1\right)}{9 N_2^2 (N_2+1)^2}\right) S(-1,N_2) S(1,N_2) \nonumber \\
	& + \left(\frac{-11 N_2^2-11 N_2+8}{18 N_2^2+18 N_2}-\frac{8}{9} S(1,N_2)\right)\ln^3\bar{N}_2 \nonumber \\
	&+ \Bigg(-\frac{4 }{9}\ln^2 2+\frac{4 (-1)^{N_2} (2 N_2+1) }{9 N_2 (N_2+1)}\ln 2-\frac{4}{9} S(-1,N_2)^2+\frac{4}{3} S(1,N_2)^2 \nonumber \\
	&+\frac{6 N_2^4+12 N_2^3-11 N_2^2-17 N_2+8}{36 N_2^2 (N_2+1)^2}+\frac{23 }{54}\pi^2+\left(\frac{4 (-1)^{N_2} (2 N_2+1)}{9 N_2 (N_2+1)}-\frac{8 }{9}\ln 2\right) S(-1,N_2) \nonumber \\
	&+\frac{\left(11 N_2^2+11 N_2-8\right) }{6 N_2 (N_2+1)}S(1,N_2)\Bigg) \ln^2\bar{N}_2 \nonumber \\
	&+ \Bigg(-\frac{8}{9} S(1,N_2)^3+\frac{\left(-11 N_2^2-11 N_2+8\right)}{6 N_2^2+6 N_2} S(1,N_2)^2+\frac{8}{9} S(-1,N_2)^2 S(1,N_2) \nonumber \\
	&+\left(\frac{8 }{9}\ln^2 2-\frac{8 (-1)^{N_2} (2 N_2+1) }{9 N_2 (N_2+1)}\ln 2-\frac{31 }{27}\pi^2-\frac{6 N_2^4+12 N_2^3-11 N_2^2+15 N_2+24}{18 N_2^2 (N_2+1)^2}\right) S(1,N_2) \nonumber \\
	&+\left(\frac{16 }{9}\ln 2-\frac{8 (-1)^{N_2} (2 N_2+1)}{9 N_2 (N_2+1)}\right) S(-1,N_2) S(1,N_2)+\frac{16}{9} S(2,N_2) S(1,N_2) \nonumber \\
	&+\frac{\left(33 N_2^2+33 N_2-8\right) }{18 N_2 (N_2+1)}S(-1,N_2)^2+\frac{33 N_2^2+33 N_2-8}{18 N_2 (N_2+1)}\ln^2 2\nonumber \\
	&+\frac{-404 N_2^6-1212 N_2^5-1050 N_2^4-206 N_2^3+21 N_2^2+33 N_2+30}{54 N_2^3 (N_2+1)^3}+\frac{\left(-21 N_2^2-21 N_2+31\right) }{9 N_2^2+9 N_2} \frac{\pi^2}{6}+\frac{65}{9} \zeta_3 \nonumber \\
	&+\left(\frac{ 33 N_2^2+33 N_2-8}{9 N_2 (N_2+1)}\ln 2-\frac{(-1)^{N_2} \left(66 N_2^3+99 N_2^2+17 N_2-8\right)}{18 N_2^2 (N_2+1)^2}\right) S(-1,N_2) \nonumber \\
	&+\frac{\left(21 N_2^2+21 N_2-8\right) S(2,N_2)}{9 N_2 (N_2+1)}-\frac{2}{9} S(3,N_2)-\frac{(-1)^{N_2} \ln 2 \left(66 N_2^3+99 N_2^2+17 N_2-8\right)}{18 N_2^2 (N_2+1)^2}\Bigg) \ln\bar{N}_2 \nonumber \\
	&+ \Bigg(-\frac{\ln^2 2}{18}+\frac{(-1)^{N_2} (2 N_2+1) }{18 N_2 (N_2+1)}\ln 2+\frac{1}{18} S(1,N_2)^2-\frac{37 }{108}\pi^2 \nonumber \\
	&-\frac{13 N_2^4+26 N_2^3+21 N_2^2+10 N_2+1}{18 N_2^2 (N_2+1)^2}\Bigg)S(-1,N_2)^2 -\frac{71}{18} S(-1,N_2)^2 S(2,N_2) -\frac{5}{18} S(1,N_2)^2 S(2,N_2) \nonumber \\
	&+\left(\frac{\ln^2 2}{9}-\frac{(-1)^{N_2} (2 N_2+1) }{2 N_2 (N_2+1)}\ln 2-\frac{\pi^2}{27}-\frac{68 N_2^4+136 N_2^3+47 N_2^2+3 N_2+18}{36 N_2^2 (N_2+1)^2}\right) S(2,N_2) \nonumber \\
	&+\left(\frac{3 (-1)^{N_2} (2 N_2+1)}{2 N_2 (N_2+1)}-\frac{22 }{9}\ln 2\right) S(-1,N_2) S(2,N_2) +\frac{\left(-33 N_2^2-33 N_2+5\right) }{18 N_2^2+18 N_2}S(1,N_2) S(2,N_2) \nonumber \\
	&+\frac{\left(10 N_2^2+10 N_2-19\right) }{18 N_2 (N_2+1)}S(3,N_2)+\frac{19}{9} S(1,N_2) S(3,N_2)-\frac{16}{9} S(4,N_2)-\frac{1}{6} S(-3,-1,N_2) \nonumber \\
	&+S(-2,-2,N_2)+\frac{\left(N_2^2+N_2-1\right) }{2 N_2 (N_2+1)}S(-2,-1,N_2)+S(1,N_2) S(-2,-1,N_2) \nonumber \\
	&+\left(\frac{8 (-1)^{N_2} (2 N_2+1)}{9 N_2 (N_2+1)}-\frac{8 }{3}\ln 2\right) S(-2,1,N_2)-\frac{29}{9} S(-1,N_2) S(-2,1,N_2) \nonumber \\
	&+\left(\frac{20 }{9}\ln^2 2+\frac{11}{9 N_2^2+9 N_2} \ln 2+\frac{(-1)^{N_2} \left(66 N_2^3+95 N_2^2+13 N_2-10\right)}{18 N_2^2 (N_2+1)^2}-\frac{\pi^2}{27}\right) S(-1,1,N_2) \nonumber \\
	&+4 S(-2,N_2) S(-1,1,N_2)+\frac{\left(-11 N_2^2-11 N_2+3\right) }{6 N_2^2+6 N_2}S(-1,N_2) S(-1,1,N_2) \nonumber \\
	&+\left(\frac{2 (-1)^{N_2} (2 N_2+1)}{N_2 (N_2+1)}-\frac{22 }{9}\ln 2\right) S(1,N_2) S(-1,1,N_2)-S(-1,N_2) S(1,N_2) S(-1,1,N_2) \nonumber \\
	&+\left(\frac{8 }{3}\ln 2-\frac{2 (-1)^{N_2} (2 N_2+1)}{N_2 (N_2+1)}\right) S(2,-1,N_2)+\frac{73}{18} S(-1,N_2) S(2,-1,N_2) \nonumber \\
	& +\frac{\left(-9 N_2^2-9 N_2+29\right) }{18 N_2^2+18 N_2}S(2,1,N_2)-\frac{29}{9} S(1,N_2) S(2,1,N_2)-\frac{19}{9} S(3,1,N_2)-S(-2,-1,1,N_2) \nonumber \\
	&+\frac{5}{9} S(-2,1,-1,N_2) +\frac{\left(33 N_2^2+33 N_2+13\right) }{18 N_2^2+18 N_2}S(-1,1,-1,N_2) -\frac{13}{9} S(1,N_2) S(-1,1,-1,N_2) \nonumber \\
	&+N_f \Bigg\{-\frac{2 }{27}\ln^3 2+\frac{\left(10 N_2^2+2 \left(5+3 (-1)^{N_2}\right) N_2+3 (-1)^{N_2}\right) }{54 N_2 (N_2+1)}\ln^2 2 \nonumber \\
	&+\frac{(-1)^{N_2} \left(-10 N_2^3-6 N_2^2+4 N_2+3\right)}{27 N_2^2 (N_2+1)^2} \ln 2+\frac{\ln^3\bar{N}_2}{27}-\frac{1}{27} S(1,N_2)^3+\frac{5}{27} S(-1,N_2)^2 \nonumber \\
	& +\frac{\left(-10 N_2^2-10 N_2+3\right) S(1,N_2)^2}{54 N_2^2+54 N_2} +\frac{1143 N_2^6+3429 N_2^5+3505 N_2^4+1175 N_2^3+4 N_2^2+48 N_2+36}{648 N_2^3 (N_2+1)^3} \nonumber \\
	&+\left(\frac{\ln 2}{9}+\frac{-106 N_2^2-106 N_2+3}{54 N_2^2+54 N_2}\right) \frac{\pi^2}{6} +\left(\frac{(-1)^{N_2} (2 N_2+1)}{9 N_2 (N_2+1)}-\frac{2}{9} \ln 2\right) S(-2,N_2) \nonumber \\
	&+\left(-\frac{\ln^2 2}{9}+\frac{10 }{27}\ln 2+\frac{(-1)^{N_2} \left(-10 N_2^3-6 N_2^2+4 N_2+3\right)}{27 N_2^2 (N_2+1)^2}\right) S(-1,N_2)-\frac{1}{9} S(-2,N_2) S(-1,N_2) \nonumber \\
	&+\left(-\frac{28 N_2^4+56 N_2^3+4 N_2^2-6 N_2+9}{81 N_2^2 (N_2+1)^2}-\frac{\pi^2}{54}\right) S(1,N_2) \nonumber \\
	& + \left(\frac{-10 N_2^2-10 N_2+3}{54 N_2^2+54 N_2}-\frac{1}{9} S(1,N_2)\right)\ln^2(\bar{N}_2) +\frac{\left(10 N_2^2+10 N_2-3\right) S(2,N_2)}{54 N_2 (N_2+1)} +\frac{1}{9} S(1,N_2) S(2,N_2)\nonumber \\
	& + \Bigg(-\frac{\ln^2 2}{9}+\frac{(-1)^{N_2} (2 N_2+1) }{9 N_2 (N_2+1)}\ln 2-\frac{1}{9} S(-1,N_2)^2+\frac{1}{9} S(1,N_2)^2+\frac{28 N_2^4+56 N_2^3+4 N_2^2-6 N_2+9}{81 N_2^2 (N_2+1)^2} \nonumber \\
	&+\frac{\pi^2}{27}+\left(\frac{(-1)^{N_2} (2 N_2+1)}{9 N_2 (N_2+1)}-\frac{2}{9} \ln 2\right) S(-1,N_2)+\frac{\left(10 N_2^2+10 N_2-3\right) S(1,N_2)}{27 N_2 (N_2+1)}-\frac{2}{9} S(2,N_2)\Bigg)\ln\bar{N}_2 \nonumber \\
	&-\frac{2}{27} S(3,N_2)-\frac{1}{9} S(-2,-1,N_2)+\frac{1}{9} S(-1,N_2) S(-1,1,N_2)+\frac{1}{9} S(2,1,N_2) \nonumber \\
	&-\frac{1}{9} S(-1,1,-1,N_2)-\frac{7 }{54}\zeta_3-\frac{(-1)^{N_2} (2 N_2+1) S(-1,1,N_2)}{9 N_2 (N_2+1)}\Bigg\} \nonumber \\
	&+\left(\frac{62 \ln 2}{9}-\frac{2 (-1)^{N_2} (2 N_2+1)}{N_2 (N_2+1)}\right) S(-1,1,1,N_2) +\frac{62}{9} S(-1,N_2) S(-1,1,1,N_2)+\frac{25}{18} S(-1,2,-1,N_2) \nonumber \\
	&+\frac{13}{3} S(2,1,1,N_2)+\frac{13}{9} S(-1,1,-1,1,N_2)-\frac{S(-1,N_2)^2 S(1,N_2)}{18 N_2 (N_2+1)} \nonumber \\
	&-\frac{7683 N_2^8+30732 N_2^7+49082 N_2^6+38508 N_2^5+14631 N_2^4+1928 N_2^3-504 N_2^2-372 N_2-120}{432 N_2^4 (N_2+1)^4}.
\end{align}}
The notation $S[a_1,...,a_n,N_2]$ denotes finite harmonic sums as
defined in Ref.~\cite{Blumlein:1998if}, and viewed as analytic
functions of the complex variable $N_2$. The numerical evaluation of the analytic continuation can be carried out using ${\tt ANCONT}$~\cite{Blumlein:2000hw}.

\sect{Transformation of distributions from $(\tau,u)$ to $(x_1,x_2)$ space}
\label{app:distr}

As discussed in Sect.~\ref{sec:nnlody}, the fixed-order NLO and NNLO
Drell-Yan cross section in
Ref.~\cite{Anastasiou:2003yy,Anastasiou:2003ds} is expressed in terms
of variables $\tau$ and $u$, which are related to the variables $x_1$,
$x_2$ relevant for the resummation by
Eqs.~(\ref{eq:uttox1}-\ref{eq:x1x2tou}). Hence the coefficient
function must be rewritten accordingly, using Eq.~(\ref{eq:ctocbar}),
which involves the jacobian of the transformation  
\begin{equation}\label{eq:invjac1}
j(x_1,x_2)=\left|\frac{\partial(\tau,u)}{\partial(x_1,x_2)}\right|=\frac{2x_1x_2(1+x_1x_2)}{(x_1+x_2)^2(1-x_1x_2)},
\end{equation}
which is singular when $\tau=x_1x_2=1$.

The coefficient function in $(\tau,u)$ space involves distributions in the soft limit
$\tau\to1$ and in the two collinear limits $u\to0$ and $u\to1$. In
$(x_1,x_2)$ space instead the soft limit corresponds to both $x_1\to1$
and $x_2\to1$, while the two collinear limits correspond to either
$x_1\to1$ or $x_2\to1$. Transformations of distributions from
$(\tau,u)$ space to $(x_1,x_2)$ space is accordingly nontrivial.

We
have to transform double distributions in $\tau$ and $u$, and single
distributions in $u$ localized at $u=1$ or $u=0$, while single
distributions in $\tau$ are not relevant because $\tau=1$ corresponds
to both $x_1=1$ and $x_2=1$ which can only be realized with $u=0$ or
$u=1$. We discuss these two cases in turn.

\subsection{Double distributions in $\tau $ and $u$}\label{app:doubledist}

Our task is facilitated by the observation that the coefficient
function in $(x_1,x_2)$ in the double-soft limit only depends on the
combination $M^2(1-x_1)(1-x_2)$, as shown in
Sect.~\ref{sec:dsoft}. Recalling Eq.~(\ref{CF}-\ref{CFn}), this is
seen to imply that only distributions generated by differentiation of
a generating function that depend on a power of $(1-x_1)(1-x_2)$ are
relevant. The strategy that we adopt is thus to start with such a
generating function, rewrite in $(\tau,u)$ space, and then work out
the relations between distributions that follow from differentiation
of the equation that expresses the equality of these two forms of
the generating function. These will turn out to be indeed all and only
the double distributions that appear in the $(\tau,u)$ space
coefficient function,

We start by expressing the relevant variable in $(\tau,u)$ space:
\begin{equation}
(1-x_1)(1-x_2)=\frac{u(1-u)(1-\tau)^2}{G(\tau,u)},
\end{equation}
where 
\begin{equation}
G(\tau,u)=\frac{\sqrt{\tau+u(1-u)(1-\tau)^2}(\sqrt{\tau+u(1-u)(1-\tau)^2}+\sqrt{\tau})}{1+\tau}.
\end{equation}
It follows that for a generic test function $T(x_1,x_2)$, regular in the range $0\leq x_1\leq 1,0\leq x_2\leq 1$, 
\begin{align}
&\int_0^1dx_1 \int_0^1dx_2 \, T(x_1,x_2)(1-x_1)^{-1+\epsilon} (1-x_2)^{-1+\epsilon} 
\nonumber\\
&= \int_0^1 d\tau \, \int_0^1 du \, t(\tau,u)\frac{J(\tau,u)G^{1-\epsilon}(\tau,u) }{1-\tau}
(1-\tau)^{-1+2\epsilon}u^{-1+\epsilon}(1-u)^{-1+\epsilon} 
\label{eq:functgen_0}
\end{align}
where $t(\tau,u)=T(x_1(\tau,u),x_2(\tau,u))$ and
\begin{equation}
J(\tau,u)=\left|\frac{\partial(x_1,x_2)}{\partial(\tau,u)}\right|=\frac{1-\tau^2}{2[1-u(1-\tau)][\tau+u(1-\tau)]}
\end{equation}
is the inverse of the jacobian  $j(x_1,x_2)$
Eq.~(\ref{eq:invjac1}). Equation~(\ref{eq:functgen_0}) provides the
expression in $(\tau,u)$ space of the  relevant  $(x_1,x_2)$ space generating function.

However, we wish to express the distributions in $(\tau,u)$ space
that appear in the coefficient function 
in terms of distributions in $(x_1,x_2)$ space, not the other way up.
To this purpose, we note that,
as already mentioned,
when $\tau=1$ it follows that  $x_1=x_2=1$, hence $t(1,u)$ is
$u$-independent, which is why the jacobian $j(x_1,x_2)$
Eq.~(\ref{eq:invjac1}) is singular as $\tau\to1$.
It follows that the function
\begin{equation}
 H(\tau,u,\epsilon) \frac{J(\tau,u)G^{1-\epsilon}(\tau,u) }{1-\tau}
=\frac{(1+x_1)(1+x_2)}{2(1+x_1x_2)}
\left[\frac{x_1x_2(1+x_1)(1+x_2)}{(x_1+x_2)^2}\right]^{-\epsilon}\equiv\frac{1}{F(x_1,x_2,\epsilon)} 
\end{equation}
that appears on the right-hand side of Eq.~(\ref{eq:functgen_0}) has the same property, and is regular at the boundaries of the
 integration range.
 We can thus divide
both sides of Eq.~(\ref{eq:functgen_0})  by
 $H(\tau,u,\epsilon)=\frac{1}{F(x_1,x_2,\epsilon)}$, 
as this just amounts to a redefinition of the test function. We thus get
\begin{align}
&\int_0^1 d\tau \, \int_0^1 du \, t(\tau,u)
(1-\tau)^{-1+2\epsilon}u^{-1+\epsilon}(1-u)^{-1+\epsilon}
\nonumber\\
&=\int_0^1dx_1 \int_0^1dx_2 \, T(x_1,x_2)(1-x_1)^{-1+\epsilon}(1-x_2)^{-1+\epsilon}F(x_1,x_2,\epsilon).
\label{eq:functgen_1}
\end{align}
This is the desired relation between generating functions and it is
our starting point.

We first express 
both sides of  Eq.~(\ref{eq:functgen_1}) in terms of distributions, using the identity 
\begin{equation}
(1-z)^{-1+\epsilon}=
\frac{\delta(1-z)}{\epsilon} + \left[(1-z)^{-1+\epsilon}\right]_+.
\end{equation}
We get
\begin{align}
& \int_0^1 d\tau \, \int_0^1du \, t(\tau,u)
 \left(\frac{\delta(1-\tau)}{2\epsilon} + \left[(1-\tau)^{-1+2\epsilon}\right]_+\right) \nonumber\\ 
&\times \left(\frac{\delta(u)+\delta(1-u)}{\epsilon}+ \left[u^{-1+\epsilon}\right]_+ + \left[(1-u)^{-1+\epsilon}\right]_+ 
+ (u(1-u))^{-1+\epsilon}- u^{-1+\epsilon} - (1-u)^{-1+\epsilon}\right) 
\nonumber\\
&=\int _0^1dx_1 \int _0^1dx_2 \, T(x_1,x_2)
\Bigg\{
\frac{\delta(1-x_1)\delta(1-x_2)}{\epsilon^2}
\nonumber\\
&
+\frac{1}{\epsilon}\delta(1-x_1)\left[(1-x_2)^{-1+\epsilon}\right]_+ F(1,x_2,\epsilon)
+\frac{1}{\epsilon}\delta(1-x_2)\left[(1-x_1)^{-1+\epsilon}\right]_+ F(x_1,1,\epsilon)
\nonumber\\
&+\left[(1-x_1)^{-1+\epsilon}\right]_+\left[(1-x_2)^{-1+\epsilon}\right]_+ F(x_1,x_2,\epsilon)
\Bigg\}
\label{eq:functgen_2}
\end{align}

Equation~\eqref{eq:functgen_2} can be simplified by observing that
for any function or distribution $g(u)$, integrable in the range $0\leq u\leq 1$, we have
\begin{align}
&\int_0^1d\tau\int_0^1du\,\delta(1-\tau)t(\tau,u)g(u)=t(1,0)\int_0^1du\,g(u)
\nonumber\\
&\qquad=\int_0^1dx_1\int_0^1dx_2\,T(x_1,x_2)\delta(1-x_1)\delta(1-x_2)\int_0^1du\,g(u).
\end{align}
In particular,
\begin{align}
&\int_0^1d\tau\int_0^1du\,\delta(1-\tau)t(\tau,u)\frac{\delta(u)+\delta(1-u)}{2}
=\int_0^1dx_1\int_0^1dx_2\,T(x_1,x_2)\delta(1-x_1)\delta(1-x_2)
\nonumber\\
&\int_0^1d\tau\int_0^1du\,\delta(1-\tau)t(\tau,u)\left(\left[u^{-1+\epsilon}\right]_++\left[(1-u)^{-1+\epsilon}\right]_+\right)
=0
\nonumber\\
&\int_0^1d\tau\int_0^1du\,\delta(1-\tau)t(\tau,u)
\left((u(1-u))^{-1+\epsilon}- u^{-1+\epsilon} - (1-u)^{-1+\epsilon}\right) 
\nonumber\\
&\qquad\qquad
=\frac{2}{\epsilon}\left[\frac{\Gamma^2(1+\epsilon)}{\Gamma(1+2\epsilon)}-1\right]\int_0^1dx_1\int_0^1dx_2\,T(x_1,x_2)
.\delta(1-x_1)\delta(1-x_2).
\label{deltaomtau}
\end{align}

Furthermore, the term
\begin{equation}
\int_0^1d\tau\int_0^1du\,\left[\delta(u)+\delta(1-u)\right]t(\tau,u)\left[(1-\tau)^{-1+\epsilon}\right]_+
\end{equation}
can easily be recast as an integral in $x_1,x_2$. Indeed,
\begin{align}
&d\tau du\,\delta(u) =dx_1 dx_2\, j(x_1,x_2)\frac{\delta(1-x_2)}{\left|\frac{\partial u(x_1,x_2)}{\partial x_2}\right|_{x_2=1}}
=dx_1 dx_2\delta(1-x_2)
\\
&d\tau du\,\delta(1-u) =dx_1 dx_2\, j(x_1,x_2)\frac{\delta(1-x_1)}{\left|\frac{\partial u(x_1,x_2)}{\partial x_1}\right|_{x_1=1}}
=dx_1 dx_2\delta(1-x_1).
\label{deltaomu}
\end{align}
Therefore
\begin{align}
&\int_0^1d\tau\int_0^1du\,t(\tau,u)\left[\delta(u)+\delta(1-u)\right]\left[(1-\tau)^{-1+2\epsilon}\right]_+
\nonumber\\
&=
\int_0^1dx_1\int_0^1dx_2\,T(x_1,x_2)\left\{
 \delta(1-x_1)\left[(1-x_2)^{-1+2\epsilon}\right]_+
 +\delta(1-x_2)\left[(1-x_1)^{-1+2\epsilon}\right]_+
\right\}.
\label{deltau}
\end{align}

Substituting Eqs.~\eqref{deltaomtau},~\eqref{deltau}  in Eq.~\eqref{eq:functgen_2}
we find
\begin{align}
&\int_0^1 d\tau \, \int_0^1du \, t(\tau,u)
\left[(1-\tau)^{-1+2\epsilon}\right]_+ 
\nonumber\\ 
& \left( \left[u^{-1+\epsilon}\right]_+ + \left[(1-u)^{-1+\epsilon}\right]_+ 
+ (u(1-u))^{-1+\epsilon}- u^{-1+\epsilon} - (1-u)^{-1+\epsilon}\right)
\nonumber\\
&=\int _0^1dx_1 \int _0^1dx_2 \, T(x_1,x_2)\Bigg\{
\frac{1}{\epsilon^2}\left[1-\frac{\Gamma^2(1+\epsilon)}{\Gamma(1+2\epsilon)}\right]\delta(1-x_1)\delta(1-x_2)
\nonumber\\
&+\frac{1}{\epsilon}\delta(1-x_1)\left(\left[(1-x_2)^{-1+\epsilon}\right]_+F(1,x_2,\epsilon)-\left[(1-x_2)^{-1+2\epsilon}\right]_+\right)
\nonumber\\
&+\frac{1}{\epsilon}\delta(1-x_2)\left(\left[(1-x_1)^{-1+\epsilon}\right]_+F(x_1,1,\epsilon)-\left[(1-x_1)^{-1+2\epsilon}\right]_+\right)
\nonumber\\
&+\left[(1-x_1)^{-1+\epsilon}\right]_+\left[(1-x_2)^{-1+\epsilon}\right]_+ F(x_1,x_2,\epsilon)
\Bigg\}
\label{eq:functgen_3}
\end{align}
This is the final form of the relation between generating functions that we use
in order to generate the identities that we need, which follow by 
expanding both sides of  Eq.~\eqref{eq:functgen_3} in powers of
$\epsilon$ around $\epsilon=0$. The distributions that appear up to
NNLO are found expanding up to order $\epsilon^2$.

For $\epsilon=0$ we get
\begin{align}
&\int_0^1 d\tau \, \int_0^1du \, t(\tau,u)
\left[\frac{1}{1-\tau}\right]_+ \left( \left[\frac{1}{u}\right]_+ + \left[\frac{1}{1-u}\right]_+ \right)
\nonumber\\
&=\int _0^1dx_1 \int _0^1dx_2 \, T(x_1,x_2)\Bigg\{\zeta_2\delta(1-x_1)\delta(1-x_2)
\nonumber\\
&-\delta(1-x_1)\left(\left[\frac{\ln(1-x_2)}{1-x_2}\right]_+-\frac{1}{1-x_2}\ln\frac{2x_2}{1+x_2}\right)
-\delta(1-x_2)\left(\left[\frac{\ln(1-x_1)}{1-x_1}\right]_+-\frac{1}{1-x_1}\ln\frac{2x_1}{1+x_1}\right)
\nonumber\\
&+\left[\frac{1}{1-x_1}\right]_+\left[\frac{1}{1-x_2}\right]_+ +\frac{1}{(1+x_1)(1+x_2)}
\Bigg\}.
\label{ordertwo}
\end{align}
This identity was already obtained in Refs.~\cite{Lustermans:2019cau} and~\cite{Bonvini:2023mfj} by different methods.

The order-$\epsilon$ term of the expansion  gives
\begin{align}
&\int_0^1 d\tau \, \int_0^1du \, t(\tau,u)
\Bigg\{
2\left[\frac{\ln(1-\tau)}{1-\tau}\right]_+\left( \left[\frac{1}{u}\right]_+ + \left[\frac{1}{1-u}\right]_+ \right)
\nonumber\\
&+\left[\frac{1}{1-\tau}\right]_+
\left(
\left[\frac{\ln u}{u}\right]_+ + \left[\frac{\ln(1-u)}{1-u}\right]_++\frac{\ln u}{1-u}+\frac{\ln(1-u)}{u}
\right)\Bigg\}
\nonumber\\
&=\int_0^1dx_1\int_0^1dx_2\,T(x_1,x_2)\Bigg\{-2\zeta_3\delta(1-x_1)\delta(1-x_2)
\nonumber\\
&+\frac{1}{2}\delta(1-x_2)\left(\frac{1}{1-x_1}\ln^2\frac{2x_1}{1+x_1}
+2\frac{\ln(1-x_1)}{1-x_1}\ln\frac{2x_1}{1+x_1}-3\left[\frac{\ln^2(1-x_1)}{1-x_1}\right]_+\right)
\nonumber\\
&+\frac{1}{2}\delta(1-x_1)\left(\frac{1}{1-x_2}\ln^2\frac{2x_2}{1+x_2}
+2\frac{\ln(1-x_2)}{1-x_2}\ln\frac{2x_2}{1+x_2}-3\left[\frac{\ln^2(1-x_2)}{1-x_2}\right]_+\right)
\nonumber\\
&+\frac{2(1+x_1x_2)}{(1+x_1)(1+x_2)}\Bigg(
\left[\frac{\ln(1-x_1)}{1-x_1}\right]_+\left[\frac{1}{1-x_2}\right]_+
+\left[\frac{1}{1-x_1}\right]_+\left[\frac{\ln(1-x_2)}{1-x_2}\right]_+
\nonumber\\
&+\left[\frac{1}{1-x_1}\right]_+\left[\frac{1}{1-x_2}\right]_+\ln\frac{x_1x_2(1+x_1)(1+x_2)}{(x_1+x_2)^2}
\Bigg)
\Bigg\}.
\label{orderthree}
\end{align}

The order-$\epsilon^2$ leads to
\begin{align}
&\int_0^1 d\tau \, \int_0^1du \, t(\tau,u)
\Bigg\{
2\left[\frac{\ln^2(1-\tau)}{1-\tau}\right]_+\left( \left[\frac{1}{u}\right]_+ + \left[\frac{1}{1-u}\right]_+ \right)
\nonumber\\
&
+2\left[\frac{\ln(1-\tau)}{1-\tau}\right]_+
\left( \left[\frac{\ln u}{u}\right]_+ + \left[\frac{\ln(1-u)}{1-u}\right]_+ 
+\frac{\ln u}{1-u}+\frac{\ln(1-u)}{u}\right)
\nonumber\\
&+\frac{1}{2}\left[\frac{1}{1-\tau}\right]_+
\left(
\left[\frac{\ln^2u}{u}\right]_+ + \left[\frac{\ln^2(1-u)}{1-u}\right]_+
+\frac{\ln^2 u}{1-u}+\frac{\ln^2(1-u)}{u}+\frac{2\ln u\ln(1-u)}{u(1-u)}
\right)\Bigg\}
\nonumber\\
&=\int_0^1dx_1\int_0^1dx_2\,T(x_1,x_2)\Bigg\{\frac{9}{4}\zeta_4\delta(1-x_1)\delta(1-x_2)
\nonumber\\
&+\frac{\delta(1-x_2)}{6}\left(
\frac{1}{1-x_1}\ln^3\frac{2x_1}{1+x_1}
+3\frac{\ln(1-x_1)}{1-x_1}\ln^2\frac{2x_1}{1+x_1}
+3\frac{\ln^2(1-x_1)}{1-x_1}\ln\frac{2x_1}{1+x_1}
-7\left[\frac{\ln^3(1-x_1)}{1-x_1}\right]_+
\right)
\nonumber\\
&+\frac{\delta(1-x_1)}{6}\left(\frac{1}{1-x_2}\ln^3\frac{2x_2}{1+x_2}
+3\frac{\ln(1-x_2)}{1-x_2}\ln^2\frac{2x_2}{1+x_2}
+3\frac{\ln^2(1-x_2)}{1-x_2}\ln\frac{2x_2}{1+x_2}
-7\left[\frac{\ln^3(1-x_2)}{1-x_2}\right]_+
\right)
\nonumber\\
&+\frac{1}{2}\frac{2(1+x_1x_2)}{(1+x_1)(1+x_2)}
\nonumber\\
&\Bigg(
\left[\frac{\ln^2(1-x_1)}{1-x_1}\right]_+\left[\frac{1}{1-x_2}\right]_+
+2\left[\frac{\ln(1-x_1)}{1-x_1}\right]_+\left[\frac{\ln(1-x_2)}{1-x_2}\right]_+
+\left[\frac{1}{1-x_1}\right]_+\left[\frac{\ln^2(1-x_2)}{1-x_2}\right]_+
\nonumber\\
&
+2\left(\left[\frac{\ln(1-x_1)}{1-x_1}\right]_+\left[\frac{1}{1-x_2}\right]_+
+\left[\frac{1}{1-x_1}\right]_+\left[\frac{\ln(1-x_2)}{1-x_2}\right]_+\right)\ln\frac{x_1x_2(1+x_1)(1+x_2)}{(x_1+x_2)^2}
\nonumber\\
&+\left[\frac{1}{1-x_1}\right]_+\left[\frac{1}{1-x_2}\right]_+\ln^2\frac{x_1x_2(1+x_1)(1+x_2)}{(x_1+x_2)^2}
\Bigg)
\Bigg\}.
\label{orderfour}
\end{align}

Note that some single distributions in $\tau$ multiplied by ordinary
functions of $\tau$ and $u$  do appear on the left-hand side of
Eqs.~(\ref{ordertwo}-\ref{orderfour}). However, we can conclude that no other single distributions in $\tau$ other than these can appear in the coefficient function, because, as already
discussed, the limit $\tau\to1$ corresponds to the double-soft limit,
in which the coefficient function only depends on the variable
$(1-x_1)(1-x_2)$ which appears in the generating function
Eq.~(\ref{eq:functgen_3}). This expectation is borne out by comparing
to the explicit form of the NNLO coefficient function Eq.~(\ref{eq:fctudecomp}).

\subsection{Single distributions in $u$}\label{app:singledist}

We now turn to single distributions  in  $u$. A relevant
observation here is that the coefficient function is necessarily
symmetric upon the interchange of $u$ and $1-u$. Hence these
distribution always appears in pairs. This said, all single
distributions in $u$ can be transformed into distributions in $x_1$
and $x_2$ by simply performing the change of variables with the
jacobian Eq.~(\ref{eq:invjac1}).

We illustrate the procedure with a simple example.
Consider
\begin{equation}
\int_0^1d\tau\int_0^1du\,\left[\frac{1}{u}\right]_+t(\tau,u)=\int_0^1d\tau\int_0^1du\,\frac{1}{u}[t(\tau,u)-t(\tau,0)].
\end{equation}
Changing integration variables to $x_1,x_2$ we get
\begin{align}
&\int_0^1d\tau\int_0^1du\,\left[\frac{1}{u}\right]_+t(\tau,u)
\nonumber\\
&=\int_0^1dx_1\int_0^1dx_2\,\frac{2x_1x_2(1+x_1x_2)}{(x_1+x_2)^2(1-x_1x_2)}
\frac{(x_1+x_2)(1-x_1x_2)}{x_2(1-x_1^2)}\left[T(x_1,x_2)-T(1,x_2)\right]
\nonumber\\
&=\int_0^1dx_2\int_0^1dx_1\,
\frac{h(x_1,x_2)}{1-x_1}\left[T(x_1,x_2)-T(1,x_2)\right],
\label{ut}
\end{align}
where 
\begin{equation}
h(x_1,x_2)=\frac{2x_1(1+x_1x_2)}{(x_1+x_2)(1+x_1)}
\end{equation}
is non-singular at $x_1=1$ for all values of $x_2$.

Equation~\eqref{ut} can be written as
\begin{align}
&\int_0^1d\tau\int_0^1du\,\left[\frac{1}{u}\right]_+t(\tau,u)
\nonumber\\
&=\int_0^1dx_2\int_0^1dx_1\,\frac{1}{1-x_1}\left\{\left[h(x_1,x_2)T(x_1,x_2)-h(1,x_2)T(1,x_2)\right]
-T(1,x_2)\left[h(x_1,x_2)-h(1,x_2)\right]\right\}
\nonumber\\
&=\int_0^1dx_2\int_0^1dx_1\,
\left\{h(x_1,x_2)\left[\frac{1}{1-x_1}\right]_+-\delta(1-x_1)\int_0^1dx\,\frac{h(x,x_2)-h(1,x_2)}{1-x}\right\}T(x_1,x_2).
\end{align}
The same argument is easily extended to distributions containing powers of $\ln u$, and to distributions in $1-u$.

\subsection{Distributional identities up to
NNLO}\label{sec:distresults}
We now list all distributional identities that we have derived using
the methods  described above, and that we have uses in order to  transform the
coefficient function from $(\tau,u)$ space to $(x_1,x_2)$ space.
All identities written below should be understood with the left-hand
side multiplied by a measure of integration $d\tau\,du$, and the
right-hand side multiplied by  an integration measure $dx_1\,dx_2$.
\begin{align}
    &\delta(1-\tau) \frac{\delta(u)+\delta(1-u)}{2} = \delta(1-x_1) \delta(1-x_2) \\
    &\left[\frac{1}{1-\tau}\right]_+ \left(\delta(u)+\delta(1-u)\right) = \delta(1-x_1)\left[\frac{1}{1-x_2}\right]_+ + \delta(1-x_2) \left[\frac{1}{1-x_1}\right]_+\\
    &\left[\frac{1}{1-\tau}\right]_+ \left(\left[\frac{1}{u}\right]_+ + \left[\frac{1}{1-u}\right]_+\right) = \left[\frac{1}{1-x_1}\right]_+ \left[\frac{1}{1-x_2}\right]_+ - \delta(1-x_1) \left(\left[\frac{\ln(1-x_2)}{1-x_2}\right]_++ \frac{\ln\frac{2x_2}{1+x_2}}{1-x_2}\right) \nonumber \\
    &-\delta(1-x_2) \left(\left[\frac{\ln(1-x_1)}{1-x_1}\right]_++ \frac{\ln\frac{2x_1}{1+x_1}}{1-x_1}\right) +\frac{\pi^2}{6} \delta(1-x_1) \delta(1-x_2) + \frac{1}{(1+x_1)(1+x_2)} \\
    &2 \left[\frac{\ln(1-\tau)}{1-\tau}\right]_+ \left(\left[\frac{1}{u}\right]_+ + \left[\frac{1}{1-u}\right]_+\right) + \left[\frac{1}{1-\tau}\right]_+ \left(\left[\frac{\ln u}{u}\right]_+ + \left[\frac{\ln(1-u)}{1-u}\right]_+ + \frac{\ln u}{1-u} + \frac{\ln(1-u)}{u}\right) \nonumber \\
    &=\left[\frac{1}{1-x_1}\right]_+ \left(\left[\frac{\ln(1-x_2)}{1-x_2}\right]_+
    + \frac{\ln\frac{2x_2}{1+x_2}}{1-x_2}\right)
    + \left[\frac{1}{1-x_2}\right]_+ \left(\left[\frac{\ln(1-x_1)}{1-x_1}\right]_+
    + \frac{\ln\frac{2x_1}{1+x_1}}{1-x_1}\right) \nonumber \\    
    &+ \delta(1-x_1) \left(-\frac{3}{2}\left[\frac{\ln^2(1-x_2)}{1-x_2}\right]_+ + \frac{1}{2} \ln \frac{2x_2(1-x_2)^2}{1+x_2} \frac{\ln\frac{2x_2}{1+x_2}}{1-x_2} \right) \nonumber \\
    &+ \delta(1-x_2) \left(-\frac{3}{2}\left[\frac{\ln^2(1-x_1)}{1-x_1}\right]_+ + \frac{1}{2} \ln\frac{2x_1(1-x_1)^2}{1+x_1}
    \frac{\ln\frac{2x_1}{1+x_1}}{1-x_1} \right) \nonumber \\
    &-2\zeta_3 \delta(1-x_1) \delta(1-x_2) + \frac{1}{(1+x_1)(1+x_2)} \ln \frac{x_1 x_2 (1+x_1)(1+x_2)(1-x_1)(1-x_2)}{(x_1+x_2)^2} \nonumber \\
    &+ \frac{2}{(1-x_1)(1-x_2)} \ln\frac{(1+x_1)(1+x_2)}{2(x_1+x_2)}
\end{align}
\begin{align}
    &2 \left[\frac{\ln^2(1-\tau)}{1-\tau}\right]_+ \left(\left[\frac{1}{u}\right]_+ + \left[\frac{1}{1-u}\right]_+\right) \nonumber \\
    &+2 \left[\frac{\ln(1-\tau)}{1-\tau}\right]_+ \left( \left[\frac{\ln u}{u}\right]_+ + \left[\frac{\ln(1-u)}{1-u}\right]_+ + \frac{\ln u}{1-u} + \frac{\ln(1-u)}{u} \right) \nonumber \\
    &+\frac{1}{2} \left[\frac{1}{1-\tau}\right]_+ \left(\left[\frac{\ln^2 u}{u}\right]_+ + \left[\frac{\ln^2(1-u)}{1-u}\right]_+ + \frac{\ln^2u}{1-u} + \frac{\ln^2(1-u)}{u} + 2\frac{ \ln u \ln(1-u)}{u(1-u)}\right) \nonumber \\
    &=\left[\frac{\ln(1-x_1)}{1-x_1}\right]_+ \left[\frac{\ln(1-x_2)}{1-x_2}\right]_+ + \frac{1}{2} \left[\frac{1}{1-x_1}\right]_+ \left[\frac{\ln^2(1-x_2)}{1-x_2}\right]_+ + \frac{1}{2} \left[\frac{1}{1-x_2}\right]_+ \left[\frac{\ln^2(1-x_1)}{1-x_1}\right]_+ \nonumber \\
    &+ \left[\frac{\ln(1-x_1)}{1-x_1}\right]_+ \frac{\ln\frac{2x_2}{1+x_2}}{1-x_2} 
    + \left[\frac{\ln(1-x_2)}{1-x_2}\right]_+ \frac{\ln\frac{2x_1}{1+x_1}}{1-x_1} \nonumber \\
    &+ \frac{1}{2} \left[\frac{1}{1-x_1}\right]_+ \ln\frac{2x_2 (1-x_2)^2}{1+x_2}\frac{\ln\frac{2x_2}{1+x_2}}{1-x_2} + \frac{1}{2} \left[\frac{1}{1-x_2}\right]_+ \ln\frac{2x_1 (1-x_1)^2}{1+x_1}\frac{\ln\frac{2x_1}{1+x_1}}{1-x_1} \nonumber \\
    &+\delta(1-x_1) \left(-\frac{7}{6} \left[\frac{\ln^3(1-x_2)}{1-x_2}\right]_+ + \left(\frac{1}{2} \ln\frac{2x_2(1-x_2)}{1+x_2}\ln(1-x_2) 
    + \frac{1}{6} \ln^2\frac{2x_2}{1+x_2}\right)\frac{\ln\frac{2x_2}{1+x_2}}{1-x_2}\right) \nonumber \\
    &+\delta(1-x_2) \left(-\frac{7}{6} \left[\frac{\ln^3(1-x_1)}{1-x_1}\right]_+ + \left(\frac{1}{2} \ln\frac{2x_1(1-x_1)}{1+x_1}\ln(1-x_1) 
    + \frac{1}{6} \ln^2 \frac{2x_1}{1+x_1}\right)\frac{\ln\frac{2x_1}{1+x_1}}{1-x_1}\right) \nonumber \\
    &+\frac{\pi^4}{40} \delta(1-x_1) \delta(1-x_2) + \frac{1}{2(1+x_1)(1+x_2)} \ln\frac{x_1 x_2 (1+x_1)(1+x_2)(1-x_1)(1-x_2)}{(x_1+x_2)^2}\nonumber \\
    &+\frac{\ln\frac{2x_1}{1+x_1}}{1-x_1}\frac{\ln\frac{2x_2}{1+x_2}}{1-x_2} + 2 \frac{\ln(1-x_1)+\ln(1-x_2)}{(1-x_1)(1-x_2)} 
    \ln\frac{(1+x_1)(1+x_2)}{2(x_1+x_2)}\nonumber \\
    &+ 2 \frac{1}{(1-x_1)(1-x_2)} \ln \frac{2x_1 x_2}{x_1+x_2} \ln \frac{(1+x_1)(1+x_2)}{2(x_1+x_2)}
\end{align}

\begin{align}
    &\delta(u) + \delta(1-u)=\delta(1-x_1) + \delta(1-x_2) \\
    &\left[\frac{1}{u}\right]_+ + \left[\frac{1}{1-u}\right]_+ = \left[\frac{1}{1-x_1}\right]_+ + \left[\frac{1}{1-x_2}\right]_+ 
    + \delta(1-x_1) \ln\frac{2x_2}{(1+x_2)(1-x_2)} \nonumber \\
    &+ \delta(1-x_2) \ln\frac{2x_1}{(1+x_1)(1-x_1)} - \frac{x_1+x_2+2 x_1 x_2}{(1+x_1)(1+x_2)} \\
    &\left[\frac{\ln u}{u}\right]_+ + \left[\frac{\ln(1-u)}{1-u}\right]_+ = \left[\frac{\ln(1-x_1)}{1-x_1}\right]_+ + \left[\frac{\ln(1-x_2)}{1-x_2}\right]_+ + \left[\frac{1}{1-x_1}\right]_+ \ln\frac{2x_2}{(1+x_2)(1-x_2)}\nonumber \\
    &+ \left[\frac{1}{1-x_2}\right]_+ \ln \frac{2x_1}{(1+x_1)(1-x_1)}+ \delta(1-x_1) \frac{1}{2} \ln^2\frac{2x_2}{(1+x_2)(1-x_2)}\nonumber \\
    &+ \delta(1-x_2) \frac{1}{2} \ln^2 \frac{2x_1}{(1+x_1)(1-x_1)} + \frac{\ln\frac{1-x_1}{1-x_1 x_2}}{1-x_2} + \frac{\ln\frac{1-x_2}{1-x_1 x_2}}{1-x_1} \nonumber \\
    &+\frac{2x_1(1+x_1 x_2)}{(x_1+x_2)(1+x_1)} \frac{\ln\frac{(1+x_1)(1+x_2)}{2(1+x_1 x_2)}}{1-x_1}+\frac{2x_2(1+x_1 x_2)}{(x_1+x_2)(1+x_2)} \frac{\ln\frac{(1+x_1)(1+x_2)}{2(1+x_1 x_2)}}{1-x_2} \nonumber \\
    &+ \frac{x_1-x_2 -2 x_1 x_2}{(1+x_1)(x_1+x_2)}\ln \frac{2x_2(1-x_1)}{(1+x_2)(1-x_1 x_2)} + \frac{x_2-x_1 -2 x_1 x_2}{(1+x_2)(x_1+x_2)}\ln\frac{2x_1(1-x_2)}{(1+x_1)(1-x_1 x_2)} \\
    &\left[\frac{\ln^2 u}{u}\right]_+ + \left[\frac{\ln^2(1-u)}{1-u}\right]_+ = \left[\frac{\ln^2(1-x_1)}{1-x_1}\right]_+ + \left[\frac{\ln^2(1-x_2)}{1-x_2}\right]_+ \nonumber \\
    &+ 2 \left[\frac{\ln(1-x_1)}{1-x_1}\right]_+ \ln\frac{2x_2}{(1+x_2)(1-x_2)}+ 2 \left[\frac{\ln(1-x_2)}{1-x_2}\right]_+ 
    \ln \frac{2x_1}{(1+x_1)(1-x_1)} \nonumber \\
    &+\left[\frac{1}{1-x_1}\right]_+ \ln^2 \frac{2x_2}{(1+x_2)(1-x_2)} +\left[\frac{1}{1-x_2}\right]_+ \ln^2 \frac{2x_1}{(1+x_1)(1-x_1)} \nonumber \\
    &+ \delta(1-x_1) \frac{1}{3} \ln^3 \frac{2x_2}{(1+x_2)(1-x_2)} + \delta(1-x_2) \frac{1}{3} \ln^3 \frac{2x_1}{(1+x_1)(1-x_1)} \nonumber \\
    &+\frac{\ln^2(1-x_1 x_2)-\ln^2(1-x_2)}{1-x_1}+\frac{\ln^2(1-x_1 x_2)-\ln^2(1-x_1)}{1-x_2} \nonumber \\
    &+ 2 \ln\frac{2x_2(1-x_1)}{1+x_2} \frac{\ln\frac{1-x_2}{1-x_1 x_2}}{1-x_1} + 2 \ln \frac{2x_1(1-x_2)}{1+x_1}\frac{\ln\frac{1-x_1}{1-x_1 x_2}}{1-x_2} \nonumber \\
    &+\frac{2x_1(1+x_1 x_2)}{(1+x_1)(x_1+x_2)}\ln \frac{2x_2^2(1+x_1)(1-x_1)^2}{(1+x_2)(x_1+x_2)(1-x_1 x_2)^2}
    \frac{\ln \frac{2(x_1+x_2)}{(1+x_1)(1+x_2)}}{1-x_1} \nonumber \\
    &+\frac{2x_2(1+x_1 x_2)}{(1+x_2)(x_1+x_2)}\ln\frac{2x_1^2(1+x_2)(1-x_2)^2}{(1+x_1)(x_1+x_2)(1-x_1 x_2)^2}
    \frac{\ln \frac{2(x_1+x_2)}{(1+x_2)(1+x_1)}}{1-x_2} \nonumber \\
    &+\frac{x_1-x_2-2x_1x_2}{(x_1+x_2)(1+x_1)} \ln^2 \frac{2x_2(1-x_1)}{(1+x_2)(1-x_1 x_2)}+\frac{x_2-x_1-2x_1x_2}{(x_1+x_2)(1+x_2)} \ln^2 \frac{2x_1(1-x_2)}{(1+x_1)(1-x_1 x_2)}
\end{align}

\sect{Distributions and anomalous dimensions in SCET}

We collect here some definitions and results used in the SCET
resummation discussed in Sect.~\ref{sec_dQCD_vs_SCET}. First, we
derive some distributional identities; then we list several anomalous
dimensions that enter the renormalization-group equations whose
solution leads to resummation in SCET.

\subsection{Distributional identities}
\label{appendix_SCET_defs}
In Ref.~\cite{Lustermans:2019cau}, the distributions that contain soft logs are written in
terms of a variable $w$, which vanishes in the soft limit:
\begin{equation}\label{eq:ldef}
    \calL_n(w)  = \left[\frac{\ln^n w}{w}\right]_+ \,,
    \qquad n \ge 0 \,.
\end{equation}
Their action over a space of  regular test functions $f(w)$ is defined by
\begin{equation}
    \int_0^1dw\, \calL_n(w)f(w)=\int_0^1dw\,\frac{\ln^n w}{w} \big[ f(w)-f(0)\big] \,.
\end{equation}
The distributions Eq.~\eqref{eq:ldef} can be obtained by differentiation of a generating function:
\begin{equation}
    \calL_n(w) = \left[\dv[n]{}{\eta} \calL^\eta(w)\right]_{\eta = 0} \,,
\end{equation}
where
\begin{equation}
    \label{eq:letadef}
    \calL^\eta(w)  = \left[\frac{1}{w^{1-\eta}}\right]_+ 
\end{equation}
and
\begin{equation}
\label{eq:letadef2}
    \int_0^1dw\, \calL_n(w)f(w)=\int_0^1dw\,w^{-1+\eta} \big[ f(w)-f(0)\big].
\end{equation}
This is useful because, for $\eta>-1$, the two integrals in Eq.~\eqref{eq:letadef2} are separately convergent.

Upon rescaling  of its argument  by a positive constant $\lambda >
0$, the generating function Eq.~\eqref{eq:letadef} transforms according to
\begin{equation}
\label{eq:letaresc}
    \lambda \calL^\eta(\lambda x) = \lambda^\eta \calL^\eta(x) + \frac{\lambda^\eta - 1}{\eta} \delta(x) \,.
\end{equation}
Indeed, for any regular function $f(\lambda x)$,
\begin{align}
    \int_0^1 dx\,\lambda\calL^\eta(\lambda x) f(\lambda x)
    & =\int_0^1d(\lambda x)\,(\lambda x)^{-1+\eta}f(\lambda x)-f(0)\int_0^1 d(\lambda x)\,(\lambda x)^{-1+\eta} \nonumber \\
    & = \lambda^\eta \int_0^1dx\, x^{-1+\eta}f(\lambda x)-\frac{1}{\eta}f(0) \,,
\end{align}
from which we obtain
\begin{align}
    \int_0^1dx\,\left[\lambda^\eta \calL^\eta(x) + \frac{\lambda^\eta - 1}{\eta} \delta(x)\right]f(\lambda x)
    & = \lambda^\eta \int_0^1dx\,x^{-1+\eta}f(\lambda x)-f(0)\frac{\lambda^\eta}{\eta}+ \frac{\lambda^\eta - 1}{\eta} f(0)
    \nonumber\\
    & = \lambda^\eta \int_0^1dx\,x^{-1+\eta}f(\lambda x)-\frac{1}{\eta} f(0) \,.
\end{align}
As a consequence,
\begin{equation}\label{eq:lresc}
    \lambda \calL_n(\lambda x) = \sum_{k=0}^{n} \binom{n}{k} \ln^k\lambda\, \calL_{n-k}(x) + \frac{\ln^{n+1}\lambda}{n+1} \delta(x) \,.
\end{equation}
Accordingly, a scale-dependent distribution is also defined:  
\begin{equation}\label{eq:lscal}
    \hat\calL^{\eta}\left(t, \mu^2\right) 
    \equiv
    \frac{1}{\mu^2} \, \calL^{\eta} \bigg(\frac{t}{\mu^2}\bigg) \,.
\end{equation}

\subsection{Anomalous dimensions}
\label{app_subsec_anom_dim}

We list the anomalous dimensions used in Sect.~\ref{sec_dQCD_vs_SCET} up to the order required to achieve NNLL resummation.  
These anomalous dimensions are expanded in powers of $\frac{\as}{4\pi}$:
\begin{equation}
    \gamma(\as)
    =  \gamma_0 \left(\frac{\as}{4\pi}\right)+ \gamma_1 \left(\frac{\as}{4\pi}\right)^2+\dots .
\end{equation}
The first expansion coefficients of the quark anomalous dimension $\gamma_H^q$ of the hard function, appearing in \eq\eqref{eq_UH_as_BN}, are given by~\cite{Stewart:2010qs}  
\begin{align}
    \label{eq_gamma_H_q}
    \gamma_{H,0}^q & = - 6 \Cf 
    \\ \label{eq_gamma_H_q1}
    \gamma_{H,1}^q & = - \Cf \bigg[\bigg(\frac{82}{9} - 52 \zeta_3\bigg)\Ca + (3-4\pi^2 + 48 \zeta_3) \Cf + \bigg(\frac{65}{9} + \pi^2\bigg) (4\pi\beta_0)\bigg],
\end{align}
with the beta-function coefficients defined as in
Eqs.~(\ref{eq_beta_0}-\ref{eq_beta_1}).

The expansion coefficients of the anomalous dimension of the quark
beam function $\gamma_B^q$ (see
Eqs.~(\ref{eq_TK_tilde_B_solution_of_RGE}--\ref{eq_KB_def})) are
\begin{align}
    \label{eq_gamma_B_q}
    \gamma_{B,0}^q & = 6 \Cf \\ \label{eq_gamma_B_q1}
    \gamma_{B,1}^q & = \Cf \bigg[\bigg(\frac{146}{9} - 80\zeta_3\bigg)\Ca + (3-4\pi^2 + 48 \zeta_3) \Cf + \left(\frac{121}{9} + \frac{2\pi^2}{3}\right) (4\pi\beta_0) \bigg]
\end{align}
while the corresponding coefficients of
the anomalous dimension $\gamma_\phi^q$ are
\begin{align}
    \label{eq_gamma_phi_q}
    \gamma_{\phi,0}^q & = 3 \Cf \,, \\
    \gamma_{\phi,1}^q & = \Cf^2 \left(\frac{3}{2} - 2\pi^2 + 24 \zeta_3\right) + \Cf\Ca \left(\frac{17}{6} + \frac{22\pi^2}{9} - 12\zeta_3\right) - \Cf \TR\nf \left(\frac{2}{3} + \frac{8\pi^2}{9}\right) .
    \label{eq_gamma_phi_q_1}
\end{align}

Using these results we find that the expansion coefficients of the
anomalous dimension $\gamma_W^q$ of \eq\eqref{eq_gamma_W_j_def}
\begin{equation}
    \gamma_W^q(\as) = \gamma_H^q(\as) + 2\gamma_\phi^q(\as)
\end{equation}
are
given by~\cite{Catani:2001ic, Becher:2007ty}
\begin{align}
    \gamma_{W,0}^q & = 0 \\
    \gamma_{W,1}^q & = \Ca \Cf \bigg(-\frac{808}{27} + \frac{11 \pi^2}{9} + 28\zeta_3 \bigg) + \Cf \TR \nf \bigg(\frac{224}{27} - \frac{4\pi^2}{9}\bigg). 
    \label{eq_gamma_W_j}
\end{align}

\bibliographystyle{UTPstyle}
\bibliography{rapres}

\end{document}